\documentclass[twocolumn]{aastex631}
\received{May 27, 2024}
\revised{July 23, 2024}
\accepted{July 29, 2024}
\submitjournal{ApJ}

\shorttitle{Revising the $z\sim 5$ AGN Luminosity Function}
\shortauthors{Grazian et al.}
\graphicspath{{./}{figures/}}

\begin{document}

\title{What are the Pillars of Reionization? Revising the AGN Luminosity Function at $z\sim 5$}

\correspondingauthor{Andrea Grazian}
\email{andrea.grazian@inaf.it}

\author[0000-0002-5688-0663]{Andrea Grazian}
\affil{INAF--Osservatorio Astronomico di Padova, 
Vicolo dell'Osservatorio 5, I-35122, Padova, Italy}

\author[0000-0003-0734-1273]{Emanuele Giallongo}
\affil{INAF--Osservatorio Astronomico di Roma, Via Frascati 33, I-00078,
Monte Porzio Catone, Italy}

\author[0000-0003-4432-5037]{Konstantina Boutsia}
\affil{Cerro Tololo Inter-American Observatory/NSFs NOIRLab, Casilla 603, La Serena, Chile}

\author[0000-0002-2115-5234]{Stefano Cristiani}
\affil{INAF--Osservatorio Astronomico di Trieste, 
Via G.B. Tiepolo, 11, I-34143, Trieste, Italy}
\affiliation{INFN-National Institute for Nuclear Physics,  
via Valerio 2, I-34127 Trieste}
\affil{IFPU--Institute for Fundamental Physics of the Universe,
via Beirut 2, I-34151, Trieste, Italy}

\author[0000-0003-4744-0188]{Fabio Fontanot}
\affil{INAF--Osservatorio Astronomico di Trieste, 
Via G.B. Tiepolo, 11, I-34143, Trieste, Italy} 
\affil{IFPU--Institute for Fundamental Physics of the Universe,
via Beirut 2, I-34151, Trieste, Italy}

\author[0000-0002-4314-021X]{Manuela Bischetti}
\affil{Dipartimento di Fisica, Sezione di Astronomia,
Universit\`a di Trieste, via G.B. Tiepolo 11, I-34131, Trieste, Italy}
\affil{INAF--Osservatorio Astronomico di Trieste, 
Via G.B. Tiepolo, 11, I-34143, Trieste, Italy} 

\author[0000-0003-0492-4924]{Laura Bisigello}
\affiliation{INAF--Istituto di Radioastronomia, Via Piero Gobetti 101, I-40129 Bologna, Italy}

\author[0000-0002-0101-6624]{Angela Bongiorno}
\affil{INAF--Osservatorio Astronomico di Roma, Via Frascati 33, I-00078,
Monte Porzio Catone, Italy}

\author[0000-0002-7738-5389]{Giorgio Calderone}
\affil{INAF--Osservatorio Astronomico di Trieste, 
Via G.B. Tiepolo, 11, I-34143, Trieste, Italy}

\author[0009-0001-7950-2892]{Francesco Chiti Tegli}
\affiliation{Universit\`a degli Studi di Firenze, Via G. Sansone, 1, I-50019 Sesto Fiorentino, Italy}

\author[0000-0002-6830-9093]{Guido Cupani}
\affil{INAF--Osservatorio Astronomico di Trieste, 
Via G.B. Tiepolo, 11, I-34143, Trieste, Italy}
\affil{IFPU--Institute for Fundamental Physics of the Universe,
via Beirut 2, I-34151, Trieste, Italy}

\author[0000-0002-6220-9104]{Gabriella De Lucia}
\affil{INAF--Osservatorio Astronomico di Trieste, 
Via G.B. Tiepolo, 11, I-34143, Trieste, Italy}
\affil{IFPU--Institute for Fundamental Physics of the Universe,
via Beirut 2, I-34151, Trieste, Italy}

\author[0000-0003-3693-3091]{Valentina D'Odorico}
\affil{INAF--Osservatorio Astronomico di Trieste, 
Via G.B. Tiepolo, 11, I-34143, Trieste, Italy} 
\affil{IFPU--Institute for Fundamental Physics of the Universe,
via Beirut 2, I-34151, Trieste, Italy}
\affil{Scuola Normale Superiore, P.zza dei Cavalieri, 7 I-56126 Pisa, Italy}

\author[0000-0002-4227-6035]{Chiara Feruglio}
\affil{INAF--Osservatorio Astronomico di Trieste, 
Via G.B. Tiepolo, 11, I-34143, Trieste, Italy} 
\affil{IFPU--Institute for Fundamental Physics of the Universe,
via Beirut 2, I-34151, Trieste, Italy}

\author[0000-0002-4031-4157]{Fabrizio Fiore}
\affil{INAF--Osservatorio Astronomico di Trieste, 
Via G.B. Tiepolo, 11, I-34143, Trieste, Italy} 
\affil{IFPU--Institute for Fundamental Physics of the Universe,
via Beirut 2, I-34151, Trieste, Italy}

\author[0000-0003-3248-5666]{Giovanni Gandolfi}
\affiliation{Department of Physics and Astronomy, Universit\`a degli Studi di Padova, Vicolo
dell’Osservatorio 3, I-35122, Padova, Italy}
\affil{INAF--Osservatorio Astronomico di Padova, 
Vicolo dell'Osservatorio 5, I-35122, Padova, Italy}

\author[0009-0005-6156-4066]{Giorgia Girardi}
\affiliation{Department of Physics and Astronomy, Universit\`a degli Studi di Padova, Vicolo
dell’Osservatorio 3, I-35122, Padova, Italy}
\affil{INAF--Osservatorio Astronomico di Padova, 
Vicolo dell'Osservatorio 5, I-35122, Padova, Italy}

\author[0000-0003-4740-9762]{Francesco Guarneri}
\affil{Hamburger Sternwarte, Universitat Hamburg, Gojenbergsweg 112, D-21029 Hamburg, Germany}
\affil{INAF--Osservatorio Astronomico di Trieste, 
Via G.B. Tiepolo, 11, I-34143, Trieste, Italy}

\author[0000-0002-3301-3321]{Michaela Hirschmann}
\affiliation{Institute for Physics, Laboratory for Galaxy Evolution, EPFL, Observatoire de
Sauverny, Chemin Pegasi 51, 1290 Versoix, Switzerland}
\affil{INAF--Osservatorio Astronomico di Trieste, 
Via G.B. Tiepolo, 11, I-34143, Trieste, Italy}

\author[0009-0004-2597-6146]{Matteo Porru}
\affil{Dipartimento di Fisica, Sezione di Astronomia, Universit\`a di Trieste,
via G.B. Tiepolo 11, I-34131, Trieste, Italy}
\affil{INAF--Osservatorio Astronomico di Trieste, 
Via G.B. Tiepolo, 11, I-34143, Trieste, Italy} 

\author[0000-0002-9415-2296]{Giulia Rodighiero}
\affiliation{Department of Physics and Astronomy, Universit\`a degli Studi di Padova, Vicolo 
dell’Osservatorio 3, I-35122, Padova, Italy}
\affil{INAF--Osservatorio Astronomico di Padova, 
Vicolo dell'Osservatorio 5, I-35122, Padova, Italy}

\author[0000-0003-1174-6978]{Ivano Saccheo}
\affil{Dipartimento di Matematica e Fisica, Universit\`a Roma Tre,
Via della Vasca Navale 84, I-00146, Roma, Italy}
\affil{INAF--Osservatorio Astronomico di Roma, Via Frascati 33, I-00078,
Monte Porzio Catone, Italy}

\author[0000-0002-8830-1259]{Matteo Simioni}
\affil{INAF--Osservatorio Astronomico di Padova, 
Vicolo dell'Osservatorio 5, I-35122, Padova, Italy}

\author[0000-0002-5959-5964]{Andrea Trost}
\affil{Dipartimento di Fisica, Sezione di Astronomia, Universit\`a di Trieste,
via G.B. Tiepolo 11, I-34131, Trieste, Italy}
\affil{INAF--Osservatorio Astronomico di Trieste, 
Via G.B. Tiepolo, 11, I-34143, Trieste, Italy} 

\author[0000-0001-9383-786X]{Akke Viitanen}
\affil{INAF--Osservatorio Astronomico di Roma, Via Frascati 33, I-00078,
Monte Porzio Catone, Italy}

\begin{abstract}
In the past, high-z AGNs were given a minor role as possible
drivers of reionization, despite initial evidences in favor
of their large space densities at low luminosities by Chandra and
HST. Recent observations from JWST are finding relatively large
numbers of faint AGNs at $z>4$, convincingly confirming these early
results. We present a sample of $z\sim 5$ AGNs (both from wide,
shallow ground-based surveys and from deep, pencil-beam observations
from JWST), allowing to estimate their space densities with
unprecedented accuracy. The bright end ($M_{1450}<-26$) of the $z\sim
5$ AGN luminosity function is well constrained, with a rather steep
slope. The faint end ($M_{1450}\ge -22$) indicates a high space
density, the scatter is significant and the knee ($M_{1450}\sim -24$)
is mostly undetermined. Comparisons with state-of-the-art models
find reasonable agreement with the observed AGN luminosity function at
$z=5$, while the predicted space density evolution at higher redshifts
appears to be too fast with respect to observational
constraints. Given the large variance at the faint end, we consider
different options in fitting the luminosity functions and deriving the
ionizing emissivity. Even in the most conservative scenario, the
photo-ionization rate produced by $z\sim 5$ AGNs is consistent with
the UV background measurements. A slow evolution of the space density
of faint AGNs is observed, indicating that active SMBHs are
probably producing large amounts of ionizing photons at $z>6$, well
into the epoch of reionization. This is an important
indication that high-z AGNs could be the major contributors to the
reionization of the Universe.
\end{abstract}

%https://astrothesaurus.org/thesaurus/alphabetical-browse/

\section{Introduction}
\label{sec:intro}

The Cosmic Dawn marks the Renaissance of light, when the first sources
of ultraviolet (UV) radiation, after the so-called Dark
Ages, were able to separate again the electrons from the protons,
leaving them as free particles. This cosmological phase transition, marking the end
of the Dark Ages, has been identified as the Epoch of Reionization
\citep[EoR, e.g.][]{barkana01,dayal18}.

Robust constraints on the beginning of the reionization process have been
derived through the analysis of the optical depth $\tau_e$ of the
Cosmic Microwave Background (CMB) by WMAP and Planck, which indicate a
relatively late ($z_{reion}<8$) and quick ($\Delta z_{reion}<2.8$)
reionization transition \citep{planck20,george15,reichardt20}.
Additional evidences for a late end (at $z\sim 5.3$) to
reionization have been obtained by the fact that excess
fluctuations in the Lyman-$\alpha$ forest persisted until 1.1 Gyr
after the Big Bang \citep{bosman22,zhu22,jin23,spina24,zhu24}.

One of the most important questions in modern astrophysics is which
sources drove cosmic reionization process in the early
Universe. In general, the astronomical community is currently embarked in
finding the long-sought probes that star forming galaxies are actually
the main responsible of the hydrogen reionization. The active
players of cosmic reionization of hydrogen have been mostly identified with
star-forming galaxies at high redshift \citep[e.g.][]{jung24,atek24} or looking for their
low-z analogs \citep[e.g.][]{dayal18,flury22a,flury22b}. So far, the discussion
has been mainly focused on the role played by star-forming galaxies
\citep[see e.g.][and references therein]{mascia24}, and
only a few authors have investigated in detail the possible contribution of
bright QSOs and faint AGNs to the reionization process
\citep[e.g.][]{haardt96,haardtMadau12,giallongo12,fontanot12,haardt15,madau15,chardin17,puchwein19,madau24},
despite some early evidence in favour
of a consistent space density of faint AGNs, based on HST and Chandra observations
\citep{giallongo15,giallongo19}.
Very few exceptions \citep[e.g.][]{boutsia21,grazian22,grazian23,fontanot23}
to this mainstream scenario have been recently provided, proposing the AGN
population as a possible alternative in the cosmic reionization scenario.

This pictures has been substantially revised recently, thanks to two observational
results. New wide ground based surveys of QSOs
\citep[e.g. QUBRICS, RUBICON;][]{grazian22,grazian23} provide substantial
evidences in favour of a relatively large population of bright
($-28.5\le M_{1450}\le -26.5$) QSOs at $z\sim 5$.
In addition, early JWST observations prompt a revolution in our understanding of the
early Universe, by highlighting a large population of faint AGNs at $z>4$
\citep{kocevski23,harikane23,matthee24,maiolino23lf,larson23,furtak23,greene24,scholtz23}.
A significant space density of variability selected AGN candidates
at $z>6$ has been found by \citet{hayes24}.
Few exceptional detections of AGNs at $z>10$ have been provided
by \citet{maiolino23gnz11,goulding23,bogdan24} and \citet{kovacs24}, suggestive of a
high space density of active SMBHs even at these early epochs.
A straightforward estimation of the contribution of these AGNs to the reionization process, mainly based on the AGN fraction in galaxies derived by recent JWST surveys, has been provided by \citet{madau24}.
A first indication has been given by
\citet{harikane23,maiolino23lf} that these SMBHs are over-massive with
respect to the local $M_{BH}-M_{star}$ relations,
although these results may be biased by selection effects \citep[see e.g.][]{li24}.

The emergence of an unprecedented population of faint objects at
$z>4$ with broad Balmer lines, compact morphology, and V-shaped
spectral energy distribution (i.e. flat in the UV rest frame and steep
in the optical rest frame wavelengths), the so called ``Little Red
Dots'' or ``Little Red Monsters'' \citep[and interpreted as probable
AGNs at high-z; see e.g.][]{kocevski23,kokorev24,kocevski24},
added further hints that these faint high-z AGNs could be indeed
widespread in the early Universe.
The nature of these ``Little Red Dots'' was not fully understood
before the launch of JWST \citep[see e.g.][]{fiore08}, and a few
of them were typically misidentified as
$z>12$ galaxies in the first surveys \citep[e.g.][]{labbe22}.
Evidences in favor of
unobscured (type 1) AGNs powering these ``Little Red Dots'' come
from the fact that they show broad Balmer lines, blue (flat) UV
continuum, compact morphology without any visible hosting galaxy,
and small attenuation \citep[e.g.][]{killi23} given the
observed Balmer decrement of the narrow line component \citep[but see][for a different
interpretation on the nature of the Little Red Dots based on deep MIRI observations with JWST]{perezgonzalez24}.
Interestingly, the contribution of a large population of faint AGNs at $z>4$, reaching
till $z\sim 10$ and even beyond
\citep[e.g.][]{fujimoto23,maiolino23gnz11,goulding23,hegde24,kovacs24,habouzit24} 
could also contribute to explain the excess of bright ($M_{UV}\sim -20$) primeval ($z>8$) galaxies
recently found by JWST at these high-z
by e.g. \citet{naidu22,finkelstein24,chemerynska24,adams23},
relaxing the claims for possible tensions of these observations
with the $\Lambda$ cold dark matter ($\Lambda$-CDM) standard cosmological model.
Indeed, \citet{chworowsky23} showed that, taking into account the contamination
by AGNs and with a modest change to star
formation efficiency, the tension with modern cosmology can be
significantly reduced, without invoking ad-hoc scenarios
\citep[e.g.][]{ferrara23} or photon budget crisis \citep[e.g.][]{munoz24}.
Hidden high-z AGNs in
the form of Little Red Dots could also explain the excess of massive galaxies
in the Galaxy Stellar Mass Functions at $z\sim 10$ \citep{Harvey24}.

Constraining the AGN luminosity function at $z>4$ is of utmost
importance not only for reionization, but also to find hints about the
origin of the large SMBHs discovered at the highest redshifts and the
processes governing their growth \citep[e.g.][]{trinca22,fontanot23}.
Different predictions for the AGN space density as a function of luminosity
at $z>4$ have been derived from theoretical works. They
suggest that the volume density of active SMBHs is relevant
at high-z, with several predictions lying above the observed luminosity
function of AGNs at these redshifts. For example, predictions from
GAEA \citep{fontanot20,DeLucia24} or the Rome SAM \citep{menci14} are above the
observed fit of the $z=4$ AGN luminosity function at the faint end, as
shown in Fig. 4 of \citet{boutsia21}, but are consistent with the
observed bright-end. In a similar way, the CAT
semi-analytical model \citep{trinca22,trinca23} predicts more
bright QSOs than the ones observed at $z>4$. At such high redshifts,
these results have deep implication for the proposed SMBH seeding
mechanisms (which somehow represent the initial conditions for gas
accretions in these models). Mismatch between predictions and observations
suggests that seeding mechanisms should be revised accordingly.
On the opposite sides,
results from the DELPHI semi-analytic model \citep{dayal24} seem to
indicate that AGN are able to only provide about 23\% of the total
reionization budget and that their space densities is relevant only at
$z<7$, at the end stages of the reionization process. Recent results
from \citet{fujimoto23} at $z\sim 10$ are one
order of magnitude above the predictions of DELPHI model, possibly
indicating that their model cannot easily reproduce the observed
AGN luminosity function at high-z. It is worth noting here that a simple
model, aimed at reproducing the $z\sim 6$ BH mass-$\sigma$ relation with
overmassive SMBHs with respect to the local relations, would naturally
predict abundant AGNs at $z>7$, as it has been recently observed
by e.g. \citet{harikane23,greene24,furtak23,fujimoto23,kokorev24,kocevski24}.

In order to assess the role of active SMBHs at
high-z in the cosmic reionization, we revisit in this work the AGN
luminosity function at $z\sim 5$ by combining the results on
the bright side from ground-based wide surveys \citep{grazian22,grazian23}
with the deep pencil-beam surveys from JWST at the faint side.

The structure of this paper is the following: in
Section \ref{sec:data} we describe the recent AGN luminosity functions
at the bright end from wide ground-based surveys and at faint
luminosities from JWST. In Section \ref{sec:method} we describe the
method we adopt to fit the AGN luminosity functions at
$z\sim 5$ and to compute the ionizing emissivity and
photo-ionization rate from active SMBHs. In Section \ref{sec:resu} we
present our best-fit luminosity function of $z\sim 5$ AGNs, its redshift evolution,
a comparison with state-of-the-art semi-analytic models,
the resulting ionizing emissivity and photo-ionization rate. We critically discuss
the outcome of our analysis in Section \ref{sec:discussion}, providing
a summary and the concluding remarks in Section \ref{sec:conclusion}.
Throughout the paper, we adopt H$_{0}$=70 km s$^{-1}$ Mpc$^{-1}$,
$\Omega_{M}$=0.3, and $\Omega_{\Lambda}$=0.7, in agreement with the
widely adopted $\Lambda$-CDM concordance
cosmological model. All magnitudes are in the AB photometric system.

%%%%%%%%%%%%%%%%%%%%%%%%%%%%%%%%%%%%%%%%%%%%%%%%%%%%%%%%%%%%%%%%%%%%%

\section{Data}
\label{sec:data}

We collect all recent data on $z\sim 5$ QSO and AGN space density estimates
appeared in the astronomical literature in Table \ref{tab:qlfz5}. At
the bright end ($M_{1450}\le -26$), we consider the results of
\citet{grazian22} and \citet{grazian23} from the QUBRICS and RUBICON
surveys, respectively, and the luminosity function derived from the
survey of \citet{glikman11}. In the latter case, we have computed the
space density of $z\sim 5$ QSOs by simply dividing the observed number
of confirmed QSOs with $4.5\le z_{spec}\le 5.2$ and $-26.5\le M_{1450}\le -25.5$
to the cosmological volume of their survey. In an area of 3.76 sq. deg.,
a total of 3 QSOs have been found by \citet{glikman11} in the redshift
and luminosity intervals listed above. We assume here a conservative
completeness of 100\% for this survey, which is slightly higher than
the actual values derived by \citet{glikman11} through simulations,
of 66-80\%. The errors on the space density have been derived assuming
\citet{gehrels86} for low-number statistics. The resulting space density from
\citet{glikman11} is thus $\Phi=7.65^{+10.2}_{-5.06}\times 10^{-8}
cMpc^{-3} mag^{-1}$, as summarized in Table \ref{tab:qlfz5}.

At faint luminosities ($M_{1450}\ge -22.5$), we take into account the
data of \citet{giallongo19} and \citet{grazian20} from the CANDELS
survey with HST and all the results from recent JWST surveys
\citep{kocevski23,harikane23,matthee24,maiolino23lf,greene24}
for broad line (type 1) AGNs. We do
not consider here the two brightest points of the \citet{giallongo19}
luminosity function, since they have been derived with only one object per bin.
Similarly, we do not take into account here
the $z\sim 5$ AGN luminosity functions of
\citet{yang16,McGreer18,kulkarni19,niida20,kim20,kimim21,pan22,FB22,onken22,jiang22,shin22,schindler23,Matsuoka23},
since they could be severely affected by strong incompleteness, as
shown by \citet{grazian23}.

Recently, \citet{scholtz23} showed that the space density of narrow
line (type 2) AGNs at $z>4$ is a factor of 2 times larger than the
observed density of broad line (type 1) AGNs derived by
\citet{maiolino23lf}. The type2/type1 ratio is around 3 for the brightest AGNs ($M_{1450}=-20.15$)
in the JADES survey \citep{scholtz23}. These narrow line AGNs, however,
could have a negligible emission of HI ionizing photons (i.e. low
values of the escape fraction at $\lambda_{rest}\le 912$ {\AA} along their line of sight).
Moreover, their luminosities at 1450 {\AA} rest frame are
probably contaminated by the host galaxy light \citep[e.g.][]{harikane23},
so it is difficult to
properly take into account the contribution of these components to
the ionizing emissivity. In the following, we will
include the \citet{scholtz23} results in our best fit of the AGN luminosity
function at $z\sim 5$, but we impose a conservatively low value for the Lyman Continuum (LyC)
escape fraction of
this faint population of narrow-line AGNs, as we will discuss in the next sections.

\begin{table*}
\caption{The QSO space densities at $z\sim 5$ adopted in this work.}
\label{tab:qlfz5}
\begin{center}
\begin{tabular}{l c c c c c c}
\hline
Survey & z$_{spec}$ & $M_{1450}$ & $\Phi$ & $+1\sigma_\Phi$ & $-1\sigma_\Phi$ & Reference \\
\hline
QUBRICS & 4.75 & -28.6 & 3.12E-10 & 1.08E-10 & 8.25E-11 & \citet{grazian22} \\
RUBICON & 4.85 & -27.0 & 1.32E-8 & 1.29E-8 & 7.31E-9 & \citet{grazian23} \\
RUBICON & 4.85 & -27.1 & 8.80E-9 & 4.76E-9 & 3.27E-9 & \citet{grazian23} \\
Glikman & 4.85 & -26.0 & 7.65E-8 & 1.02E-7 & 5.06E-8 & This paper \\
CANDELS & 5.50 & -22.3 & 1.29E-6 & 1.72E-6 & 0.85E-6 & \citet{grazian20} \\
G19 & 5.60 & -19.0 & 7.27E-6 & 7.12E-6 & 4.02E-6 & \citet{giallongo19} \\
G19 & 5.60 & -20.0 & 4.77E-6 & 3.79E-6 & 2.31E-6 & \citet{giallongo19} \\
CEERS & 5.00 & -19.8 & 1.07E-5 & 2.48E-5 & 9.25E-6 & \citet{kocevski23} \\
H23 & 4.91 & -21.5 & 1.20E-5 & 0.00E-5 & 1.20E-5 & \citet{harikane23} \\
H23 & 4.91 & -18.5 & 1.40E-4 & 1.90E-4 & 1.40E-4 & \citet{harikane23} \\
H23 & 5.76 & -21.5 & 3.50E-6 & 8.20E-6 & 3.30E-6 & \citet{harikane23} \\
H23 & 5.76 & -18.5 & 1.50E-4 & 3.50E-4 & 1.50E-4 & \citet{harikane23} \\
M24 & 5.00 & -20.0 & 0.87E-5 & 0.39E-5 & 0.37E-5 & \citet{matthee24} \\
M24 & 5.00 & -19.0 & 1.66E-5 & 0.58E-5 & 0.54E-5 & \citet{matthee24} \\
M24 & 5.00 & -18.0 & 1.05E-5 & 0.43E-5 & 0.43E-5 & \citet{matthee24} \\
G24 & 5.50 & -19.5 & 3.00E-5 & 0.60E-5 & 0.60E-5 & \citet{greene24} \\
G24 & 5.50 & -19.0 & 2.10E-5 & 0.70E-5 & 0.70E-5 & \citet{greene24} \\
G24 & 5.50 & -18.5 & 2.10E-5 & 0.80E-5 & 0.80E-5 & \citet{greene24} \\
Mo23 & 5.00 & -18.0 & 8.71E-4 & 4.47E-4 & 3.92E-4 & \citet{maiolino23lf} \\
Mo23 & 5.00 & -19.5 & 1.48E-4 & 1.03E-4 & 0.94E-4 & \citet{maiolino23lf} \\
S24 & 5.50 & -20.15 & 3.76E-4 & 2.41E-4 & 1.57E-4 & \citet{scholtz23} \\
S24 & 5.50 & -18.80 & 7.72E-4 & 3.85E-4 & 2.80E-4 & \citet{scholtz23} \\
S24 & 5.50 & -17.59 & 2.17E-3 & 9.39E-4 & 7.22E-4 & \citet{scholtz23} \\
\hline
\end{tabular}
\tablecomments{
\\
The AGN space density of \citet{kocevski23} adopted
here has the errors on $\Phi$ derived through \citet{gehrels86} statistics.
}
\end{center}
\end{table*}

%%%%%%%%%%%%%%%%%%%%%%%%%%%%%%%%%%%%%%%%%%%%%%%%%%%%%%%%%%%%%%%%%%%%%

\section{Method}
\label{sec:method}

\subsection{The best fit of the AGN luminosity function at $z\sim 5$}

The observed space densities of $z\sim 5$ AGNs (summarised in Table
\ref{tab:qlfz5}) have been fitted with a two-power-law function,
leaving free the four parameters, i.e. the faint-end and bright-end slopes $\alpha$ and
$\beta$, the absolute magnitude break $M^*$ and its normalization
$\Phi^*$. The parametric best-fit values have been obtained through a minimization of
the $\chi^2$ of the observed AGN space densities at $z\sim 5$.
Since the values collected in Table \ref{tab:qlfz5} are not fully
consistent among each others, due to the large variance of the AGN space
densities, especially at the faint end, we identified three main groups
of results, called ``options'' in the following.

The different best-fit options to the observed luminosity functions,
summarised in Table \ref{tab:qlfz5}, are shown in Table
\ref{tab:options}. We identify three different options for combining
the surveys available in the literature:
a first conservative option implying the lowest number density of faint type 1 AGNs,
according to the values found in the literature;
the second option is intermediate, and a third option maximizes the space
densities of faint AGNs, considering both
type 1 and type 2 AGNs, adopting the results of \citet{scholtz23}.
In all these options, we consider in the
bright side ($M_{1450}\le -25$) of the observed luminosity functions
the results from the QUBRICS, RUBICON, and \citet{glikman11} surveys,
shown in Table \ref{tab:qlfz5}. For the conservative luminosity
function (option 1), we take into account the results of CANDELS
\citep{giallongo19,grazian20}, CEERS \citep{kocevski23},
\citet{harikane23,matthee24}, and \citet{greene24}.
For the intermediate luminosity function (option 2), we only
consider the recent results of \citet{maiolino23lf} for broad
line (type 1) AGNs, while for option 3 (high luminosity function) we
take into account the space density of \citet{scholtz23}.

\begin{table*}
\caption{The different options for fitting the AGN Luminosity Functions}
\label{tab:options}
\begin{center}
\begin{tabular}{l l l}
\hline
Option & Surveys & Comments \\
\hline
Option 1 & QUBRICS, RUBICON, Glikman, CANDELS, CEERS, G19, H23, M24, G24 & Low Luminosity Function \\
Option 2 & QUBRICS, RUBICON, Glikman, Mo23 & Intermediate Luminosity Function \\
Option 3 & QUBRICS, RUBICON, Glikman, S24 & High Luminosity Function \\
\hline
\end{tabular}
\tablecomments{
\\
For Option 3, the LyC escape fraction of faint AGNs has been fixed to 5\%.
}
\end{center}
\end{table*}

\subsection{Comparison with theoretical models}

A comparison of the observed $z\sim 5$ AGN luminosity function with model predictions
is useful in order to understand the physical properties of the active SMBHs at high-z
(e.g. seeding, accretion, feedback, ionization). Moreover, since observations have
limited ranges both in redshifts and in luminosities, theoretical models are useful
to fill these gaps and to make quantitative predictions for the AGN population at $z>6$,
where the observations are still limited.
We compare here the observed data of the AGN luminosity function at $z\sim 5$ with the recent
model predictions by \citet{DeLucia24,dayal24,trinca22}.

The GAlaxy Evolution and Assembly \citep[{\sc gaea},][]{DeLucia24}
semi-analytic model follows the evolution of the different
galaxy populations over cosmological volumes and a wide redshift
range, bridging the information of high-z sources to the properties of
local galaxies. In this paper we adopt the latest {\sc gaea} version, which
couples an explicit partition of the cold gas into an atomic and a
molecular components \citep{Xie17}, a treatment for the
non-instantaneous stripping of cold and hot gas in satellites galaxies
\citep{Xie20} and an improved modeling for cold gas accretion onto
SMBHs and AGN-driven outflows
\citep{fontanot20ff}. \citet{DeLucia24} show that this version of the
model, calibrated on the $z<3$ evolution of the stellar mass function
and AGN luminosity function, is able to reproduce the evolution of
the fraction and number densities of quenched galaxies up to $z\sim4$.
It also provides comparably large number densities for massive galaxies at
$z>3$ among similar theoretical models, predicting the existence
of quenched massive galaxies at $z\sim6-7$
\citep{Xie24}. In particular, in this work we consider predictions
from a {\sc gaea} realization (Fontanot et al. in preparation) coupled
with merger trees extracted from the {\sc p-millennium} simulation
\citep{Baugh19}, tracking the evolution of the matter distribution
over a volume box of 800 comoving Mpc on a side.
A more thorough analysis of the properties of the highest-z galaxies and AGNs,
especially in terms of their predictions UV luminosities as seen by
JWST, will be presented in forthcoming work (Cantarella et al., in
preparation).

The DELPHI semi-analytic model \citep{dayal24} recovers the physical properties of early
star-forming galaxies and AGNs by including the impact of reionization feedback in the
suppression of the baryonic content of low-mass galaxies in ionized regions. DELPHI is
able to track the build-up of dark matter halos and their baryonic components, in terms
of gas, stellar, dust, metal masses, and black holes, from $z\sim 40$ to $z=4.5$, over a
cosmological volume of 240 comoving Mpc on a side.
The free parameters of the model (including heavy seeding
mechanisms and high accretion rates) are obtained by matching the faint- and bright-ends
of the Lyman Break Galaxy (LBG) UV luminosity function at $z\sim 5-9$ and the number density of
JWST-detected AGNs at $z\sim 5-7$, in addition to the stellar and black hole mass
functions and the AGN bolometric luminosity function.
By construction, the DELPHI semi-analytic model
provides an AGN luminosity function at 1450 {\AA} rest frame that is in good agreement with the
observations at $z\sim 5-7$ (including the recent outcome from JWST).

The CAT semi-analytic model \citep{trinca22,trinca23} follows the co-evolution of star formation
and nuclear accretion at $z\ge 4$, starting from the formation of the first stars and BHs in a
self-consistent way. The free parameters of the CAT model have been tuned to reproduce
the observed properties of high-redshift QSOs at $z\ge 5$ (SMBH mass and bolometric luminosity
functions), the cosmic star formation history and stellar mass density from $z=4$ to $z=8$.
In the following, we will compare the observed properties of $z\sim 5$ AGNs with the
''reference model'' of \citet{trinca22}, with Eddington limit for BH accretion, especially
tuned to guarantee a good agreement with the observed QSO population at $z>5$.

\subsection{Derivation of ionizing emissivity and photo-ionization rate
produced by $z\sim 5$ AGN}

The HI ionizing emissivity at 912 {\AA} rest frame $\epsilon_{912}$
is computed starting from the best fit of the luminosity
function of $z\sim 5$ AGN derived above. Following \citet{lusso15}, we
assume a non-ionizing continuum at $\lambda_{rest}\ge 912$ {\AA} with
a spectral index $\alpha_\nu=-0.61$ and a softening at shorter
wavelengths ($\lambda_{rest}<912$ {\AA}) of $\alpha_\nu=-1.70$. The
non-ionizing emissivity $\epsilon_{1450}$ is derived by
integrating the luminosity function of AGNs, multiplied by the
luminosity, at 1450 {\AA} rest frame. The integration interval runs
from $M_{1450}=-30$ to $M_{1450}=-18$, following the method of
\citet{giallongo15,giallongo19,boutsia21}.

The ionizing emissivity $\epsilon_{912}$ is derived from
the non-ionizing emissivity $\epsilon_{1450}$ by taking into account the
spectral energy distribution of \citet{lusso15}.
The HI photo-ionization rate $\Gamma_{HI}$ is derived by
adopting equation 11 of \citet{lusso15}, multiplied by an additional
factor of 1.2 in order to take into account the radiative
recombination effect, as explained by \citet{DAloisio18}. This
formula assumes the evolution of the mean free path for ionizing
photons obtained by \citet{worseck14}.

In the above calculations of $\epsilon_{912}$ and $\Gamma_{HI}$, an
escape fraction of Lyman continuum photons of 100\% is
assumed, together with a spectral slope of $\alpha_\nu=-1.70$
at $\lambda_{rest}<912$ {\AA}. Recent results by \citet{cristiani16,grazian18,romano19}
indicate that the escape fraction is around 75\%, when assuming a spectral
slope for the AGN continuum emission of $\alpha_\nu=-0.7$, without any
softening term. This is almost equivalent to assume an escape fraction of
100\% with a spectral softening similar to the one found by \citet{lusso15}.
In the following, the latter assumption from \citet{lusso15} is followed,
i.e. $\alpha_\nu=-1.70$ at $\lambda_{rest}<912$ {\AA} and LyC escape fraction of 100\%.
For Option 3 only, an escape fraction of 5\% is assumed for AGNs
fainter than $M_{1450}=-23.0$ since, at these magnitudes, type 2 AGNs are
probably dominating.

%%%%%%%%%%%%%%%%%%%%%%%%%%%%%%%%%%%%%%%%%%%%%%%%%%%%%%%%%%%%%%%%%%%%%

\section{Results}
\label{sec:resu}

\subsection{The best-fit luminosity function of $z\sim 5$ AGNs}

\begin{figure}
\includegraphics[width=\linewidth]{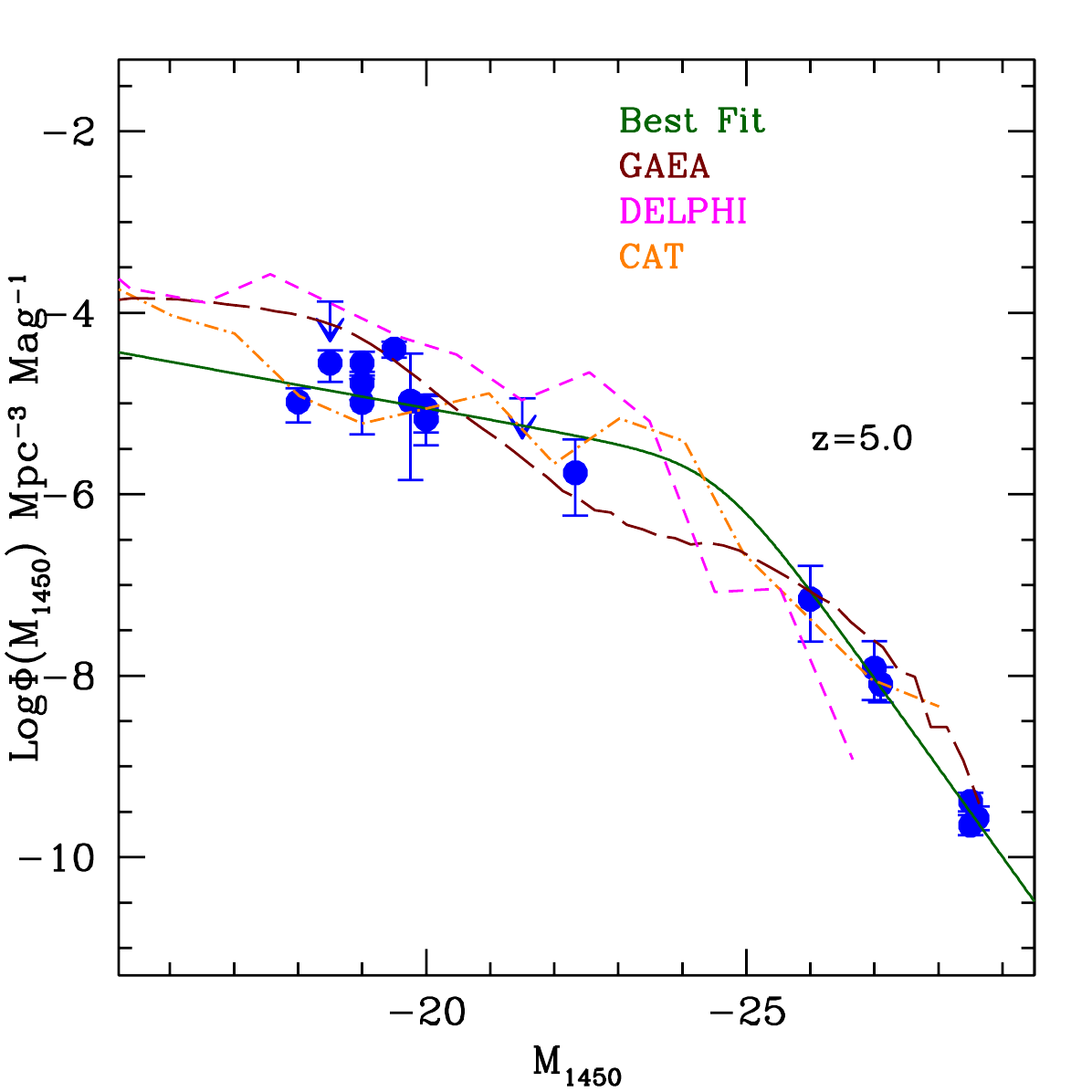}
\caption{The observed luminosity function of AGNs at z=5 (blue points and arrows),
with the best fit (dark green solid line) for Option 1. The comparisons with model
predictions by GAEA \citep{DeLucia24}, DELPHI \citep{dayal24}, and CAT \citep{trinca22}
are shown by dark-red long-dashed, magenta short-dashed, and orange dot-dashed curves, respectively.}
\label{fig:lfresu1}
\end{figure}

Table \ref{tab:bestfit} summarises the parameters of the best-fit
luminosity functions for the different options identified in Table
\ref{tab:options}. The associated error bars correspond to 68.3\%
confidence limits, or equivalently 1$\sigma$ level. Table \ref{tab:bestfit}
also summarizes the best fit and 1$\sigma$
uncertainties for the non-ionizing and ionizing emissivities,
$\epsilon_{1450}$ and $\epsilon_{912}$, respectively, and the
photo-ionization rate $\Gamma_{HI}$.
Fig. \ref{fig:lfresu1} shows the observed space densities (blue
points and arrows) of $z\sim 5$ AGNs for Option 1 of Table \ref{tab:options},
with the derived best-fit luminosity function (thick solid dark green line).

\begin{figure}
\includegraphics[width=\linewidth]{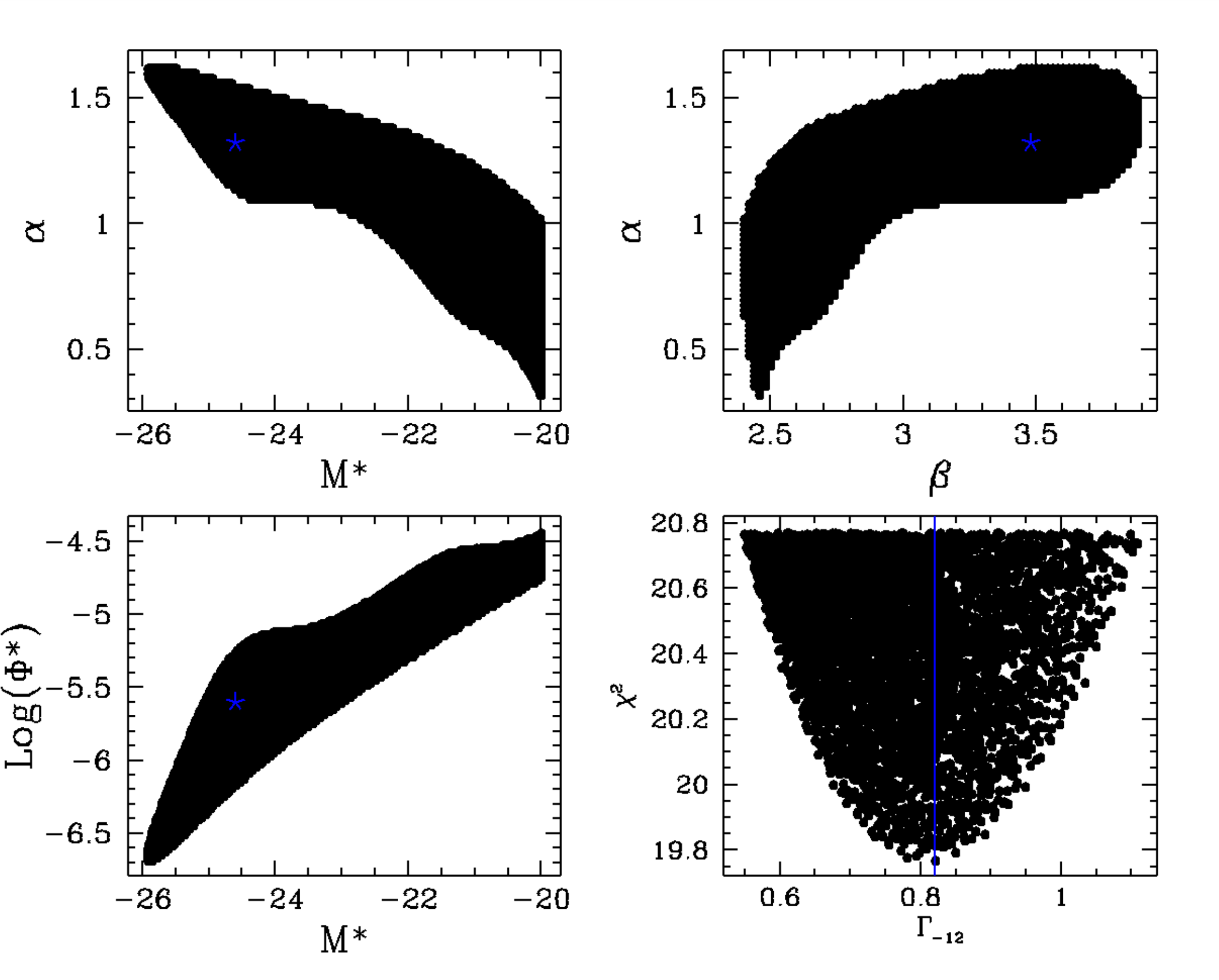}
\caption{The best-fit luminosity function parameters of AGNs at z=5
for Option 1. The dark areas show the confidence interval at
1$\sigma$ (68.3\% c.l.) for the parameters of the AGN luminosity
function at $z\sim 5$. The bottom-right plot shows the permitted
interval for the photo-ionization rate $\Gamma_{HI}$ at 1$\sigma$ level,
while the best-fit value is indicated by the blue vertical line.}
\label{fig:lfparam1}
\end{figure}

The covariances of the luminosity function parameters at 1$\sigma$ level for Option 1 are
shown in Fig. \ref{fig:lfparam1} (black areas in the top and bottom-left panels). The
best-fit values are identified by a blue asterisk. The 1$\sigma$
uncertainties for the photo-ionization rate $\Gamma_{HI}$ are shown
in the bottom-right panel of Fig. \ref{fig:lfparam1}, where the
best-fit value is indicated by a blue vertical line.
Similar versions of Fig. \ref{fig:lfresu1} and \ref{fig:lfparam1} for
options 2 and 3 are shown in the Appendix (Fig. \ref{fig:lfresu3}, \ref{fig:lfparam3},
\ref{fig:lfresu4}, and \ref{fig:lfparam4}, respectively).

\begin{table*}
\caption{The best fit of the z=5 AGN luminosity function
and the corresponding emissivities and photo-ionization rate}
\label{tab:bestfit}
\begin{center}
\begin{tabular}{l c c c c c c c}
\hline
Option & $\log(\Phi^*)$ & $M^*$ & $\alpha$ & $\beta$ & $\epsilon_{1450}$ &
$\epsilon_{912}$ & $\Gamma_{HI}$ \\
 & & & & & $10^{24} erg~s^{-1}$ & $10^{24} erg~s^{-1}$ & $10^{-12} s^{-1}$ \\
\hline
Option 1 & $-5.60^{+1.15}_{-1.08}$ & $-24.57^{+4.56}_{-1.37}$ & $-1.32^{+1.01}_{-0.31}$ & $-3.46^{+0.43}_{-1.07}$ & $12.71^{+4.99}_{-4.01}$ & $7.18^{+2.82}_{-2.26}$ & $0.80^{+0.31}_{-0.25}$ \\
Option 2 & $-7.14^{+4.44}_{-3.16}$ & $-26.40^{+6.40}_{-3.60}$ & $-2.22^{+2.22}_{-0.32}$ & $-3.70^{+1.08}_{-1.30}$ & $22.38^{+34.81}_{-9.00}$ & $12.65^{+19.68}_{-5.09}$ & $1.41^{+2.19}_{-0.57}$ \\
Option 3 & $-4.22^{+1.52}_{-5.12}$ & $-23.16^{+3.16}_{-5.52}$ & $-1.76^{+1.26}_{-0.80}$ & $-3.46^{+0.68}_{-1.54}$ & $23.69^{+41.54}_{-21.01}$ & $17.86^{+31.31}_{-15.83}$ & $1.99^{+3.49}_{-1.76}$ \\
\hline
\end{tabular}
\tablecomments{
\\
All the quoted errors are at 1$\sigma$ level (68.3\% confidence level).
For Option 3, the LyC escape fraction of faint AGNs at $M_{1450}\ge -23.0$
has been fixed to 5\%, while it is 100\% for brighter sources.
}
\end{center}
\end{table*}

In Fig. \ref{fig:lfresu1} we also compare both observational determinations
and our best fit luminosity function against the predictions of semi-analytic
models GAEA (dark-red long-dashed curve), DELPHI (magenta short-dashed curve),
and CAT (orange dot-dashed curve).
At the bright end ($M_{1450}\le -25$), the GAEA and CAT models \citep{DeLucia24,trinca22}
recovers the observed space densities of bright QSOs. The DELPHI model \citep{dayal24},
instead, seems to predict a steeper bright-end slope compared to the observations, but it
should be noted that their simulations are limited to $M_{1450}\gtrsim -27$.
Moreover, the non monotonic trend of these models is possibly
indicating that their
uncertainties are of the same order of magnitude of the differences with the
observed luminosity function. It is thus difficult to conclude that one model
is better than the others.
On the faint-end side ($M_{1450}\ge -23$), all the three
models are able to reproduce the observed space densities of AGNs (for Option 1, i.e. our
conservative luminosity function). In case of an even higher space density of $z\sim 5$ AGNs
at low luminosities, in agreement with \citet{maiolino23lf} or \citet{scholtz23},
i. e. our options 2 and 3, then the predictions of GAEA, DELPHI, and CAT models will
underestimate the faint end of the luminosity function.

The HI photo-ionization rates $\Gamma_{HI}$ in Table \ref{tab:bestfit} cover the
interval $0.80-1.99\times 10^{-12} s^{-1}$, according to the different options in Table
\ref{tab:options}. In these calculations, we have assumed a LyC escape
fraction of 100\% for the z=5 AGNs. In Option 3 we assume fesc=5\% for
AGNs fainter than $M_{1450}=-23.0$ and 100\% at higher luminosities.
These values of $\Gamma_{HI}$ are
in rough agreement with the ones derived by the proximity effects or
by the fit of the Lyman-$\alpha$ forest observed in high-z QSOs. In
particular, a recent estimate by \citet{davies23} from XQR30 survey \citep{dodorico23}
gives a photo-ionization rate of
$\Gamma_{HI}=0.599_{-0.085}^{+0.109}$
(in units of $10^{-12} s^{-1}$), which is
in agreement with our conservative estimate in Table
\ref{tab:bestfit} (Option 1). If a lower value of the LyC escape
fraction is assumed \citep[e.g.][]{micheva17,iwata21}, then also
Options 2 is not far from the \citet{davies23}
determination. Option 3 gives a large value of $\Gamma_{HI}$, three times
larger than the XQR30 measurements, but it is worth mentioning that the
best fit of the AGN luminosity function has been derived taking into
account also the type 2 AGNs by \citet{scholtz23}, and that an arbitrary
escape fraction of 5\% has been adopted at $M_{1450}\ge -23.0$ (similar
to the values assumed for star-forming galaxies in recent papers). A
different (lower) choice for LyC escape fraction or $M_{1450}$ threshold could result
into a better agreement with the \citet{davies23} $\Gamma_{HI}$.
In addition, the uncertainties on the determination of the photo-ionization rate
in the literature are still large, and it is not excluded that $\Gamma_{HI}$ could
be higher than the present values derived by \citet{Gaikwad23} or \citet{davies23}
at $z\sim 5$ by $\sim 0.2$ dex, reaching a consistency with options 2 and 3.

This indicates that with conservative assumptions on the shape of the
$z\sim 5$ AGN luminosity function and with the hypothesis that all
these AGNs, from $M_{1450}=-30$ to $M_{1450}=-18$, have a substantial
escape fraction of $\sim 100\%$, then AGNs only could sustain
the ionization of the universe close to the reionization epoch.

\subsection{The redshift evolution of the AGN luminosity function}
\label{sec:redevol}

\begin{figure}
\includegraphics[width=\linewidth]{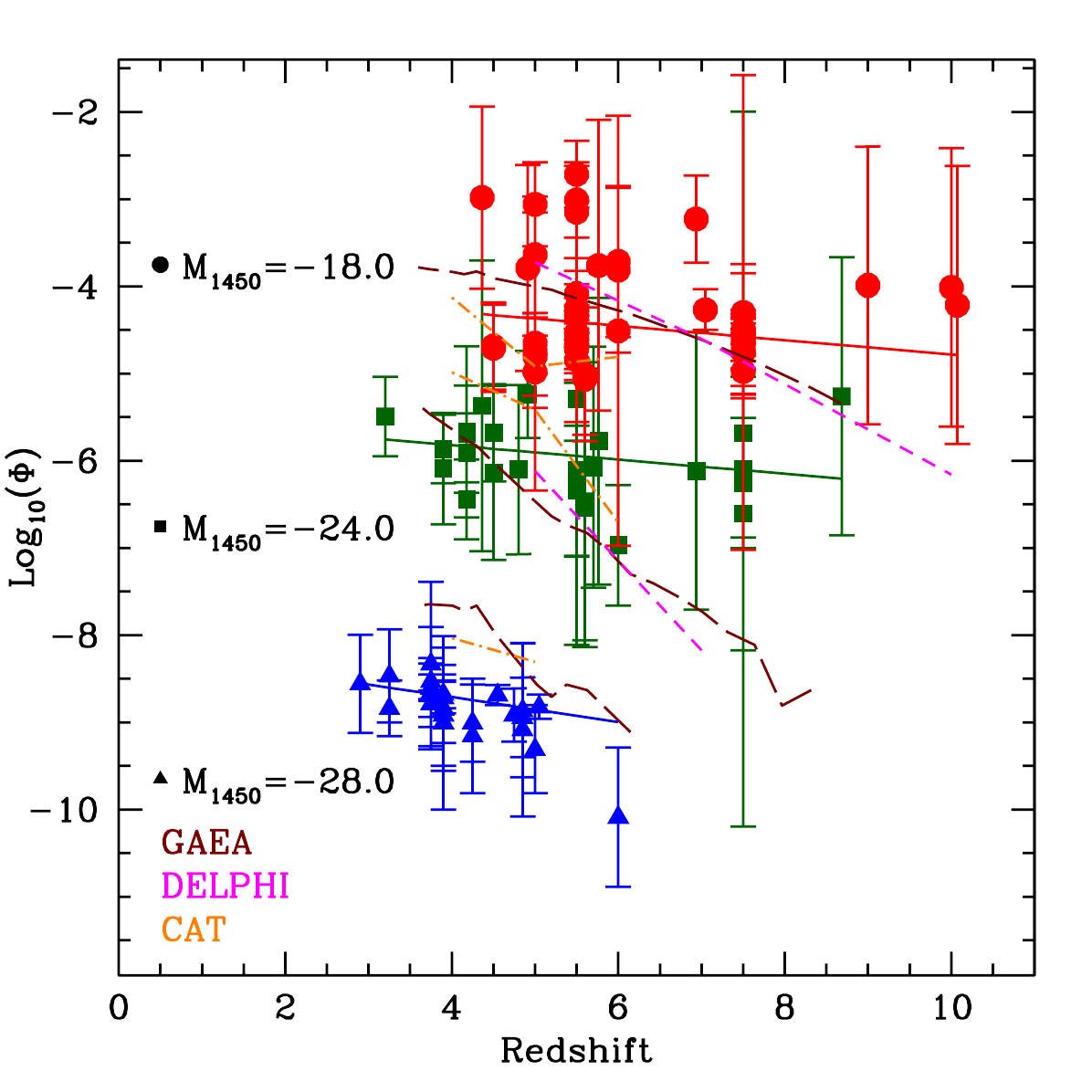}
\caption{The evolution of the space densities of bright QSOs and faint
AGNs at high-z, collected from different surveys in the literature.
The points at $z\sim 5$ are summarized in Table \ref{tab:qlfz5}.
The blue, green, and red solid lines are the best fit
weighted linear relations to the observed data points.
The dark-red long-dashed lines, the magenta short-dashed
and orange dot-dashed lines are the model predictions
by GAEA \citep{DeLucia24}, DELPHI \citep{dayal24}, and CAT \citep{trinca22},
respectively.
}
\label{fig:evolphiz}
\end{figure}

Fig. \ref{fig:evolphiz} shows the evolution of the space densities of
bright QSOs and faint AGNs at high-z, recently appeared in the
literature. The blue triangles are for UV absolute magnitudes
$M_{1450}\le -26.0$ and these space densities have been translated to a
magnitude $M_{1450}=-28.0$ following the observed luminosity function,
by assuming a bright-end slope of
$\beta=-3.46$, as derived from the best fit in Table \ref{tab:bestfit}
(for Option 1). Following the same procedures, the green squares are
the space densities in the range $-26.0 \le M_{1450}\le -21.0$, and
have been translated all to $M_{1450}= -24.0$ assuming a faint end
slope of $\alpha=-1.32$, as found in Table \ref{tab:bestfit}
(for Option 1). The red circles are the
space densities of AGNs with $M_{1450}>-21.0$ and have been translated
all to $M_{1450}= -18.0$ assuming the same faint-end slope of
$\alpha=-1.32$.

For each group of space densities ($M_{1450}=-28.0$, $M_{1450}=-24.0$,
and $M_{1450}=-18.0$), we compute the best fit evolution
assuming a linear relation, $\log(\Phi)=\log(\Phi_0)+\gamma(z-z_0)$.
The best fit values for $\gamma$ are $-0.32\pm 0.12$, $-0.05\pm 0.10$,
and $-0.09\pm 0.08$ for
$M_{1450}\le -26.0$, $-26.0 \le M_{1450}\le -21.0$, and $M_{1450}>-21.0$,
respectively. The observed trend indicates that the evolution of the
AGN luminosity function at $z>3$ is not evolving as a pure space
density, as suggested in \citet{grazian23} based on the bright side ($M_{1450}\le -28$) of
the QSO luminosity functions at z=4 and z=5, but has a differential
evolution. While the bright end of the luminosity function is evolving relatively
rapidly from $z=3$ to $z=6$, at
intermediate and faint luminosities the AGN density evolution is slightly milder,
consistent with no evolution from z=4 to z=10. If confirmed by future surveys,
this differential trend can give stringent constraints to the models for the
formation and evolution of the high-z SMBHs and their seeding process
\citep[e.g.,][]{trinca22}.

In Fig. \ref{fig:evolphiz}, we present a comparison with the predictions of
the semi-analytic models of \citet{DeLucia24,dayal24,trinca22}. In general, the three models
are able to fairly reproduce the normalization of the AGN space densities at different
luminosities for $z\sim 5$. The models considered in this study
are facing some difficulties to recover the redshift evolution of the
AGN space densities for the intermediate ($M_{1450}\sim -24.0$) luminosities, since they
all show a steeper evolution
with respect to the observations.
If these discrepancies will be confirmed
by future observations of faint AGNs at high-z, then a general revision of these theoretical
models is probably required. In particular, the BH seeding prescription of
the models is particularly sensitive to the space density of high-z AGNs at the faint end.
It is also worth pointing out that at the faint end of the AGN luminosity function the
contamination from the host galaxy is more probable to have an impact.

%%%%%%%%%%%%%%%%%%%%%%%%%%%%%%%%%%%%%%%%%%%%%%%%%%%%%%%%%%%%%%%%%%%%%

\section{Discussion}
\label{sec:discussion}

\subsection{Differential evolution of the AGN Luminosity Function ?}

As shown in Fig. \ref{fig:evolphiz}, the evolution of the faint side
of the AGN luminosity function is milder than the one on the bright end. It is
possible that this differential evolution is mainly driven by the point at $z=6$ and
$M_{1450}=-28$ by \citet{Jiang2016}, where the completeness correction
could be underestimated, as also found at lower redshifts by
\citet{Sch19a,Sch19b,boutsia21} and \citet{grazian23}. If we exclude from the fit
of the blue triangles in Fig. \ref{fig:evolphiz} the point at $z=6$, the resulting slope is then
$\gamma=-0.28\pm 0.13$, similar to the value derived by
\citet{grazian22} of $\gamma=-0.25$, and still compatible with the value derived here of
$\gamma=-0.32\pm 0.12$, when the latter point by \citet{Jiang2016} is included.
This result probably hints at a differential evolution of the AGN
Luminosity Function between the bright side, evolving with redshift, and
the faint side, where it remains almost constant from z=4 to z=10, but better data are needed before supporting these conclusions.

As discussed in Section \ref{sec:redevol}, the points in Fig. \ref{fig:evolphiz}
have been shifted to $M_{1450}=-28$, $-24$ and $-18$ by adopting the
bright- and faint-end slopes of the best-fit Luminosity Function at $z=5$
derived here for Option 1, i.e. $\alpha=-1.32$ and $\beta=-3.46$, respectively.
We have checked that the above results on the space density evolution
of the bright and faint AGNs do not change if we assume the
best-fit values for Options 2 or 3.
The possible differential evolution of the bright and faint
side of the AGN space density does not depend on the
assumptions on the exact shape of the AGN Luminosity Function at $z=5$.

It is worth stressing here that the space densities of faint AGNs
have large error bars and show a non-negligible variance, indicating
that more observations are needed in this luminosity regime.
If it will be confirmed in the future that the space density of
faint AGNs is relatively constant from z=4 to z=10, then this implies
that the AGN occupation fraction among distant galaxies is an increasing
function of redshift, due to the relatively rapid evolution in redshift of the space density of star-forming galaxies \citep[e.g.,][]{oesch20}.
The numerous presence of faint AGNs among bright high-z galaxies recently found
by JWST spectroscopy up to $z=10$
\citep{kocevski23,harikane23,maiolino23gnz11,labbe22,xu23,furtak23,greene24,scholtz23,kokorev24,andika24,kovacs24}
seems to go in this direction, confirming previous results by
\citet{giallongo15,giallongo19} and \citet{grazian20,grazian22,grazian23}.
The almost ubiquitous presence of active SMBHs at the center of star-forming galaxies
at very high redshifts could explain the surprisingly slow evolution of the space
density of the bright galaxies at $z>8$ recently derived by deep JWST observations
\citep[e.g.][]{finkelstein24}. Indeed, the excess of relatively bright galaxies found by JWST
at very high-z with respect to the predictions of $\Lambda$-CDM models \citep{yung24}
could be due to the presence of an AGN at their center, making them brighter, more
compact, and also causing the derivation of stellar masses and star formation rates
from spectro-photometric codes to be less reliable.
Not surprisingly, the space density of the high-z bright galaxies by \citet{finkelstein24}
is of the order of $10^{-4} Mpc^{-3}$ at $M_{1450}\sim -19$
and $10^{-5} Mpc^{-3}$ at $M_{1450}\sim -22$, in very good agreement with the AGN space densities
plotted in Fig. \ref{fig:evolphiz}. In the future, a careful
investigation of the physical engine (star formation vs AGN) powering
these bright galaxies at $z>8$ will be important to check this hypothesis.
Vice versa, the host galaxy light can contaminate the UV
luminosity of the central AGN leading to an 
overestimate of the faint-end of the AGN luminosity function.

\subsection{Implications on ionizing emissivity}

\begin{figure}
\includegraphics[width=\linewidth]{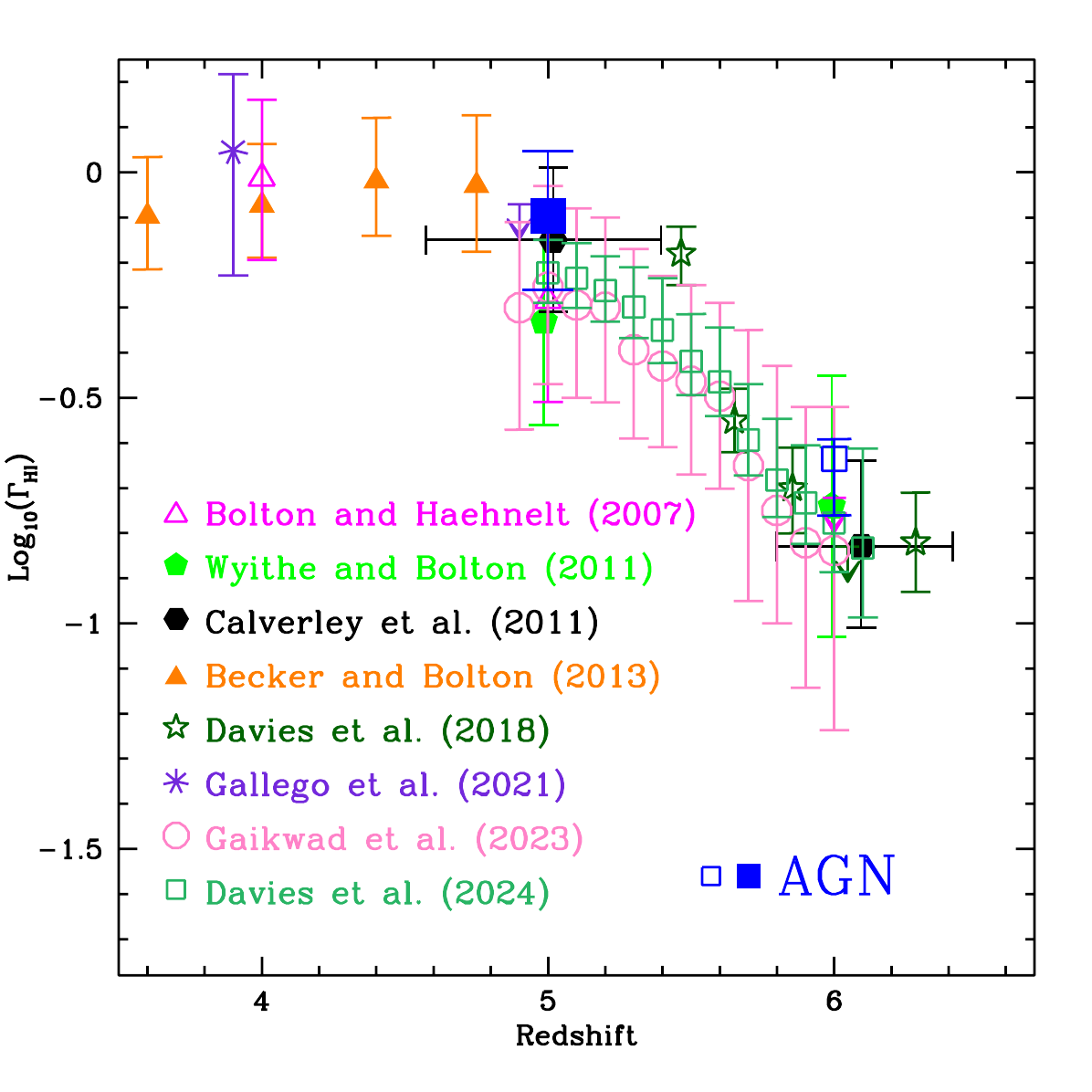}
\caption{The photo-ionization rate $\Gamma_{HI}$ (in units of $10^{-12} s^{-1}$)
produced by bright QSOs and faint AGNs at z=5 (filled blue square, option 1)
and at z=6 (open blue square). These values have been obtained from the
best fit AGN luminosity function assuming an escape fraction of Lyman Continuum photons
of 100\%. All the other points are measurements of the photo-ionization rate from
Lyman-$\alpha$ forest fitting or from the proximity effect.
}
\label{fig:gamma}
\end{figure}

If confirmed by future observations, the mild evolution in redshift of the faint
side of the AGN luminosity function has deep implications both for the
reionization of hydrogen and the search for the progenitors of the SMBHs
detected at $z\gtrsim 6$ with masses of $10^9-10^{10} M_\odot$. As shown in the
results above, the population of bright QSOs and faint AGNs at $z\sim 5$
are able to produce an amount of HI ionizing photons comparable with
recent estimates of the photo-ionization rate by the Lyman forest or by the
Proximity effect at the end of the reionization epoch, as shown in
Fig. \ref{fig:gamma}. This agreement suggests a relevant role of AGNs in the
process of reionization, as recently proposed by \citet{madau24} on the basis of 
the latest JWST results.
This result is in agreement with
what has been recently concluded by \citet{fontanot23},
confirming previous results by
\citet{giallongo15,giallongo19}. The shape and redshift evolution of
the AGN luminosity function at $z\ge 5$, moreover, could provide strong
constraints on the different scenarios proposed for the formation of
the SMBH seeds \citep{trinca22,trinca23,volonteri23,fontanot23,dayal24}.

In the calculations of Section \ref{sec:resu},
we have assumed that the escape fraction of bright QSOs and
faint AGNs down to $M_{1450}=-18$ is always 100\%. This assumption
is not unrealistic, since
the observed AGNs at $z>4$ found by JWST at these luminosities show
broad and symmetric H$\alpha$ lines \citep[e.g.][]{kocevski23,harikane23,greene24,maiolino23lf,kokorev24},
typical of type 1 AGNs, thus not obscured. Moreover, these objects are showing
compact morphologies and blue UV continuum, both typical of unobscured (type 1) AGNs.
This is at odds with the conclusions of \citet[e.g.][]{matthee24},
hinting at a negligible contribution of these faint AGNs to the ionizing emissivity, due to
their possible low LyC escape fraction. Their conclusions about the
small fraction of ionizing photons able to escape from these objects are
mainly based on their steep slopes in the rest-frame optical, possibly
indicating large extinction. However, their spectral slopes in the UV
rest frame are relatively flat, hinting at a negligible amount of dust
extinction.

At low redshift ($z<1.5$), the escape fraction of faint
($M_{1450}\sim -19$) AGNs is close to 100\%, as shown e.g. by
\citet[][]{Stenans14}.
The escape fraction of faint ($M_{1450}\lesssim -23$) AGNs at $z\sim 4$
has been studied by \citet{grazian18}, showing that it is substantial,
i.e. $\sim 70\%$, over a wide magnitude range. This result has been
questioned by \citet{iwata21}, based on deep u-band imaging of faint AGNs
at $z\sim 3-4$. Some of their AGNs, where they estimate a negligible
LyC escape fraction, are however in common with \citet{grazian18}, where
the latter instead find substantial escape fraction of ionizing photons
from deep spectroscopy. This mismatch leaves still open the question on the
exact value of the escape fraction of faint AGNs at high-z. A more careful 
investigations of the escape fraction of accreting SMBHs at low luminosity
is necessary in the future in order to determine this crucial parameter for
estimating the amount of ionizing photons reaching the IGM from high-z,
faint AGNs.

Fig. \ref{fig:gamma} shows the strong evolution of the photo-ionization
rate $\Gamma_{HI}$ produced by AGNs at $z=5$ and $z=6$. The result at $z=6$
has been derived by a fit of the AGN space density observed at these redshifts
(as shown in Fig. \ref{fig:evolphiz}), which could
be affected by strong incompleteness or large uncertainties. Moreover, we have used the
mean free path of \citet{worseck14} at $z=6$ \citep[almost consistent with the one
by][]{Bolton07} to obtain the photo-ionization rate
from the ionizing emissivity $\epsilon_{912}$, but there are possible indications
\citep{becker21,davies23,Gaikwad23} that the mean free path at $z=6$ is possibly lower than the
one by \citet{worseck14}. 
The strong drop of the photo-ionization rate produced by AGNs from $z=5$ to
$z=6$ is in slight agreement with the measurements of the ionizing background
derived from Lyman-$\alpha$ forest fitting or from the proximity effect
\citep{Bolton07,calverley11,wyithe11,BeckerBolton13,davies18,Gallego21,Gaikwad23,davies23}.

\begin{figure}
\includegraphics[width=\linewidth]{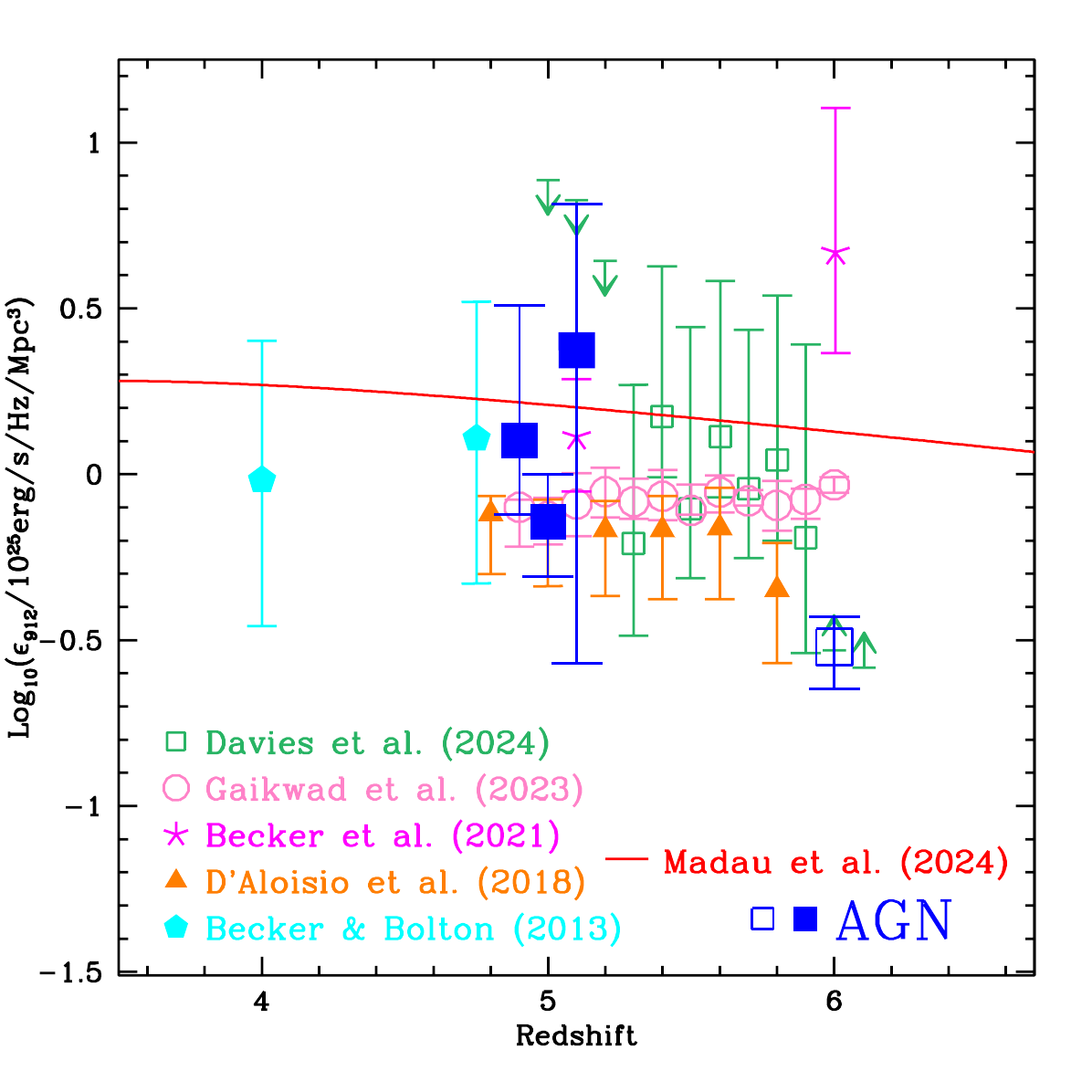}
\caption{The ionizing emissivity $\epsilon_{912}$
(in units of $10^{25} erg~ s^{-1} Hz^{-1} Mpc^{-3}$)
produced by bright QSOs and faint AGNs at z=5 (filled blue squares; lower point represents
option 1, intermediate value is for option 2, while upper point shows the emissivity for option 3)
and at z=6 (open blue square). The emissivities for options 2 and 3 have been shown
by the blue filled squares at z=4.9 and z=5.1, respectively,
in order to avoid overlaps in the plot,
but they have been computed all at z=5. The red line shows the emissivity
adopted by \citet{madau24}.
The other points have been taken from \citet{BeckerBolton13,DAloisio18,becker21,Gaikwad23,davies23}.
}
\label{fig:eps912}
\end{figure}

In Fig. \ref{fig:eps912}, we show the redshift evolution of the
ionizing emissivity $\epsilon_{912}$ from a collection of recent works \citep{BeckerBolton13,DAloisio18,becker21,Gaikwad23,davies23}.
For comparison, we add in the plot the emissivity produced
by the AGN population at z=5 (for options 1, 2, and 3)
and at z=6. The point at z=6 is still uncertain, due to
possible incompleteness in the bright side of the AGN luminosity
functions at this redshift.
From the recent results by \citet{Gaikwad23}, it seems that
the ionizing emissivity is flat from z=4.5 to z=6.
This implies that the strong drop in the photo-ionization rate
$\Gamma_{HI}$ is mainly due to a drop of the mean free path of ionizing
photons. Moreover, the constancy of $\epsilon_{912}$ is almost
consistent with an ionizing
emissivity produced by AGNs, where their space density, at least at
the faint luminosities, is almost flat, as shown in Fig. \ref{fig:evolphiz}.
This is another hint in favour of AGNs as possible responsible
for reionization.

\subsection{The contribution of the host galaxy to the ionizing emission of AGNs}

As discussed in \citet{maiolino23lf} and \citet{scholtz23},
their observed fluxes are most likely a combination of AGN plus host galaxy light,
thus providing an upper limit for the AGN luminosity function for Options 2 and 3.
There are however hints that the contribution by the host galaxy of these high-z AGNs
is negligible. For example, \citet{harikane23} find that for 90\% of their sample (9 objects
out of 10) the point-like central source is a necessary component of the 2-dimensional fit
of the JWST images, with a typical stellar mass of the host galaxies of
$M_{host}\sim 10^{9.5} M_\odot$. In a similar way, the luminosity of the broad component is
dominating (by a factor of 2-5) that of the narrow component in 40\% of their sample, it is
almost comparable (factor 0.6-0.9) for 20\% of their sample, while it is only a factor of 0.2-0.3
for the remaining 40\% of their sample. The small sample by
\citet{harikane23} of 10 objects only, however, could be affected by large variance 
due to small number statistics. In the future, larger samples of AGN hosts are 
required to address this issue.

In addition, their estimated SMBH masses are 0.01-0.1 times the stellar mass of their host
galaxies, thus implying that the central engine is accreting close to its Eddington limit. 
In the case of the two AGNs by \citet{kocevski23}, the light is dominated by the
central point-like sources. All these observations indicate that the emissivity of the
AGN host galaxy is not the dominant contribution even at the faint side of the AGN luminosity
function studied in this work, i.e. at $M_{1450}\sim -18$. Moreover, almost all of
the Little Red Dots
recently found by JWST \citep[e.g.][]{kocevski23,kokorev24,perezgonzalez24,barro24,kocevski24}
show compact morphologies, blue slopes of their UV rest-frame SEDs, and Balmer lines
dominated by the broad component. These are all strong indications that these AGNs are
of type 1, thus their UV light is by far dominated by the nuclear component, with a small
contribution by their host galaxies.

It is indeed possible that the central SMBHs are cleaning the
surrounding medium during their active phases, leaving also the sub-dominant
stellar ionizing light from the
host galaxy free to escape from the inter-stellar medium (ISM),
with a different (softer) spectral shape.
This possibility is mentioned by \citet{fiore23} to explain the great number of blue
``super-Eddington'' objects recently discovered by JWST.

\subsection{The possible issues of a reionization process driven by AGNs}

A reionization driven by faint AGNs and bright QSOs has three main
apparent issues:
\begin{itemize}
\item
  if AGNs are able to produce the whole ionizing radiation to ionize
  the Universe, what is the role of star-forming galaxies? They are
  much more numerous than AGNs, and even assuming a low escape
  fraction ($\sim 10\%$), their additional contribution can exceed the
  observed emissivity at high-z.
\item
  A large number of AGNs at high-z would over-produce the X-ray background
  \citep{padmanabhan23}.
\item
  A dominant population of AGNs at high-z would ionize the HeII too
  early \citep[coeval with HI and HeI reionization, see e.g.][]{garaldi19},
  while the observed epoch for HeII
  reionization is at $z\sim 3$ \citep{worseck19,makan21,makan22}.
\end{itemize}

The first item is not a serious problem. As we have discussed in the
above sections, the star-forming galaxies that have been confirmed as
LyC emitters at $z>3$ are relatively rare, of the order of $\sim 1\%$
of the total galaxy population at that epochs \citep[e.g.][]{kerutt24},
and they have large values of the ionizing escape fraction, of the
order of 100\%. It is thus not surprising if these few leakers would
be confirmed to be AGNs, as happened for peculiar sources like GDS3073
\citep{grazian20,barchiesi23,ubler23} or for Green Pea analogues,
which seem to be powered by luminous X-ray sources \citep{singha24}.
The recent finding of broad Balmer lines
on typical star-forming galaxies, thus hinting at their AGN nature
\citep[e.g.][]{harikane23,maiolino23gnz11,labbe22,furtak23,greene24}
may point to the fact that all the few leakers found among the
star-forming galaxy population could be in fact powered by an AGN.
Further spectroscopic analysis with JWST of the confirmed leakers would
clarify their nature and could corroborate or refute this hypothesis.

The other two issues, i.e. over production of the X-ray background and
too early HeII reionization, have been extensively discussed in \citet{madau24},
where it has been shown that the internal absorption
expected for AGNs at 4 Rydberg will delay the HeII reionization until
$z\sim 3$, as also found by \citet{madau15,basu24} and in agreement with
recent observations of the HeII Lyman forest \citep[e.g.][]{worseck19,makan21,makan22}.
The faint AGN
population recently detected by JWST does not produce an excess of the
X-ray background, since they show very few detections in the
rest-frame hard X-ray, as shown by e.g. \citet[][]{kocevski24,maiolino24xray}.

Summarizing, the three apparent issues above are not critical
show-stoppers for the AGN driven reionization scenario we are
proposing in this work.

\subsection{Future Prospects}

As shown in Table \ref{tab:qlfz5}, the present-day determination of
the QSO/AGN luminosity functions at $\sim 5$ does not cover the
intermediate luminosities ($-25.5\le M_{1450}\le -22.5$), where
the bulk of ionizing photons are expected to be produced
\citep[see e.g.][]{giallongo19}. Several
attempts to fill this gap resulted in very low space densities of AGNs
\citep[e.g.][]{yang16,McGreer18,niida20,kim20,kimim21,FB22,jiang22,shin22,schindler23,Matsuoka23},
but as discussed in \citet{grazian23}, it is possible that these
surveys are strongly affected by severe incompletenesses. The fact
that even the most conservative results from JWST
\citep[e.g.][]{matthee24} are finding an order of magnitude more AGNs
than the above surveys, provides another indication that the former
are strongly incomplete. In the future, it is fundamental to extend
the search for $z\ge 5$ AGNs in the intermediate luminosities
regime ($M_{1450}\sim -24$). This will be possible both by extending the present wide
surveys \citep[e.g., RUBICON][]{grazian23} to two or three magnitudes fainter
and by enlarging the sky area covered by deep JWST
spectroscopy of the $H\alpha$ line with NIRSpec, thus extending the
statistics of relatively bright AGNs ($M_{1450}\sim -24$).

Moreover, a leap forward to the AGN space density at high-z
will be provided by the Euclid mission \citep{laureijs20,barnett19,mellier24}. 
It is expected that the deep near infrared and optical imaging of the Euclid
Wide Survey will be instrumental in finding approximately 100 QSOs at $7<z<9$
and much more at $z\sim 5$ and at $M_{1450}\sim -24$. With these new QSOs,
the break of the AGN luminosity function will be well constrained,
leaving the LyC escape fraction as the only unknown parameter to
determine the contribution of AGNs to the ionizing backgroud.
Dedicated observations in the rest-frame UV
will thus be fundamental to close this exercise at $z\sim 5$.

%%%%%%%%%%%%%%%%%%%%%%%%%%%%%%%%%%%%%%%%%%%%%%%%%%%%%%%%%%%%%%%%%%%%%

\section{Summary and Conclusions}
\label{sec:conclusion}

The recent measurements of the space densities of bright QSOs and faint
AGNs at $z\sim 5$ trigger a revision of the AGN luminosity function at
these cosmological epochs, with a refinement of their contribution to
the hydrogen reionization.

We have collected all the recent QSO/AGN luminosity functions at
$z\sim 5$, discarding those that are severely affected by incompleteness.
At the bright end, the space density of $M_{1450}\le -26$ QSOs is well
determined, thanks to large area surveys \citep{grazian22},
deep imaging \citep{grazian23}, and efficient
selection criteria, possibly based on Machine Learning techniques
\citep[e.g.][]{guarneri22,calderone23}. On the faint side, the space
densities of $M_{1450}\ge -22$ AGNs have been constrained by JWST deep
spectroscopy, finding that the number of active SMBHs at $z\sim 5$ is
much larger than previous estimates \citep[e.g.][]{kocevski23,maiolino23lf}.
The scatter on
the number densities at such low luminosities is still large, mainly due to
the uncertainties on the identification of faint AGNs among the sample of
high-z galaxies and to the scanty number statistics of the first JWST surveys.
Future works with JWST would hopefully reduce the variance statistics
on the faint side of the AGN luminosity function at $z>5$.
The knee of the $z\sim 5$ QSO luminosity function ($M_{1450}\sim -24$)
is presently undetermined and wide/deep imaging surveys (e.g. Euclid,
Rubin, Roman) will be instrumental in filling this gap.

We have carried out a comparison of the observed AGN luminosity
function at $z=5$ with three state-of-the-art theoretical models of
galaxy and AGN evolution, i.e. GAEA \citep{DeLucia24}, DELPHI
\citep{dayal24}, and CAT \citep{trinca22}. All the three models are
able to reproduce the observed space densities of faint AGNs at $M_{1450}\ge
-23$, while on the bright side ($M_{1450}\le -25$) only GAEA and CAT
are able to recover the observed bright-end slope of the QSO luminosity function
at $z=5$ (Fig. \ref{fig:lfresu1}). The predicted AGN space density
evolution from $z\sim 3$ to $z\sim 10$ by GAEA, DELPHI, and CAT (Fig. \ref{fig:evolphiz})
appears to be too fast with respect to the observational constraints
at $M_{1450}\sim -24$ \citep[e.g.][]{lai24}.

We consider three options for fitting the AGN luminosity function at
$z\sim 5$. A first, conservative, option includes the lowest
space densities from JWST at $M_{1450}\ge -22$, namely
\citet{grazian20,giallongo19,kocevski23,harikane23,matthee24,greene24}.
Option 2 results in an intermediate estimate for the AGN luminosity function, adopting
the larger space density of AGNs by \citet{maiolino23lf}. Finally, only for a consistency
check, we consider the high luminosity function of
\citet{scholtz23} (option 3), which includes also type 2 AGNs.
At the bright end, we always consider the luminosity functions of
\citet{grazian22,grazian23,glikman11} for all the three options.

The fit to the AGN space densities has been carried out with a
two-power-law parameterization of the luminosity function. The
covariance between the four parameters of the AGN luminosity function
has been taken into account, in order to produce realistic error bars
for the output physical quantities. The emissivity at 1450 {\AA}
rest-frame $\epsilon_{1450}$ has been derived by integrating the
luminosity function in the interval $-30<M_{1450}<-18$, where the bulk
of the production of the luminosity density is expected. The ionizing
emissivity at 912 {\AA} rest-frame $\epsilon_{912}$ has been
calculated starting from $\epsilon_{1450}$ by assuming an AGN spectral
slope of $\alpha_\nu=-1.7$, in agreement with \citet{lusso15}, and a
LyC escape fraction of 100\%. This is fully equivalent to the
assumption of an harder spectral slope of $\alpha_\nu=-0.7$
\citep{cristiani16} with a LyC escape fraction of $\sim 70\%$
\citep{grazian18,romano19}. The photo-ionization rate $\Gamma_{HI}$
has been obtained from $\epsilon_{912}$ and from the mean free path of
HI ionizing photons at $z\sim 5$ by \citet{worseck14}, which is consistent with
the one derived by \citet{Gaikwad23}. The errors on
these parameters have been propagated from the covariances of the AGN
luminosity functions, as shown in Fig. \ref{fig:lfparam1}.

The resulting best-fit to the AGN luminosity function for the three options has
been summarized in Table \ref{tab:bestfit}. The ionizing emissivity
$\epsilon_{912}$ and photo-ionization rate $\Gamma_{HI}$ have been
compared to the UV background measured in the Lyman-$\alpha$ forest or
by the proximity effects and it turns out to be fully consistent at
$z\sim 5$ even if we adopt the most conservative luminosity function
(option 1, see Fig. \ref{fig:gamma} and Fig. \ref{fig:eps912}). A preliminary derivation of
$\epsilon_{912}$ and $\Gamma_{HI}$ for AGNs at $z\sim 6$ is
consistent with the UV background measurements at this redshift,
even if more data are required to confirm this result.
The agreement of the photo-ionization rate
produced by AGNs at $z\sim 5-6$ with the literature values indicates
that faint AGNs (and bright QSOs) could be the main drivers of the
reionization event that is rapidly taking place at these redshifts.
The agreement of $\Gamma_{HI}$ produced by AGNs at $z\sim 5$ indicates
that most probably the contribution of star-forming galaxies to the
ionizing background could be sub-dominant at this redshift
\citep[see e.g.][]{madau24}.

A possible major role of high-z AGNs to the reionization process
could also be
corroborated by the differential redshift evolution of the AGN space
densities at different luminosities, as shown in Fig. \ref{fig:evolphiz}. If future
data would confirm that the space density of faint AGNs ($M_{1450}\sim
-18$) is almost constant from $z\sim 4$ to 10 and beyond, then it is
plausible that these AGNs have a key role in reionizing the
cosmological neutral hydrogen. This constant space density can also
naturally explain the excess of $M_{1450}\ge -22$ galaxies at very high-z observed
by JWST \citep[e.g.][]{finkelstein24}, without any crisis for the
$\Lambda$-CDM model.

In the near future, the faint end of the AGN luminosity function at
$z>5$ will be addressed by deep JWST spectroscopy, while the knee of
the AGN luminosity function will be constrained by upcoming wide and deep
surveys, e.g. Euclid Wide/Deep Surveys, Vera Rubin LSST, and Roman
Space Telescope. With these next generation surveys it will
be possible to check whether AGNs could be the long-sought pillars
of reionization. Time is thus promising for huge progresses in the
fields of cosmic reionization and SMBH evolution.

%---------------------------------------------------

\begin{acknowledgments}

We warmly thank the referee for the comments that allow us to
significantly improve the quality of this paper.

We acknowledge the support of the INAF GO/GTO Grant 2023 ``Finding
the Brightest Cosmic Beacons in the Universe with QUBRICS'' (PI Grazian).

AG, AB, and IS acknowledge the support of the INAF Mini Grant
2022 ``Learning Machine Learning techniques to dig up
high-z AGN in the Rubin-LSST Survey''.

AG acknowledges the PRIN 2022 project 2022ZSL4BL INSIGHT,
funded by the European Union – NextGenerationEU RFF M4C2 1.1.

We acknowledge financial contribution from the grant PRIN
INAF 2019 (RIC) 1.05.01.85.09: ``New light on the Intergalactic
Medium (NewIGM)''.

A.G. and F.F. acknowledge support from PRIN MIUR project ``Black Hole winds
and the Baryon Life Cycle of Galaxies: the stone-guest at the galaxy
evolution supper'', contract 2017-PH3WAT.

\end{acknowledgments}

%\facilities{Subaru:Hyper Suprime Cam, Gaia, Magellan:Baade (IMACS)}

%------------------------------------------------------------------------------

\appendix

\section{The AGN luminosity functions and their best fit parameterization}

\begin{figure}
\includegraphics[width=\linewidth]{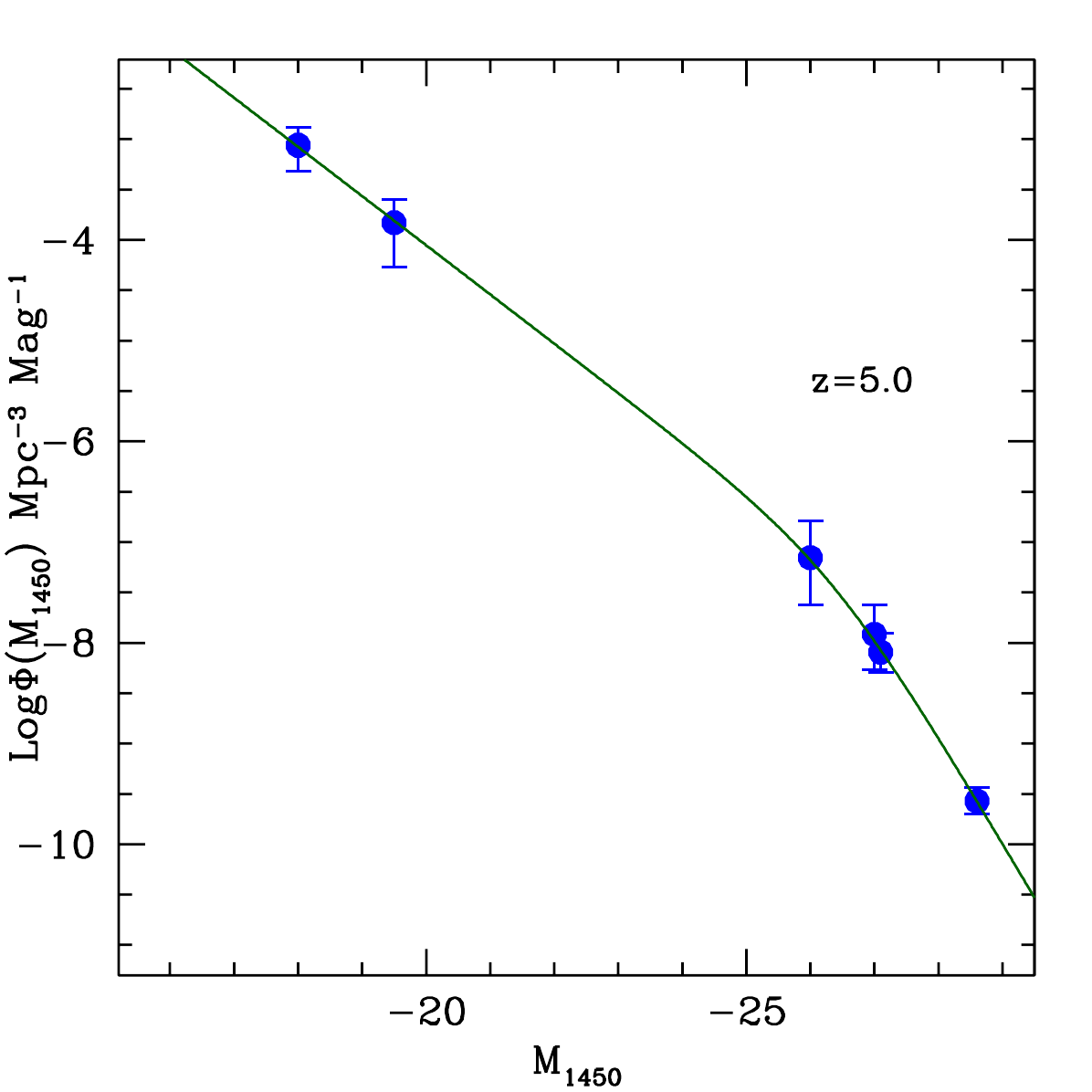}
\caption{The observed luminosity function of AGNs at z=5 (blue points),
with the best fit (green line) for Option 2.}
\label{fig:lfresu3}
\end{figure}

\begin{figure}
\includegraphics[width=\linewidth]{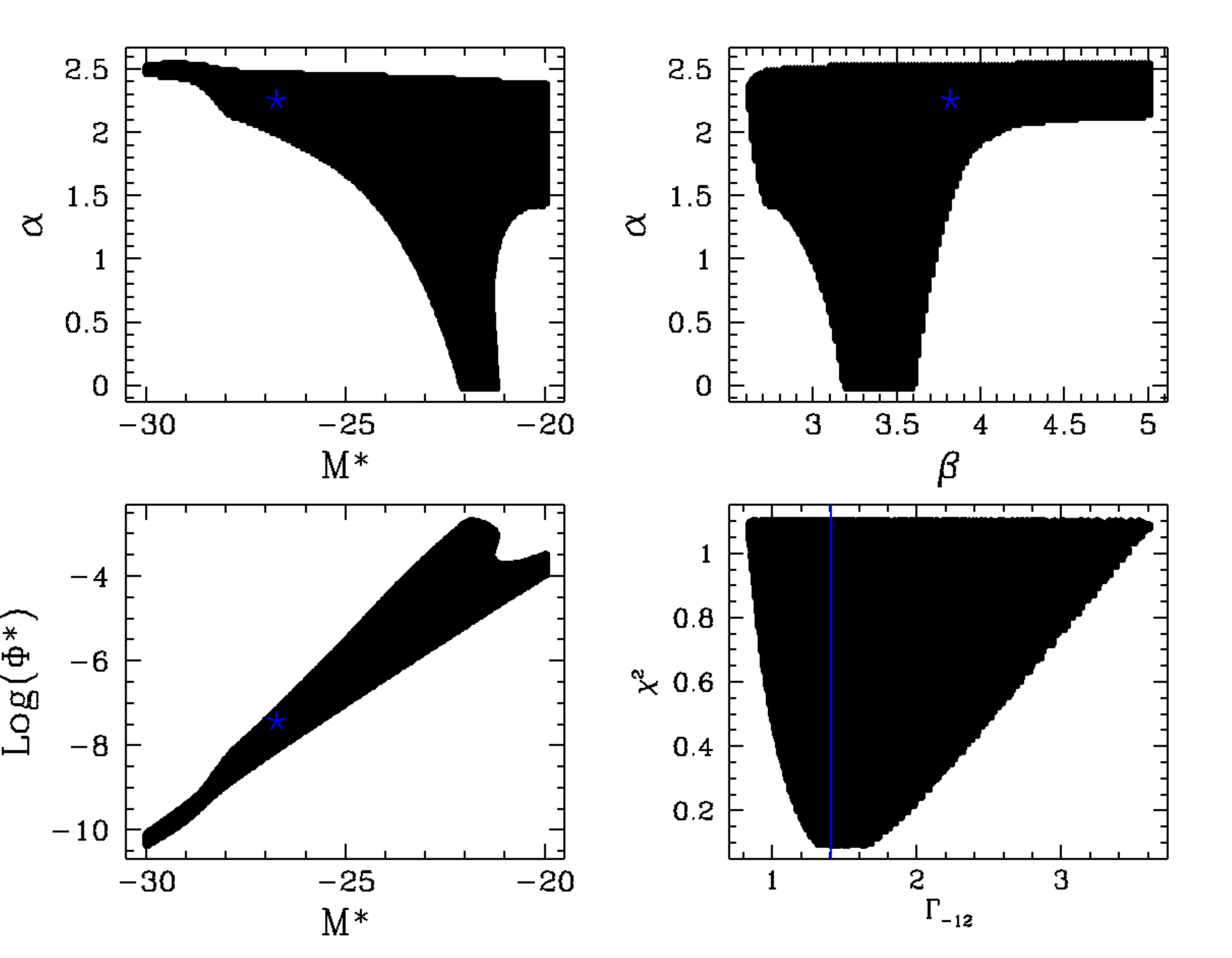}
\caption{The best-fit luminosity function parameters of AGNs at z=5
for Option 2. The dark areas show the confidence interval at
1$\sigma$ (68.3\% c.l.) for the parameters of the AGN luminosity
function at $z\sim 5$. The bottom right plot shows the permitted
interval for the photo-ionization rate $\Gamma_{-12}$ at 1$\sigma$ level.}
\label{fig:lfparam3}
\end{figure}

\begin{figure}
\includegraphics[width=\linewidth]{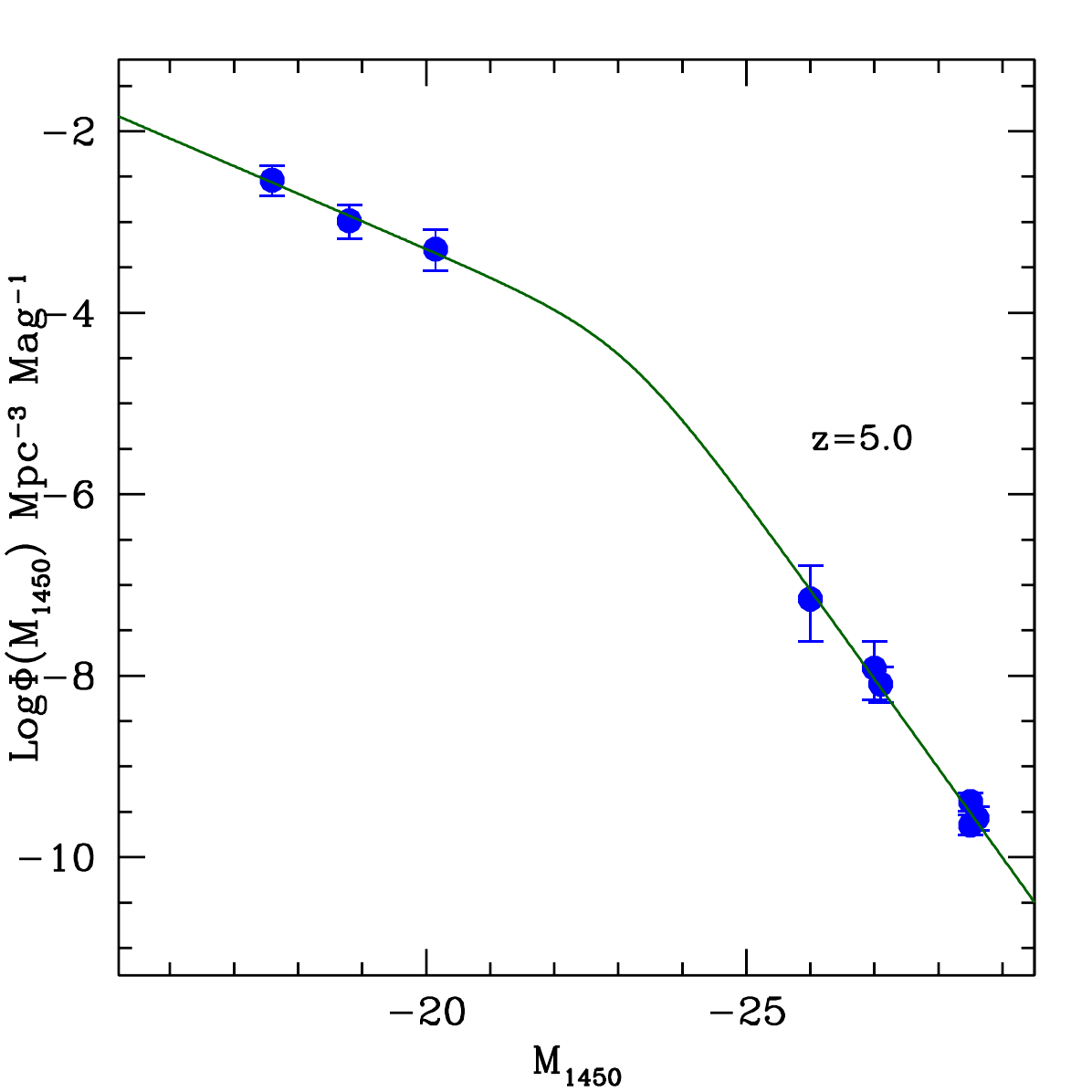}
\caption{The observed luminosity function of AGNs at z=5 (blue points),
with the best fit (dark green line) for Option 3.}
\label{fig:lfresu4}
\end{figure}

\begin{figure}
\includegraphics[width=\linewidth]{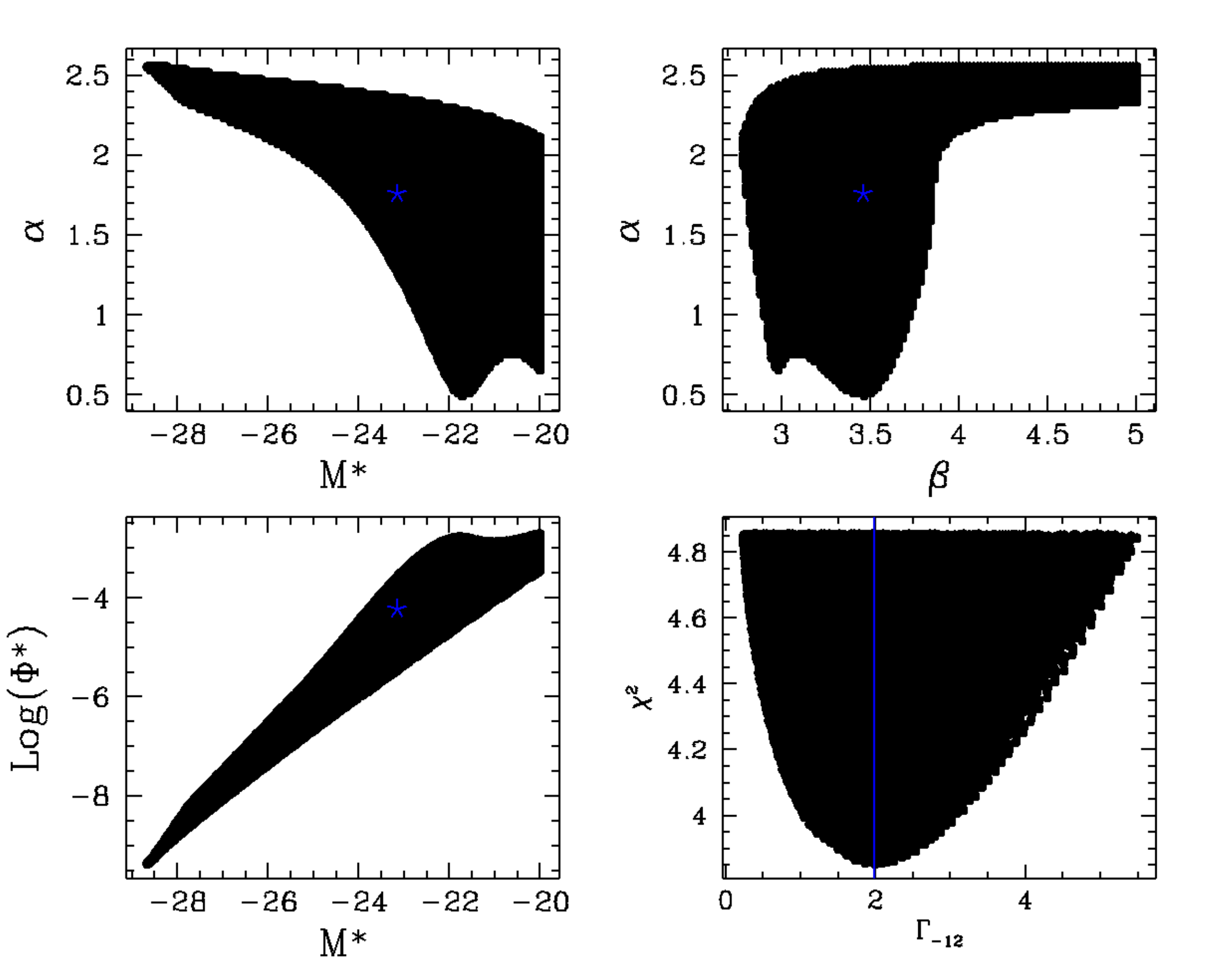}
\caption{The best-fit luminosity function parameters of AGNs at z=5
for Option 3. The dark areas show the confidence interval at
1$\sigma$ (68.3\% c.l.) for the parameters of the AGN luminosity
function at $z\sim 5$. The bottom right plot shows the permitted
interval for the photo-ionization rate $\Gamma_{-12}$ at 1$\sigma$ level.}
\label{fig:lfparam4}
\end{figure}

Fig. \ref{fig:lfresu3} and \ref{fig:lfresu4} show
the collections of the $z\sim 5$ AGN luminosity functions for options 2 and 3,
as summarised in Table \ref{tab:options}. The resulting best-fit parameters
and their covariance errors are shown in Fig. \ref{fig:lfparam3}
and \ref{fig:lfparam4}, respectively.

%------------------------------------------------------------------------------

\bibliography{References}{}

\begin{thebibliography}{}
\expandafter\ifx\csname natexlab\endcsname\relax\def\natexlab#1{#1}\fi
\providecommand{\url}[1]{\href{#1}{#1}}
\providecommand{\dodoi}[1]{doi:~\href{http://doi.org/#1}{\nolinkurl{#1}}}
\providecommand{\doeprint}[1]{\href{http://ascl.net/#1}{\nolinkurl{http://ascl.net/#1}}}
\providecommand{\doarXiv}[1]{\href{https://arxiv.org/abs/#1}{\nolinkurl{https://arxiv.org/abs/#1}}}

\bibitem[{{Adams} {et~al.}(2024){Adams}, {Conselice}, {Austin}, {Harvey},
  {Ferreira}, {Trussler}, {Juod{\v{z}}balis}, {Li}, {Windhorst}, {Cohen},
  {Jansen}, {Summers}, {Tompkins}, {Driver}, {Robotham}, {D'Silva}, {Yan},
  {Coe}, {Frye}, {Grogin}, {Koekemoer}, {Marshall}, {Pirzkal}, {Ryan},
  {Maksym}, {Rutkowski}, {Willmer}, {Hammel}, {Nonino}, {Bhatawdekar},
  {Wilkins}, {Bradley}, {Broadhurst}, {Cheng}, {Dole}, {Hathi}, \&
  {Zitrin}}]{adams23}
{Adams}, N.~J., {Conselice}, C.~J., {Austin}, D., {et~al.} 2024, \apj, 965,
  169, \dodoi{10.3847/1538-4357/ad2a7b}

\bibitem[{{Andika} {et~al.}(2024){Andika}, {Jahnke}, {Onoue}, {Silverman},
  {Fitriana}, {Bongiorno}, {Brinch}, {Casey}, {Faisst}, {Gillman}, {Gozaliasl},
  {Hayward}, {Hirschmann}, {Kocevski}, {Koekemoer}, {Kokorev}, {Lambrides},
  {Lee}, {Michael Rich}, {Trakhtenbrot}, {Megan Urry}, {Wilkins}, \&
  {Vijayan}}]{andika24}
{Andika}, I.~T., {Jahnke}, K., {Onoue}, M., {et~al.} 2024, \aap, 685, A25,
  \dodoi{10.1051/0004-6361/202349025}

\bibitem[{{Atek} {et~al.}(2024){Atek}, {Labb{\'e}}, {Furtak}, {Chemerynska},
  {Fujimoto}, {Setton}, {Miller}, {Oesch}, {Bezanson}, {Price}, {Dayal},
  {Zitrin}, {Kokorev}, {Weaver}, {Brammer}, {Dokkum}, {Williams}, {Cutler},
  {Feldmann}, {Fudamoto}, {Greene}, {Leja}, {Maseda}, {Muzzin}, {Pan},
  {Papovich}, {Nelson}, {Nanayakkara}, {Stark}, {Stefanon}, {Suess}, {Wang}, \&
  {Whitaker}}]{atek24}
{Atek}, H., {Labb{\'e}}, I., {Furtak}, L.~J., {et~al.} 2024, \nat, 626, 975,
  \dodoi{10.1038/s41586-024-07043-6}

\bibitem[{{Barchiesi} {et~al.}(2023){Barchiesi}, {Dessauges-Zavadsky},
  {Vignali}, {Pozzi}, {Marques-Chaves}, {Feltre}, {Faisst}, {B{\'e}thermin},
  {Cassata}, {Charlot}, {Fudamoto}, {Ginolfi}, {Ibar}, {Jones}, {Romano},
  {Schaerer}, {Vallini}, {Vanzella}, \& {Yan}}]{barchiesi23}
{Barchiesi}, L., {Dessauges-Zavadsky}, M., {Vignali}, C., {et~al.} 2023, \aap,
  675, A30, \dodoi{10.1051/0004-6361/202244838}

\bibitem[{{Barkana} \& {Loeb}(2001)}]{barkana01}
{Barkana}, R., \& {Loeb}, A. 2001, \physrep, 349, 125,
  \dodoi{10.1016/S0370-1573(01)00019-9}

\bibitem[{{Barro} {et~al.}(2024){Barro}, {P{\'e}rez-Gonz{\'a}lez}, {Kocevski},
  {McGrath}, {Trump}, {Simons}, {Somerville}, {Yung}, {Arrabal Haro}, {Akins},
  {Bagley}, {Cleri}, {Costantin}, {Davis}, {Dickinson}, {Finkelstein},
  {Giavalisco}, {G{\'o}mez-Guijarro}, {Hathi}, {Hirschmann}, {Holwerda},
  {Huertas-Company}, {Kartaltepe}, {Koekemoer}, {Lucas}, {Papovich}, {Pirzkal},
  {Seill{\'e}}, {Tacchella}, {Wuyts}, {Wilkins}, {de la Vega}, {Yang}, \&
  {Zavala}}]{barro24}
{Barro}, G., {P{\'e}rez-Gonz{\'a}lez}, P.~G., {Kocevski}, D.~D., {et~al.} 2024,
  \apj, 963, 128, \dodoi{10.3847/1538-4357/ad167e}

\bibitem[{{Basu} {et~al.}(2024){Basu}, {Garaldi}, \& {Ciardi}}]{basu24}
{Basu}, A., {Garaldi}, E., \& {Ciardi}, B. 2024, \mnras, 532, 841,
  \dodoi{10.1093/mnras/stae1488}

\bibitem[{{Baugh} {et~al.}(2019){Baugh}, {Gonzalez-Perez}, {Lagos}, {Lacey},
  {Helly}, {Jenkins}, {Frenk}, {Benson}, {Bower}, \& {Cole}}]{Baugh19}
{Baugh}, C.~M., {Gonzalez-Perez}, V., {Lagos}, C. D.~P., {et~al.} 2019, \mnras,
  483, 4922, \dodoi{10.1093/mnras/sty3427}

\bibitem[{{Becker} \& {Bolton}(2013)}]{BeckerBolton13}
{Becker}, G.~D., \& {Bolton}, J.~S. 2013, \mnras, 436, 1023,
  \dodoi{10.1093/mnras/stt1610}

\bibitem[{{Becker} {et~al.}(2021){Becker}, {D'Aloisio}, {Christenson}, {Zhu},
  {Worseck}, \& {Bolton}}]{becker21}
{Becker}, G.~D., {D'Aloisio}, A., {Christenson}, H.~M., {et~al.} 2021, \mnras,
  508, 1853, \dodoi{10.1093/mnras/stab2696}

\bibitem[{{Bogd{\'a}n} {et~al.}(2024){Bogd{\'a}n}, {Goulding}, {Natarajan},
  {Kov{\'a}cs}, {Tremblay}, {Chadayammuri}, {Volonteri}, {Kraft}, {Forman},
  {Jones}, {Churazov}, \& {Zhuravleva}}]{bogdan24}
{Bogd{\'a}n}, {\'A}., {Goulding}, A.~D., {Natarajan}, P., {et~al.} 2024, Nature
  Astronomy, 8, 126, \dodoi{10.1038/s41550-023-02111-9}

\bibitem[{{Bolton} \& {Haehnelt}(2007)}]{Bolton07}
{Bolton}, J.~S., \& {Haehnelt}, M.~G. 2007, \mnras, 382, 325,
  \dodoi{10.1111/j.1365-2966.2007.12372.x}

\bibitem[{{Bosman} {et~al.}(2022){Bosman}, {Davies}, {Becker}, {Keating},
  {Davies}, {Zhu}, {Eilers}, {D'Odorico}, {Bian}, {Bischetti}, {Cristiani},
  {Fan}, {Farina}, {Haehnelt}, {Hennawi}, {Kulkarni}, {Mesinger}, {Meyer},
  {Onoue}, {Pallottini}, {Qin}, {Ryan-Weber}, {Schindler}, {Walter}, {Wang}, \&
  {Yang}}]{bosman22}
{Bosman}, S. E.~I., {Davies}, F.~B., {Becker}, G.~D., {et~al.} 2022, \mnras,
  514, 55, \dodoi{10.1093/mnras/stac1046}

\bibitem[{{Boutsia} {et~al.}(2021){Boutsia}, {Grazian}, {Fontanot},
  {Giallongo}, {Menci}, {Calderone}, {Cristiani}, {D'Odorico}, {Cupani},
  {Guarneri}, \& {Omizzolo}}]{boutsia21}
{Boutsia}, K., {Grazian}, A., {Fontanot}, F., {et~al.} 2021, \apj, 912, 111,
  \dodoi{10.3847/1538-4357/abedb5}

\bibitem[{{Calderone} {et~al.}(2024){Calderone}, {Guarneri}, {Porru},
  {Cristiani}, {Grazian}, {Nicastro}, {Bischetti}, {Boutsia}, {Cupani},
  {D'Odorico}, {Feruglio}, \& {Fontanot}}]{calderone23}
{Calderone}, G., {Guarneri}, F., {Porru}, M., {et~al.} 2024, \aap, 683, A34,
  \dodoi{10.1051/0004-6361/202346625}

\bibitem[{{Calverley} {et~al.}(2011){Calverley}, {Becker}, {Haehnelt}, \&
  {Bolton}}]{calverley11}
{Calverley}, A.~P., {Becker}, G.~D., {Haehnelt}, M.~G., \& {Bolton}, J.~S.
  2011, \mnras, 412, 2543, \dodoi{10.1111/j.1365-2966.2010.18072.x}

\bibitem[{{Chardin} {et~al.}(2017){Chardin}, {Puchwein}, \&
  {Haehnelt}}]{chardin17}
{Chardin}, J., {Puchwein}, E., \& {Haehnelt}, M.~G. 2017, \mnras, 465, 3429,
  \dodoi{10.1093/mnras/stw2943}

\bibitem[{{Chemerynska} {et~al.}(2024){Chemerynska}, {Atek}, {Furtak},
  {Zitrin}, {Greene}, {Dayal}, {Weibel}, {Fujimoto}, {Kokorev}, {Goulding},
  {Williams}, {Nanayakkara}, {Bezanson}, {Brammer}, {Cutler}, {Labbe}, {Leja},
  {Pan}, {Price}, {van Dokkum}, {Wang}, {Weaver}, \&
  {Whitaker}}]{chemerynska24}
{Chemerynska}, I., {Atek}, H., {Furtak}, L.~J., {et~al.} 2024, \mnras, 531,
  2615, \dodoi{10.1093/mnras/stae1260}

\bibitem[{{Chworowsky} {et~al.}(2023){Chworowsky}, {Finkelstein},
  {Boylan-Kolchin}, {McGrath}, {Iyer}, {Papovich}, {Dickinson}, {Taylor},
  {Yung}, {Arrabal Haro}, {Bagley}, {Backhaus}, {Bhatawdekar}, {Cheng},
  {Cleri}, {Cole}, {Cooper}, {Costantin}, {Dekel}, {Franco}, {Fujimoto},
  {Hayward}, {Holwerda}, {Huertas-Company}, {Hirschmann}, {Hutchison},
  {Koekemoer}, {Larson}, {Li}, {Long}, {Lucas}, {Pirzkal}, {Rodighiero},
  {Somerville}, {Vanderhoof}, {de la Vega}, {Wilkins}, {Yang}, \&
  {Zavala}}]{chworowsky23}
{Chworowsky}, K., {Finkelstein}, S.~L., {Boylan-Kolchin}, M., {et~al.} 2023,
  arXiv e-prints, arXiv:2311.14804, \dodoi{10.48550/arXiv.2311.14804}

\bibitem[{{Cristiani} {et~al.}(2016){Cristiani}, {Serrano}, {Fontanot},
  {Vanzella}, \& {Monaco}}]{cristiani16}
{Cristiani}, S., {Serrano}, L.~M., {Fontanot}, F., {Vanzella}, E., \& {Monaco},
  P. 2016, \mnras, 462, 2478, \dodoi{10.1093/mnras/stw1810}

\bibitem[{{D'Aloisio} {et~al.}(2018){D'Aloisio}, {McQuinn}, {Davies}, \&
  {Furlanetto}}]{DAloisio18}
{D'Aloisio}, A., {McQuinn}, M., {Davies}, F.~B., \& {Furlanetto}, S.~R. 2018,
  \mnras, 473, 560, \dodoi{10.1093/mnras/stx2341}

\bibitem[{{Davies} {et~al.}(2018){Davies}, {Hennawi}, {Ba{\~n}ados},
  {Luki{\'c}}, {Decarli}, {Fan}, {Farina}, {Mazzucchelli}, {Rix}, {Venemans},
  {Walter}, {Wang}, \& {Yang}}]{davies18}
{Davies}, F.~B., {Hennawi}, J.~F., {Ba{\~n}ados}, E., {et~al.} 2018, \apj, 864,
  142, \dodoi{10.3847/1538-4357/aad6dc}

\bibitem[{{Davies} {et~al.}(2024){Davies}, {Bosman}, {Gaikwad}, {Nasir},
  {Hennawi}, {Becker}, {Haehnelt}, {D'Odorico}, {Bischetti}, {Eilers},
  {Keating}, {Kulkarni}, {Lai}, {Mazzucchelli}, {Qin}, {Satyavolu}, {Wang},
  {Yang}, \& {Zhu}}]{davies23}
{Davies}, F.~B., {Bosman}, S. E.~I., {Gaikwad}, P., {et~al.} 2024, \apj, 965,
  134, \dodoi{10.3847/1538-4357/ad1d5d}

\bibitem[{{Dayal} \& {Ferrara}(2018)}]{dayal18}
{Dayal}, P., \& {Ferrara}, A. 2018, \physrep, 780, 1,
  \dodoi{10.1016/j.physrep.2018.10.002}

\bibitem[{{Dayal} {et~al.}(2024){Dayal}, {Volonteri}, {Greene}, {Kokorev},
  {Goulding}, {Williams}, {Furtak}, {Zitrin}, {Atek}, {Chemerynska},
  {Feldmann}, {Glazebrook}, {Labbe}, {Nanayakkara}, {Oesch}, \&
  {Weaver}}]{dayal24}
{Dayal}, P., {Volonteri}, M., {Greene}, J.~E., {et~al.} 2024, arXiv e-prints,
  arXiv:2401.11242, \dodoi{10.48550/arXiv.2401.11242}

\bibitem[{{De Lucia} {et~al.}(2024){De Lucia}, {Fontanot}, {Xie}, \&
  {Hirschmann}}]{DeLucia24}
{De Lucia}, G., {Fontanot}, F., {Xie}, L., \& {Hirschmann}, M. 2024, \aap, 687,
  A68, \dodoi{10.1051/0004-6361/202349045}

\bibitem[{{D'Odorico} {et~al.}(2023){D'Odorico}, {Ba{\~n}ados}, {Becker},
  {Bischetti}, {Bosman}, {Cupani}, {Davies}, {Farina}, {Ferrara}, {Feruglio},
  {Mazzucchelli}, {Ryan-Weber}, {Schindler}, {Sodini}, {Venemans}, {Walter},
  {Chen}, {Lai}, {Zhu}, {Bian}, {Campo}, {Carniani}, {Cristiani}, {Davies},
  {Decarli}, {Drake}, {Eilers}, {Fan}, {Gaikwad}, {Gallerani}, {Greig},
  {Haehnelt}, {Hennawi}, {Keating}, {Kulkarni}, {Mesinger}, {Meyer},
  {Neeleman}, {Onoue}, {Pallottini}, {Qin}, {Rojas-Ruiz}, {Satyavolu},
  {Sebastian}, {Tripodi}, {Wang}, {Wolfson}, {Yang}, \&
  {Zanchettin}}]{dodorico23}
{D'Odorico}, V., {Ba{\~n}ados}, E., {Becker}, G.~D., {et~al.} 2023, \mnras,
  523, 1399, \dodoi{10.1093/mnras/stad1468}

\bibitem[{{Euclid Collaboration} {et~al.}(2019){Euclid Collaboration},
  {Barnett}, {Warren}, {Mortlock}, {Cuby}, {Conselice}, {Hewett}, {Willott},
  {Auricchio}, {Balaguera-Antol{\'\i}nez}, {Baldi}, {Bardelli}, {Bellagamba},
  {Bender}, {Biviano}, {Bonino}, {Bozzo}, {Branchini}, {Brescia}, {Brinchmann},
  {Burigana}, {Camera}, {Capobianco}, {Carbone}, {Carretero}, {Carvalho},
  {Castander}, {Castellano}, {Cavuoti}, {Cimatti}, {Cl{\'e}dassou}, {Congedo},
  {Conversi}, {Copin}, {Corcione}, {Coupon}, {Courtois}, {Cropper}, {Da Silva},
  {Duncan}, {Dusini}, {Ealet}, {Farrens}, {Fosalba}, {Fotopoulou},
  {Fourmanoit}, {Frailis}, {Fumana}, {Galeotta}, {Garilli}, {Gillard},
  {Gillis}, {Graci{\'a}-Carpio}, {Grupp}, {Hoekstra}, {Hormuth}, {Israel},
  {Jahnke}, {Kermiche}, {Kilbinger}, {Kirkpatrick}, {Kitching}, {Kohley},
  {Kubik}, {Kunz}, {Kurki-Suonio}, {Laureijs}, {Ligori}, {Lilje}, {Lloro},
  {Maiorano}, {Mansutti}, {Marggraf}, {Martinet}, {Marulli}, {Massey}, {Mauri},
  {Medinaceli}, {Mei}, {Mellier}, {Metcalf}, {Metge}, {Meylan}, {Moresco},
  {Moscardini}, {Munari}, {Neissner}, {Niemi}, {Nutma}, {Padilla}, {Paltani},
  {Pasian}, {Paykari}, {Percival}, {Pettorino}, {Polenta}, {Poncet},
  {Pozzetti}, {Raison}, {Renzi}, {Rhodes}, {Rix}, {Romelli}, {Roncarelli},
  {Rossetti}, {Saglia}, {Sapone}, {Scaramella}, {Schneider}, {Scottez},
  {Secroun}, {Serrano}, {Sirri}, {Stanco}, {Sureau}, {Tallada-Cresp{\'\i}},
  {Tavagnacco}, {Taylor}, {Tenti}, {Tereno}, {Toledo-Moreo}, {Torradeflot},
  {Valenziano}, {Vassallo}, {Wang}, {Zacchei}, {Zamorani}, {Zoubian}, \&
  {Zucca}}]{barnett19}
{Euclid Collaboration}, {Barnett}, R., {Warren}, S.~J., {et~al.} 2019, \aap,
  631, A85, \dodoi{10.1051/0004-6361/201936427}

\bibitem[{{Euclid Collaboration} {et~al.}(2024){Euclid Collaboration},
  {Mellier}, {Abdurro'uf}, {Acevedo Barroso}, {Ach{\'u}carro}, {Adamek},
  {Adam}, {Addison}, {Aghanim}, {Aguena}, {Ajani}, {Akrami}, {Al-Bahlawan},
  {Alavi}, {Albuquerque}, {Alestas}, {Alguero}, {Allaoui}, {Allen}, {Allevato},
  {Alonso-Tetilla}, {Altieri}, {Alvarez-Candal}, {Amara}, {Amendola}, {Amiaux},
  {Andika}, {Andreon}, {Andrews}, {Angora}, {Angulo}, {Annibali}, {Anselmi},
  {Anselmi}, {Arcari}, {Archidiacono}, {Aric{\`o}}, {Arnaud}, {Arnouts},
  {Asgari}, {Asorey}, {Atayde}, {Atek}, {Atrio-Barandela}, {Aubert}, {Aubourg},
  {Auphan}, {Auricchio}, {Aussel}, {Aussel}, {Avelino}, {Avgoustidis}, {Avila},
  {Awan}, {Azzollini}, {Baccigalupi}, {Bachelet}, {Bacon}, {Baes}, {Bagley},
  {Bahr-Kalus}, {Balaguera-Antolinez}, {Balbinot}, {Balcells}, {Baldi},
  {Baldry}, {Balestra}, {Ballardini}, {Ballester}, {Balogh}, {Ba{\~n}ados},
  {Barbier}, {Bardelli}, {Barreiro}, {Barriere}, {Barros}, {Barthelemy},
  {Bartolo}, {Basset}, {Battaglia}, {Battisti}, {Baugh}, {Baumont},
  {Bazzanini}, {Beaulieu}, {Beckmann}, {Belikov}, {Bel}, {Bellagamba}, {Bella},
  {Bellini}, {Benabed}, {Bender}, {Benevento}, {Bennett}, {Benson},
  {Bergamini}, {Bermejo-Climent}, {Bernardeau}, {Bertacca}, {Berthe},
  {Berthier}, {Bethermin}, {Beutler}, {Bevillon}, {Bhargava}, {Bhatawdekar},
  {Bisigello}, {Biviano}, {Blake}, {Blanchard}, {Blazek}, {Blot}, {Bosco},
  {Bodendorf}, {Boenke}, {B{\"o}hringer}, {Bolzonella}, {Bonchi}, {Bonici},
  {Bonino}, {Bonino}, {Bonvin}, {Bon}, {Booth}, {Borgani}, {Borlaff},
  {Borsato}, {Bosco}, {Bose}, {Botticella}, {Boucaud}, {Bouche}, {Boucher},
  {Boutigny}, {Bouvard}, {Bouy}, {Bowler}, {Bozza}, {Bozzo}, {Branchini},
  {Brau-Nogue}, {Brekke}, {Bremer}, {Brescia}, {Breton}, {Brinchmann},
  {Brinckmann}, {Brockley-Blatt}, {Brodwin}, {Brouard}, {Brown}, {Bruton},
  {Bucko}, {Buddelmeijer}, {Buenadicha}, {Buitrago}, {Burger}, {Burigana},
  {Busillo}, {Busonero}, {Cabanac}, {Cabayol-Garcia}, {Cagliari}, {Caillat},
  {Caillat}, {Calabrese}, {Calabro}, {Calderone}, {Calura}, {Camacho Quevedo},
  {Camera}, {Campos}, {Canas-Herrera}, {Candini}, {Cantiello}, {Capobianco},
  {Cappellaro}, {Cappelluti}, {Cappi}, {Caputi}, {Cara}, {Carbone}, {Cardone},
  {Carella}, {Carlberg}, {Carle}, {Carminati}, {Caro}, {Carrasco}, {Carretero},
  {Carrilho}, {Carron Duque}, {Carry}, {Carvalho}, {Carvalho}, {Casas},
  {Casas}, {Casenove}, {Casey}, {Cassata}, {Castander}, {Castelao},
  {Castellano}, {Castiblanco}, {Castignani}, {Castro}, {Cavet}, {Cavuoti},
  {Chabaud}, {Chambers}, {Charles}, {Charlot}, {Chartab}, {Chary}, {Chaumeil},
  {Cho}, {Chon}, {Ciancetta}, {Ciliegi}, {Cimatti}, {Cimino}, {Cioni},
  {Claydon}, {Cleland}, {Cl{\'e}ment}, {Clements}, {Clerc}, {Clesse}, {Codis},
  {Cogato}, {Colbert}, {Cole}, {Coles}, {Collett}, {Collins}, {Colodro-Conde},
  {Colombo}, {Combes}, {Conforti}, {Congedo}, {Conseil}, {Conselice},
  {Contarini}, {Contini}, {Conversi}, {Cooray}, {Copin}, {Corasaniti},
  {Corcho-Caballero}, {Corcione}, {Cordes}, {Corpace}, {Correnti}, {Costanzi},
  {Costille}, {Courbin}, {Courcoult Mifsud}, {Courtois}, {Cousinou}, {Covone},
  {Cowell}, {Cragg}, {Cresci}, {Cristiani}, {Crocce}, {Cropper}, {Crouzet},
  {Csizi}, {Cuby}, {Cucchetti}, {Cucciati}, {Cuillandre}, {Cunha}, {Cuozzo},
  {Daddi}, {D'Addona}, {Dafonte}, {Dagoneau}, {Dalessandro}, {Dalton},
  {D'Amico}, {Dannerbauer}, {Danto}, {Das}, {Da Silva}, {da Silva}, {Daste},
  {Davies}, {Davini}, {de Boer}, {Decarli}, {De Caro}, {Degaudenzi}, {Degni},
  {de Jong}, {de la Bella}, {de la Torre}, {Delhaise}, {Delley}, {Delucchi},
  {De Lucia}, {Denniston}, {De Paolis}, {De Petris}, {Derosa}, {Desai},
  {Desjacques}, {Despali}, {Desprez}, {De Vicente-Albendea}, {Deville}, {Dias},
  {D{\'\i}az-S{\'a}nchez}, {Diaz}, {Di Domizio}, {Diego}, {Di Ferdinando}, {Di
  Giorgio}, {Dimauro}, {Dinis}, {Dolag}, {Dolding}, {Dole}, {Dom{\'\i}nguez
  S{\'a}nchez}, {Dor{\'e}}, {Dournac}, {Douspis}, {Dreihahn}, {Droge}, {Dryer},
  {Dubath}, {Duc}, {Ducret}, {Duffy}, {Dufresne}, {Duncan}, {Dupac}, {Duret},
  {Durrer}, {Durret}, {Dusini}, {Ealet}, {Eggemeier}, {Eisenhardt}, {Elbaz},
  {Elkhashab}, {Ellien}, {Endicott}, {Enia}, {Erben}, {Escartin Vigo},
  {Escoffier}, {Escudero Sanz}, {Essert}, {Ettori}, {Ezziati}, {Fabbian},
  {Fabricius}, {Fang}, {Farina}, {Farina}, {Farinelli}, {Farrens}, {Faustini},
  {Feltre}, {Ferguson}, {Ferrando}, {Ferrari}, {Ferr{\'e}-Mateu}, {Ferreira},
  {Ferreras}, {Ferrero}, {Ferriol}, {Ferruit}, {Filleul}, {Finelli},
  {Finkelstein}, {Finoguenov}, {Fiorini}, {Flentge}, {Focardi}, {Fonseca},
  {Fontana}, {Fontanot}, {Fornari}, {Fosalba}, {Fossati}, {Fotopoulou},
  {Fouchez}, {Fourmanoit}, {Frailis}, {Fraix-Burnet}, {Franceschi}, {Franco},
  {Franzetti}, {Freihoefer}, {Frittoli}, {Frugier}, {Frusciante}, {Fumagalli},
  {Fumagalli}, {Fumana}, {Fu}, {Gabarra}, {Galeotta}, {Galluccio}, {Ganga},
  {Gao}, {Garc{\'\i}a-Bellido}, {Garcia}, {Gardner}, {Garilli},
  {Gaspar-Venancio}, {Gasparetto}, {Gautard}, {Gavazzi}, {Gaztanaga},
  {Genolet}, {Genova Santos}, {Gentile}, {George}, {Ghaffari}, {Giacomini},
  {Gianotti}, {Gibb}, {Gillard}, {Gillis}, {Ginolfi}, {Giocoli}, {Girardi},
  {Giri}, {Goh}, {G{\'o}mez-Alvarez}, {Gonzalez}, {Gonzalez}, {Gonzalez},
  {Gouyou Beauchamps}, {Gozaliasl}, {Gracia-Carpio}, {Grandis}, {Granett},
  {Granvik}, {Grazian}, {Gregorio}, {Grenet}, {Grillo}, {Grupp}, {Gruppioni},
  {Gruppuso}, {Guerbuez}, {Guerrini}, {Guidi}, {Guillard}, {Gutierrez},
  {Guttridge}, {Guzzo}, {Gwyn}, {Haapala}, {Haase}, {Haddow}, {Hailey}, {Hall},
  {Hall}, {Hamaus}, {Haridasu}, {Harnois-D{\'e}raps}, {Harper}, {Hartley},
  {Hasinger}, {Hassani}, {Hatch}, {Haugan}, {H{\"a}u{\ss}ler}, {Heavens},
  {Heisenberg}, {Helmi}, {Helou}, {Hemmati}, {Henares}, {Herent},
  {Hern{\'a}ndez-Monteagudo}, {Heuberger}, {Hewett}, {Heydenreich},
  {Hildebrandt}, {Hirschmann}, {Hjorth}, {Hoar}, {Hoekstra}, {Holland},
  {Holliman}, {Holmes}, {Hook}, {Horeau}, {Hormuth}, {Hornstrup}, {Hosseini},
  {Hu}, {Hudelot}, {Hudson}, {Huertas-Company}, {Huff}, {Hughes}, {Humphrey},
  {Hunt}, {Huynh}, {Ibata}, {Ichikawa}, {Iglesias-Groth}, {Ilbert}, {Ili{\'c}},
  {Ingoglia}, {Iodice}, {Israel}, {Israelsson}, {Izzo}, {Jablonka}, {Jackson},
  {Jacobson}, {Jafariyazani}, {Jahnke}, {Jansen}, {Jarvis}, {Jasche}, {Jauzac},
  {Jeffrey}, {Jhabvala}, {Jimenez-Teja}, {Jimenez Mu{\~n}oz}, {Joachimi},
  {Johansson}, {Joudaki}, {Jullo}, {Kajava}, {Kang}, {Kannawadi}, {Kansal},
  {Karagiannis}, {K{\"a}rcher}, {Kashlinsky}, {Kazandjian}, {Keck},
  {Keih{\"a}nen}, {Kerins}, {Kermiche}, {Khalil}, {Kiessling}, {Kiiveri},
  {Kilbinger}, {Kim}, {King}, {Kirkpatrick}, {Kitching}, {Kluge}, {Knabenhans},
  {Knapen}, {Knebe}, {Kneib}, {Kohley}, {Koopmans}, {Koskinen}, {Koulouridis},
  {Kou}, {Kov{\'a}cs}, {Kova\{{\v{c}}\}i{\'c}}, {Kowalczyk}, {Koyama},
  {Kraljic}, {Krause}, {Kruk}, {Kubik}, {Kuchner}, {Kuijken}, {K{\"u}mmel},
  {Kunz}, {Kurki-Suonio}, {Lacasa}, {Lacey}, {La Franca}, {Lagarde}, {Lahav},
  {Laigle}, {La Marca}, {La Marle}, {Lamine}, {Lam}, {Lan{\c{c}}on}, {Landt},
  {Langer}, {Lapi}, {Larcheveque}, {Larsen}, {Lattanzi}, {Laudisio}, {Laugier},
  {Laureijs}, {Lavaux}, {Lawrenson}, {Lazanu}, {Lazeyras}, {Le Boulc'h}, {Le
  Brun}, {Le Brun}, {Leclercq}, {Lee}, {Le Graet}, {Legrand}, {Leirvik}, {Le
  Jeune}, {Lembo}, {Le Mignant}, {Lepinzan}, {Lepori}, {Lesci}, {Lesgourgues},
  {Leuzzi}, {Levi}, {Liaudat}, {Libet}, {Liebing}, {Ligori}, {Lilje}, {Lin},
  {Linde}, {Linder}, {Lindholm}, {Linke}, {Li}, {Liu}, {Lloro}, {Lobo},
  {Lodieu}, {Lombardi}, {Lombriser}, {Lonare}, {Longo}, {L{\'o}pez-Caniego},
  {Lopez Lopez}, {Alvarez}, {Loureiro}, {Loveday}, {Lusso}, {Macias-Perez},
  {Maciaszek}, {Magliocchetti}, {Magnard}, {Magnier}, {Magro}, {Mahler},
  {Mainetti}, {Maino}, {Maiorano}, {Maiorano}, {Malavasi}, {Mamon}, {Mancini},
  {Mandelbaum}, {Manera}, {Manj{\'o}n-Garc{\'\i}a}, {Mannucci}, {Mansutti},
  {Manteiga Outeiro}, {Maoli}, {Maraston}, {Marcin}, {Marcos-Arenal},
  {Margalef-Bentabol}, {Marggraf}, {Marinucci}, {Marinucci}, {Markovic},
  {Marleau}, {Marpaud}, {Martignac}, {Mart{\'\i}n-Fleitas}, {Martin-Moruno},
  {Martin}, {Martinelli}, {Martinet}, {Martin}, {Martins}, {Marulli},
  {Massari}, {Massey}, {Masters}, {Matarrese}, {Matsuoka}, {Matthew},
  {Maughan}, {Mauri}, {Maurin}, {Maurogordato}, {McCarthy}, {McConnachie},
  {McCracken}, {McDonald}, {McEwen}, {McPartland}, {Medinaceli}, {Mehta},
  {Mei}, {Melchior}, {Melin}, {M{\'e}nard}, {Mendes}, {Mendez-Abreu},
  {Meneghetti}, {Mercurio}, {Merlin}, {Metcalf}, {Meylan}, {Migliaccio},
  {Mignoli}, {Miller}, {Miluzio}, {Milvang-Jensen}, {Mimoso}, {Miquel},
  {Miyatake}, {Mobasher}, {Mohr}, {Monaco}, {Mongui{\'o}}, {Montoro}, {Mora},
  {Moradinezhad Dizgah}, {Moresco}, {Moretti}, {Morgante}, {Morisset},
  {Moriya}, {Morris}, {Mortlock}, {Moscardini}, {Mota}, {Moustakas}, {Moutard},
  {M{\"u}ller}, {Munari}, {Murphree}, {Murray}, {Murray}, {Musi}, {Nadathur},
  {Nagam}, {Nagao}, {Naidoo}, {Nakajima}, {Nally}, {Natoli}, {Navarro-Alsina},
  {Navarro Girones}, {Neissner}, {Nersesian}, {Nesseris}, {Nguyen-Kim},
  {Nicastro}, {Nichol}, {Nielbock}, {Niemi}, {Nieto}, {Nilsson}, {Noller},
  {Norberg}, {Nourizonoz}, {Ntelis}, {Nucita}, {Nugent}, {Nunes}, {Nutma},
  {Ocampo}, {Odier}, {Oesch}, {Oguri}, {Magalhaes Oliveira}, {Onoue},
  {Oosterbroek}, {Oppizzi}, {Ordenovic}, {Osato}, {Pacaud}, {Pace}, {Padilla},
  {Paech}, {Pagano}, {Page}, {Palazzi}, {Paltani}, {Pamuk}, {Pandolfi},
  {Paoletti}, {Paolillo}, {Papaderos}, {Pardede}, {Parimbelli}, {Parmar},
  {Partmann}, {Pasian}, {Passalacqua}, {Paterson}, {Patrizii}, {Pattison},
  {Paulino-Afonso}, {Paviot}, {Peacock}, {Pearce}, {Pedersen}, {Peel},
  {Peletier}, {Pellejero Ibanez}, {Pello}, {Penny}, {Percival},
  {Perez-Garrido}, {Perotto}, {Pettorino}, {Pezzotta}, {Pezzuto}, {Philippon},
  {Piersanti}, {Pietroni}, {Piga}, {Pilo}, {Pires}, {Pisani}, {Pizzella},
  {Pizzuti}, {Plana}, {Polenta}, {Pollack}, {Poncet}, {P{\"o}ntinen}, {Pool},
  {Popa}, {Popa}, {Popp}, {Porciani}, {Porth}, {Potter}, {Poulain},
  {Pourtsidou}, {Pozzetti}, {Prandoni}, {Pratt}, {Prezelus}, {Prieto}, {Pugno},
  {Quai}, {Quilley}, {Racca}, {Raccanelli}, {R{\'a}cz}, {Radinovi{\'c}},
  {Radovich}, {Ragagnin}, {Ragnit}, {Raison}, {Ramos-Chernenko}, {Ranc},
  {Raylet}, {Rebolo}, {Refregier}, {Reimberg}, {Reiprich}, {Renk}, {Renzi},
  {Retre}, {Revaz}, {Reyl{\'e}}, {Reynolds}, {Rhodes}, {Ricci}, {Ricci},
  {Riccio}, {Ricken}, {Rissanen}, {Risso}, {Rix}, {Robin}, {Rocca-Volmerange},
  {Rocci}, {Rodenhuis}, {Rodighiero}, {Rodriguez Monroy}, {Rollins},
  {Romanello}, {Roman}, {Romelli}, {Romero-Gomez}, {Roncarelli}, {Rosati},
  {Rosset}, {Rossetti}, {Roster}, {Rottgering}, {Rozas-Fern{\'a}ndez}, {Ruane},
  {Rubino-Martin}, {Rudolph}, {Ruppin}, {Rusholme}, {Sacquegna},
  {S{\'a}ez-Casares}, {Saga}, {Saglia}, {Sahl{\'e}n}, {Saifollahi}, {Sakr},
  {Salvalaggio}, {Salvaterra}, {Salvati}, {Salvato}, {Salvignol},
  {S{\'a}nchez}, {Sanchez}, {Sanders}, {Sapone}, {Saponara}, {Sarpa}, {Sarron},
  {Sartori}, {Sassolas}, {Sauniere}, {Sauvage}, {Sawicki}, {Scaramella},
  {Scarlata}, {Scharr{\'e}}, {Schaye}, {Schewtschenko}, {Schindler},
  {Schinnerer}, {Schirmer}, {Schmidt}, {Schmidt}, {Schmidt}, {Schneider},
  {Schneider}, {Schneider}, {Sch{\"o}neberg}, {Schrabback}, {Schultheis},
  {Schulz}, {Schwartz}, {Sciotti}, {Scodeggio}, {Scognamiglio}, {Scott},
  {Scottez}, {Secroun}, {Sefusatti}, {Seidel}, {Seiffert}, {Sellentin},
  {Selwood}, {Semboloni}, {Sereno}, {Serjeant}, {Serrano}, {Shankar},
  {Sharples}, {Short}, {Shulevski}, {Shuntov}, {Sias}, {Sikkema}, {Silvestri},
  {Simon}, {Sirignano}, {Sirri}, {Skottfelt}, {Slezak}, {Sluse}, {Smith},
  {Smith}, {Smith}, {Smit}, {Soldano}, {Solheim}, {Sorce}, {Sorrenti},
  {Soubrie}, {Spinoglio}, {Spurio Mancini}, {Stadel}, {Stagnaro}, {Stanco},
  {Stanford}, {Starck}, {Stassi}, {Steinwagner}, {Stern}, {Stone}, {Strada},
  {Strafella}, {Stramaccioni}, {Surace}, {Sureau}, {Suyu}, {Swindells},
  {Szafraniec}, {Szapudi}, {Taamoli}, {Talia}, {Tallada-Cresp{\'\i}},
  {Tanidis}, {Tao}, {Tarr{\'\i}o}, {Tavagnacco}, {Taylor}, {Taylor}, {Taylor},
  {Teixeira}, {Tenti}, {Teodoro Idiago}, {Teplitz}, {Tereno}, {Tessore},
  {Testa}, {Testera}, {Tewes}, {Teyssier}, {Theret}, {Thizy}, {Thomas}, {Toba},
  {Toft}, {Toledo-Moreo}, {Tolstoy}, {Tommasi}, {Torbaniuk}, {Torradeflot},
  {Tortora}, {Tosi}, {Tosti}, {Trifoglio}, {Troja}, {Trombetti}, {Tronconi},
  {Tsedrik}, {Tsyganov}, {Tucci}, {Tutusaus}, {Uhlemann}, {Ulivi}, {Urbano},
  {Vacher}, {Vaillon}, {Valdes}, {Valentijn}, {Valenziano}, {Valieri},
  {Valiviita}, {Van den Broeck}, {Vassallo}, {Vavrek}, {Venemans}, {Venhola},
  {Ventura}, {Verdoes Kleijn}, {Vergani}, {Verma}, {Vernizzi}, {Veropalumbo},
  {Verza}, {Vescovi}, {Vibert}, {Viel}, {Vielzeuf}, {Viglione}, {Viitanen},
  {Villaescusa-Navarro}, {Vinciguerra}, {Visticot}, {Voggel}, {von
  Wietersheim-Kramsta}, {Vriend}, {Wachter}, {Walmsley}, {Walth}, {Walton},
  {Walton}, {Wander}, {Wang}, {Wang}, {Weaver}, {Weller}, {Whalen}, {Wiesmann},
  {Wilde}, {Williams}, {Winther}, {Wittje}, {Wong}, {Wright}, {Yankelevich},
  {Yeung}, {Youles}, {Yung}, {Zacchei}, {Zalesky}, {Zamorani}, {Zamorano
  Vitorelli}, {Zanoni Marc}, {Zennaro}, {Zerbi}, {Zinchenko}, {Zoubian},
  {Zucca}, \& {Zumalacarregui}}]{mellier24}
{Euclid Collaboration}, {Mellier}, Y., {Abdurro'uf}, {et~al.} 2024, arXiv
  e-prints, arXiv:2405.13491.
\newblock \doarXiv{2405.13491}

\bibitem[{{Ferrara} {et~al.}(2023){Ferrara}, {Pallottini}, \&
  {Dayal}}]{ferrara23}
{Ferrara}, A., {Pallottini}, A., \& {Dayal}, P. 2023, \mnras, 522, 3986,
  \dodoi{10.1093/mnras/stad1095}

\bibitem[{{Finkelstein} \& {Bagley}(2022)}]{FB22}
{Finkelstein}, S.~L., \& {Bagley}, M.~B. 2022, \apj, 938, 25,
  \dodoi{10.3847/1538-4357/ac89eb}

\bibitem[{{Finkelstein} {et~al.}(2024){Finkelstein}, {Leung}, {Bagley},
  {Dickinson}, {Ferguson}, {Papovich}, {Akins}, {Arrabal Haro}, {Dav{\'e}},
  {Dekel}, {Kartaltepe}, {Kocevski}, {Koekemoer}, {Pirzkal}, {Somerville},
  {Yung}, {Amor{\'\i}n}, {Backhaus}, {Behroozi}, {Bisigello}, {Bromm}, {Casey},
  {Ch{\'a}vez Ortiz}, {Cheng}, {Chworowsky}, {Cleri}, {Cooper}, {Davis}, {de la
  Vega}, {Elbaz}, {Franco}, {Fontana}, {Fujimoto}, {Giavalisco}, {Grogin},
  {Holwerda}, {Huertas-Company}, {Hirschmann}, {Iyer}, {Jogee}, {Jung},
  {Larson}, {Lucas}, {Mobasher}, {Morales}, {Morley}, {Mukherjee},
  {P{\'e}rez-Gonz{\'a}lez}, {Ravindranath}, {Rodighiero}, {Rowland},
  {Tacchella}, {Taylor}, {Trump}, \& {Wilkins}}]{finkelstein24}
{Finkelstein}, S.~L., {Leung}, G. C.~K., {Bagley}, M.~B., {et~al.} 2024, \apjl,
  969, L2, \dodoi{10.3847/2041-8213/ad4495}

\bibitem[{{Fiore} {et~al.}(2023){Fiore}, {Ferrara}, {Bischetti}, {Feruglio}, \&
  {Travascio}}]{fiore23}
{Fiore}, F., {Ferrara}, A., {Bischetti}, M., {Feruglio}, C., \& {Travascio}, A.
  2023, \apjl, 943, L27, \dodoi{10.3847/2041-8213/acb5f2}

\bibitem[{{Fiore} {et~al.}(2008){Fiore}, {Grazian}, {Santini}, {Puccetti},
  {Brusa}, {Feruglio}, {Fontana}, {Giallongo}, {Comastri}, {Gruppioni},
  {Pozzi}, {Zamorani}, \& {Vignali}}]{fiore08}
{Fiore}, F., {Grazian}, A., {Santini}, P., {et~al.} 2008, \apj, 672, 94,
  \dodoi{10.1086/523348}

\bibitem[{{Flury} {et~al.}(2022{\natexlab{a}}){Flury}, {Jaskot}, {Ferguson},
  {Worseck}, {Makan}, {Chisholm}, {Saldana-Lopez}, {Schaerer}, {McCandliss},
  {Wang}, {Ford}, {Heckman}, {Ji}, {Giavalisco}, {Amorin}, {Atek}, {Blaizot},
  {Borthakur}, {Carr}, {Castellano}, {Cristiani}, {De Barros}, {Dickinson},
  {Finkelstein}, {Fleming}, {Fontanot}, {Garel}, {Grazian}, {Hayes}, {Henry},
  {Mauerhofer}, {Micheva}, {Oey}, {Ostlin}, {Papovich}, {Pentericci},
  {Ravindranath}, {Rosdahl}, {Rutkowski}, {Santini}, {Scarlata}, {Teplitz},
  {Thuan}, {Trebitsch}, {Vanzella}, {Verhamme}, \& {Xu}}]{flury22a}
{Flury}, S.~R., {Jaskot}, A.~E., {Ferguson}, H.~C., {et~al.}
  2022{\natexlab{a}}, \apjs, 260, 1, \dodoi{10.3847/1538-4365/ac5331}

\bibitem[{{Flury} {et~al.}(2022{\natexlab{b}}){Flury}, {Jaskot}, {Ferguson},
  {Worseck}, {Makan}, {Chisholm}, {Saldana-Lopez}, {Schaerer}, {McCandliss},
  {Xu}, {Wang}, {Oey}, {Ford}, {Heckman}, {Ji}, {Giavalisco}, {Amor{\'\i}n},
  {Atek}, {Blaizot}, {Borthakur}, {Carr}, {Castellano}, {De Barros},
  {Dickinson}, {Finkelstein}, {Fleming}, {Fontanot}, {Garel}, {Grazian},
  {Hayes}, {Henry}, {Mauerhofer}, {Micheva}, {Ostlin}, {Papovich},
  {Pentericci}, {Ravindranath}, {Rosdahl}, {Rutkowski}, {Santini}, {Scarlata},
  {Teplitz}, {Thuan}, {Trebitsch}, {Vanzella}, \& {Verhamme}}]{flury22b}
---. 2022{\natexlab{b}}, \apj, 930, 126, \dodoi{10.3847/1538-4357/ac61e4}

\bibitem[{{Fontanot} {et~al.}(2012){Fontanot}, {Cristiani}, \&
  {Vanzella}}]{fontanot12}
{Fontanot}, F., {Cristiani}, S., \& {Vanzella}, E. 2012, \mnras, 425, 1413,
  \dodoi{10.1111/j.1365-2966.2012.21594.x}

\bibitem[{{Fontanot} {et~al.}(2020{\natexlab{a}}){Fontanot}, {De Lucia},
  {Hirschmann}, {Xie}, {Monaco}, {Menci}, {Fiore}, {Feruglio}, {Cristiani}, \&
  {Shankar}}]{fontanot20}
{Fontanot}, F., {De Lucia}, G., {Hirschmann}, M., {et~al.} 2020{\natexlab{a}},
  \mnras, 496, 3943, \dodoi{10.1093/mnras/staa1716}

\bibitem[{{Fontanot} {et~al.}(2020{\natexlab{b}}){Fontanot}, {De Lucia},
  {Hirschmann}, {Xie}, {Monaco}, {Menci}, {Fiore}, {Feruglio}, {Cristiani}, \&
  {Shankar}}]{fontanot20ff}
---. 2020{\natexlab{b}}, \mnras, 496, 3943, \dodoi{10.1093/mnras/staa1716}

\bibitem[{{Fontanot} {et~al.}(2023){Fontanot}, {Cristiani}, {Grazian},
  {Haardt}, {D'Odorico}, {Boutsia}, {Calderone}, {Cupani}, {Guarneri},
  {Fiorin}, \& {Rodighiero}}]{fontanot23}
{Fontanot}, F., {Cristiani}, S., {Grazian}, A., {et~al.} 2023, \mnras, 520,
  740, \dodoi{10.1093/mnras/stad189}

\bibitem[{{Fujimoto} {et~al.}(2023){Fujimoto}, {Wang}, {Weaver}, {Kokorev},
  {Atek}, {Bezanson}, {Labbe}, {Brammer}, {Greene}, {Chemerynska}, {Dayal}, {de
  Graaff}, {Furtak}, {Oesch}, {Setton}, {Price}, {Miller}, {Williams},
  {Whitaker}, {Zitrin}, {Cutler}, {Leja}, {Pan}, {Coe}, {van Dokkum},
  {Feldmann}, {Fudamoto}, {Goulding}, {Khullar}, {Marchesini}, {Maseda},
  {Nanayakkara}, {Nelson}, {Smit}, {Stefanon}, \& {Weibel}}]{fujimoto23}
{Fujimoto}, S., {Wang}, B., {Weaver}, J., {et~al.} 2023, arXiv e-prints,
  arXiv:2308.11609, \dodoi{10.48550/arXiv.2308.11609}

\bibitem[{{Furtak} {et~al.}(2024){Furtak}, {Labb{\'e}}, {Zitrin}, {Greene},
  {Dayal}, {Chemerynska}, {Kokorev}, {Miller}, {Goulding}, {de Graaff},
  {Bezanson}, {Brammer}, {Cutler}, {Leja}, {Pan}, {Price}, {Wang}, {Weaver},
  {Whitaker}, {Atek}, {Bogd{\'a}n}, {Charlot}, {Curtis-Lake}, {van Dokkum},
  {Endsley}, {Feldmann}, {Fudamoto}, {Fujimoto}, {Glazebrook}, {Juneau},
  {Marchesini}, {Maseda}, {Nelson}, {Oesch}, {Plat}, {Setton}, {Stark}, \&
  {Williams}}]{furtak23}
{Furtak}, L.~J., {Labb{\'e}}, I., {Zitrin}, A., {et~al.} 2024, \nat, 628, 57,
  \dodoi{10.1038/s41586-024-07184-8}

\bibitem[{{Gaikwad} {et~al.}(2023){Gaikwad}, {Haehnelt}, {Davies}, {Bosman},
  {Molaro}, {Kulkarni}, {D'Odorico}, {Becker}, {Davies}, {Nasir}, {Bolton},
  {Keating}, {Ir{\v{s}}i{\v{c}}}, {Puchwein}, {Zhu}, {Asthana}, {Yang}, {Lai},
  \& {Eilers}}]{Gaikwad23}
{Gaikwad}, P., {Haehnelt}, M.~G., {Davies}, F.~B., {et~al.} 2023, \mnras, 525,
  4093, \dodoi{10.1093/mnras/stad2566}

\bibitem[{{Gallego} {et~al.}(2021){Gallego}, {Cantalupo}, {Sarpas}, {Duboeuf},
  {Lilly}, {Pezzulli}, {Marino}, {Matthee}, {Wisotzki}, {Schaye}, {Richard},
  {Kusakabe}, \& {Mauerhofer}}]{Gallego21}
{Gallego}, S.~G., {Cantalupo}, S., {Sarpas}, S., {et~al.} 2021, \mnras, 504,
  16, \dodoi{10.1093/mnras/stab796}

\bibitem[{{Garaldi} {et~al.}(2019){Garaldi}, {Compostella}, \&
  {Porciani}}]{garaldi19}
{Garaldi}, E., {Compostella}, M., \& {Porciani}, C. 2019, \mnras, 483, 5301,
  \dodoi{10.1093/mnras/sty3414}

\bibitem[{{Gehrels}(1986)}]{gehrels86}
{Gehrels}, N. 1986, \apj, 303, 336, \dodoi{10.1086/164079}

\bibitem[{{George} {et~al.}(2015){George}, {Reichardt}, {Aird}, {Benson},
  {Bleem}, {Carlstrom}, {Chang}, {Cho}, {Crawford}, {Crites}, {de Haan},
  {Dobbs}, {Dudley}, {Halverson}, {Harrington}, {Holder}, {Holzapfel}, {Hou},
  {Hrubes}, {Keisler}, {Knox}, {Lee}, {Leitch}, {Lueker}, {Luong-Van},
  {McMahon}, {Mehl}, {Meyer}, {Millea}, {Mocanu}, {Mohr}, {Montroy}, {Padin},
  {Plagge}, {Pryke}, {Ruhl}, {Schaffer}, {Shaw}, {Shirokoff}, {Spieler},
  {Staniszewski}, {Stark}, {Story}, {van Engelen}, {Vanderlinde}, {Vieira},
  {Williamson}, \& {Zahn}}]{george15}
{George}, E.~M., {Reichardt}, C.~L., {Aird}, K.~A., {et~al.} 2015, \apj, 799,
  177, \dodoi{10.1088/0004-637X/799/2/177}

\bibitem[{{Giallongo} {et~al.}(2012){Giallongo}, {Menci}, {Fiore},
  {Castellano}, {Fontana}, {Grazian}, \& {Pentericci}}]{giallongo12}
{Giallongo}, E., {Menci}, N., {Fiore}, F., {et~al.} 2012, \apj, 755, 124,
  \dodoi{10.1088/0004-637X/755/2/124}

\bibitem[{{Giallongo} {et~al.}(2015){Giallongo}, {Grazian}, {Fiore}, {Fontana},
  {Pentericci}, {Vanzella}, {Dickinson}, {Kocevski}, {Castellano}, {Cristiani},
  {Ferguson}, {Finkelstein}, {Grogin}, {Hathi}, {Koekemoer}, {Newman}, \&
  {Salvato}}]{giallongo15}
{Giallongo}, E., {Grazian}, A., {Fiore}, F., {et~al.} 2015, \aap, 578, A83,
  \dodoi{10.1051/0004-6361/201425334}

\bibitem[{{Giallongo} {et~al.}(2019){Giallongo}, {Grazian}, {Fiore}, {Kodra},
  {Urrutia}, {Castellano}, {Cristiani}, {Dickinson}, {Fontana}, {Menci},
  {Pentericci}, {Boutsia}, {Newman}, \& {Puccetti}}]{giallongo19}
---. 2019, \apj, 884, 19, \dodoi{10.3847/1538-4357/ab39e1}

\bibitem[{{Glikman} {et~al.}(2011){Glikman}, {Djorgovski}, {Stern}, {Dey},
  {Jannuzi}, \& {Lee}}]{glikman11}
{Glikman}, E., {Djorgovski}, S.~G., {Stern}, D., {et~al.} 2011, \apjl, 728,
  L26, \dodoi{10.1088/2041-8205/728/2/L26}

\bibitem[{{Goulding} {et~al.}(2023){Goulding}, {Greene}, {Setton}, {Labbe},
  {Bezanson}, {Miller}, {Atek}, {Bogd{\'a}n}, {Brammer}, {Chemerynska},
  {Cutler}, {Dayal}, {Fudamoto}, {Fujimoto}, {Furtak}, {Kokorev}, {Khullar},
  {Leja}, {Marchesini}, {Natarajan}, {Nelson}, {Oesch}, {Pan}, {Papovich},
  {Price}, {van Dokkum}, {Wang}, {Weaver}, {Whitaker}, \&
  {Zitrin}}]{goulding23}
{Goulding}, A.~D., {Greene}, J.~E., {Setton}, D.~J., {et~al.} 2023, \apjl, 955,
  L24, \dodoi{10.3847/2041-8213/acf7c5}

\bibitem[{{Grazian} {et~al.}(2018){Grazian}, {Giallongo}, {Boutsia},
  {Cristiani}, {Vanzella}, {Scarlata}, {Santini}, {Pentericci}, {Merlin},
  {Menci}, {Fontanot}, {Fontana}, {Fiore}, {Civano}, {Castellano}, {Brusa},
  {Bonchi}, {Carini}, {Cusano}, {Faccini}, {Garilli}, {Marchetti}, {Rossi}, \&
  {Speziali}}]{grazian18}
{Grazian}, A., {Giallongo}, E., {Boutsia}, K., {et~al.} 2018, \aap, 613, A44,
  \dodoi{10.1051/0004-6361/201732385}

\bibitem[{{Grazian} {et~al.}(2020){Grazian}, {Giallongo}, {Fiore}, {Boutsia},
  {Civano}, {Cristiani}, {Cupani}, {Dickinson}, {Fontanot}, {Menci}, \&
  {Romano}}]{grazian20}
{Grazian}, A., {Giallongo}, E., {Fiore}, F., {et~al.} 2020, \apj, 897, 94,
  \dodoi{10.3847/1538-4357/ab99a3}

\bibitem[{{Grazian} {et~al.}(2022){Grazian}, {Giallongo}, {Boutsia},
  {Calderone}, {Cristiani}, {Cupani}, {Fontanot}, {Guarneri}, \&
  {Ozdalkiran}}]{grazian22}
{Grazian}, A., {Giallongo}, E., {Boutsia}, K., {et~al.} 2022, \apj, 924, 62,
  \dodoi{10.3847/1538-4357/ac33a4}

\bibitem[{{Grazian} {et~al.}(2023){Grazian}, {Boutsia}, {Giallongo},
  {Cristiani}, {Fontanot}, {Bischetti}, {Bongiorno}, {Calderone}, {Cupani},
  {D'Odorico}, {Feruglio}, {Fiore}, {Guarneri}, {Porru}, \&
  {Saccheo}}]{grazian23}
{Grazian}, A., {Boutsia}, K., {Giallongo}, E., {et~al.} 2023, \apj, 955, 60,
  \dodoi{10.3847/1538-4357/aceb60}

\bibitem[{{Greene} {et~al.}(2024){Greene}, {Labbe}, {Goulding}, {Furtak},
  {Chemerynska}, {Kokorev}, {Dayal}, {Volonteri}, {Williams}, {Wang}, {Setton},
  {Burgasser}, {Bezanson}, {Atek}, {Brammer}, {Cutler}, {Feldmann}, {Fujimoto},
  {Glazebrook}, {de Graaff}, {Khullar}, {Leja}, {Marchesini}, {Maseda},
  {Matthee}, {Miller}, {Naidu}, {Nanayakkara}, {Oesch}, {Pan}, {Papovich},
  {Price}, {van Dokkum}, {Weaver}, {Whitaker}, \& {Zitrin}}]{greene24}
{Greene}, J.~E., {Labbe}, I., {Goulding}, A.~D., {et~al.} 2024, \apj, 964, 39,
  \dodoi{10.3847/1538-4357/ad1e5f}

\bibitem[{{Guarneri} {et~al.}(2022){Guarneri}, {Calderone}, {Cristiani},
  {Porru}, {Fontanot}, {Boutsia}, {Cupani}, {Grazian}, {D'Odorico}, {Murphy},
  {Bongiorno}, {Saccheo}, \& {Nicastro}}]{guarneri22}
{Guarneri}, F., {Calderone}, G., {Cristiani}, S., {et~al.} 2022, \mnras, 517,
  2436, \dodoi{10.1093/mnras/stac2733}

\bibitem[{{Haardt} \& {Madau}(1996)}]{haardt96}
{Haardt}, F., \& {Madau}, P. 1996, \apj, 461, 20, \dodoi{10.1086/177035}

\bibitem[{{Haardt} \& {Madau}(2012)}]{haardtMadau12}
---. 2012, \apj, 746, 125, \dodoi{10.1088/0004-637X/746/2/125}

\bibitem[{{Haardt} \& {Salvaterra}(2015)}]{haardt15}
{Haardt}, F., \& {Salvaterra}, R. 2015, \aap, 575, L16,
  \dodoi{10.1051/0004-6361/201525627}

\bibitem[{{Habouzit}(2024)}]{habouzit24}
{Habouzit}, M. 2024, arXiv e-prints, arXiv:2405.05319,
  \dodoi{10.48550/arXiv.2405.05319}

\bibitem[{{Harikane} {et~al.}(2023){Harikane}, {Zhang}, {Nakajima}, {Ouchi},
  {Isobe}, {Ono}, {Hatano}, {Xu}, \& {Umeda}}]{harikane23}
{Harikane}, Y., {Zhang}, Y., {Nakajima}, K., {et~al.} 2023, \apj, 959, 39,
  \dodoi{10.3847/1538-4357/ad029e}

\bibitem[{{Harvey} {et~al.}(2024){Harvey}, {Conselice}, {Adams}, {Austin},
  {Juodzbalis}, {Trussler}, {Li}, {Ormerod}, {Ferreira}, {Duan}, {Westcott},
  {Harris}, {Bhatawdekar}, {Coe}, {Cohen}, {Caruana}, {Cheng}, {Driver},
  {Frye}, {Furtak}, {Grogin}, {Hathi}, {Holwerda}, {Jansen}, {Koekemoer},
  {Lovell}, {Marshall}, {Nonino}, {Smail}, {Vijayan}, {Wilkins}, {Windhorst},
  {Willmer}, {Yan}, \& {Zitrin}}]{Harvey24}
{Harvey}, T., {Conselice}, C., {Adams}, N.~J., {et~al.} 2024, arXiv e-prints,
  arXiv:2403.03908.
\newblock \doarXiv{2403.03908}

\bibitem[{{Hayes} {et~al.}(2024){Hayes}, {Tan}, {Ellis}, {Young}, {Cammelli},
  {Singh}, {Runnholm}, {Saxena}, {Lunnan}, {Keller}, {Monaco}, {Laporte}, \&
  {Melinder}}]{hayes24}
{Hayes}, M.~J., {Tan}, J.~C., {Ellis}, R.~S., {et~al.} 2024, arXiv e-prints,
  arXiv:2403.16138, \dodoi{10.48550/arXiv.2403.16138}

\bibitem[{{Hegde} {et~al.}(2024){Hegde}, {Wyatt}, \& {Furlanetto}}]{hegde24}
{Hegde}, S., {Wyatt}, M.~M., \& {Furlanetto}, S.~R. 2024, arXiv e-prints,
  arXiv:2405.01629, \dodoi{10.48550/arXiv.2405.01629}

\bibitem[{{Iwata} {et~al.}(2021){Iwata}, {Sawicki}, {Inoue}, {Akiyama},
  {Micheva}, {Kawaguchi}, {Kashikawa}, {Gwyn}, {Arnouts}, {Coupon}, \&
  {Desprez}}]{iwata21}
{Iwata}, I., {Sawicki}, M., {Inoue}, A.~K., {et~al.} 2021, \mnras,
  \dodoi{10.1093/mnras/stab2742}

\bibitem[{{Jiang} {et~al.}(2016){Jiang}, {McGreer}, {Fan}, {Strauss},
  {Ba{\~n}ados}, {Becker}, {Bian}, {Farnsworth}, {Shen}, {Wang}, {Wang},
  {Wang}, {White}, {Wu}, {Wu}, {Yang}, \& {Yang}}]{Jiang2016}
{Jiang}, L., {McGreer}, I.~D., {Fan}, X., {et~al.} 2016, \apj, 833, 222,
  \dodoi{10.3847/1538-4357/833/2/222}

\bibitem[{{Jiang} {et~al.}(2022){Jiang}, {Ning}, {Fan}, {Ho}, {Luo}, {Wang},
  {Wu}, {Wu}, {Yang}, \& {Zheng}}]{jiang22}
{Jiang}, L., {Ning}, Y., {Fan}, X., {et~al.} 2022, Nature Astronomy, 6, 850,
  \dodoi{10.1038/s41550-022-01708-w}

\bibitem[{{Jin} {et~al.}(2023){Jin}, {Yang}, {Fan}, {Wang}, {Ba{\~n}ados},
  {Bian}, {Davies}, {Eilers}, {Farina}, {Hennawi}, {Pacucci}, {Venemans}, \&
  {Walter}}]{jin23}
{Jin}, X., {Yang}, J., {Fan}, X., {et~al.} 2023, \apj, 942, 59,
  \dodoi{10.3847/1538-4357/aca678}

\bibitem[{{Jung} {et~al.}(2024){Jung}, {Ferguson}, {Hayes}, {Henry}, {Jaskot},
  {Schaerer}, {Sharon}, {Amor{\'\i}n}, {Atek}, {Bayliss}, {Dahle},
  {Finkelstein}, {Grazian}, {Guaita}, {{\"O}stlin}, {Pentericci},
  {Ravindranath}, {Scarlata}, {Teplitz}, \& {Verhamme}}]{jung24}
{Jung}, I., {Ferguson}, H.~C., {Hayes}, M.~J., {et~al.} 2024, arXiv e-prints,
  arXiv:2403.02388.
\newblock \doarXiv{2403.02388}

\bibitem[{{Kerutt} {et~al.}(2024){Kerutt}, {Oesch}, {Wisotzki}, {Verhamme},
  {Atek}, {Herenz}, {Illingworth}, {Kusakabe}, {Matthee}, {Mauerhofer},
  {Montes}, {Naidu}, {Nelson}, {Reddy}, {Schaye}, {Simmonds}, {Urrutia}, \&
  {Vitte}}]{kerutt24}
{Kerutt}, J., {Oesch}, P.~A., {Wisotzki}, L., {et~al.} 2024, \aap, 684, A42,
  \dodoi{10.1051/0004-6361/202346656}

\bibitem[{{Killi} {et~al.}(2023){Killi}, {Watson}, {Brammer}, {McPartland},
  {Antwi-Danso}, {Newshore}, {Coe}, {Allen}, {Fynbo}, {Gould}, {Heintz},
  {Rusakov}, \& {Vejlgaard}}]{killi23}
{Killi}, M., {Watson}, D., {Brammer}, G., {et~al.} 2023, arXiv e-prints,
  arXiv:2312.03065, \dodoi{10.48550/arXiv.2312.03065}

\bibitem[{{Kim} \& {Im}(2021)}]{kimim21}
{Kim}, Y., \& {Im}, M. 2021, \apjl, 910, L11, \dodoi{10.3847/2041-8213/abed58}

\bibitem[{{Kim} {et~al.}(2020){Kim}, {Im}, {Jeon}, {Kim}, {Pak}, {Hyun},
  {Taak}, {Shin}, {Lim}, {Paek}, {Paek}, {Jiang}, {Choi}, {Hong}, {Ji}, {Jun},
  {Karouzos}, {Kim}, {Kim}, {Kim}, {Kim}, {Lee}, {Lee}, {Park}, {Yoon},
  {Byeon}, {Hwang}, {Kim}, {Kim}, \& {Park}}]{kim20}
{Kim}, Y., {Im}, M., {Jeon}, Y., {et~al.} 2020, \apj, 904, 111,
  \dodoi{10.3847/1538-4357/abc0ea}

\bibitem[{{Kocevski} {et~al.}(2023){Kocevski}, {Onoue}, {Inayoshi}, {Trump},
  {Arrabal Haro}, {Grazian}, {Dickinson}, {Finkelstein}, {Kartaltepe},
  {Hirschmann}, {Aird}, {Holwerda}, {Fujimoto}, {Juneau}, {Amor{\'\i}n},
  {Backhaus}, {Bagley}, {Barro}, {Bell}, {Bisigello}, {Calabr{\`o}}, {Cleri},
  {Cooper}, {Ding}, {Grogin}, {Ho}, {Hutchison}, {Inoue}, {Jiang}, {Jones},
  {Koekemoer}, {Li}, {Li}, {McGrath}, {Molina}, {Papovich},
  {P{\'e}rez-Gonz{\'a}lez}, {Pirzkal}, {Wilkins}, {Yang}, \&
  {Yung}}]{kocevski23}
{Kocevski}, D.~D., {Onoue}, M., {Inayoshi}, K., {et~al.} 2023, \apjl, 954, L4,
  \dodoi{10.3847/2041-8213/ace5a0}

\bibitem[{{Kocevski} {et~al.}(2024){Kocevski}, {Finkelstein}, {Barro},
  {Taylor}, {Calabr{\`o}}, {Laloux}, {Buchner}, {Trump}, {Leung}, {Yang},
  {Dickinson}, {P{\'e}rez-Gonz{\'a}lez}, {Pacucci}, {Inayoshi}, {Somerville},
  {McGrath}, {Akins}, {Arrabal Haro}, {Bagley}, {Bowler}, {Carnall}, {Casey},
  {Cheng}, {Cleri}, {Costantin}, {Cullen}, {Davis}, {Donnan}, {Dunlop},
  {Ellis}, {Ferguson}, {Fujimoto}, {Fontana}, {Giavalisco}, {Grazian},
  {Grogin}, {Hathi}, {Hirschmann}, {Huertas-Company}, {Holwerda},
  {Illingworth}, {Juneau}, {Kartaltepe}, {Koekemoer}, {Li}, {Lucas}, {Magee},
  {Mason}, {McLeod}, {McLure}, {Napolitano}, {Papovich}, {Pirzkal},
  {Rodighiero}, {Santini}, {Wilkins}, \& {Yung}}]{kocevski24}
{Kocevski}, D.~D., {Finkelstein}, S.~L., {Barro}, G., {et~al.} 2024, arXiv
  e-prints, arXiv:2404.03576.
\newblock \doarXiv{2404.03576}

\bibitem[{{Kokorev} {et~al.}(2024){Kokorev}, {Caputi}, {Greene}, {Dayal},
  {Trebitsch}, {Cutler}, {Fujimoto}, {Labb{\'e}}, {Miller}, {Iani},
  {Navarro-Carrera}, \& {Rinaldi}}]{kokorev24}
{Kokorev}, V., {Caputi}, K.~I., {Greene}, J.~E., {et~al.} 2024, \apj, 968, 38,
  \dodoi{10.3847/1538-4357/ad4265}

\bibitem[{{Kov{\'a}cs} {et~al.}(2024){Kov{\'a}cs}, {Bogd{\'a}n}, {Natarajan},
  {Werner}, {Azadi}, {Volonteri}, {Tremblay}, {Chadayammuri}, {Forman},
  {Jones}, \& {Kraft}}]{kovacs24}
{Kov{\'a}cs}, O.~E., {Bogd{\'a}n}, {\'A}., {Natarajan}, P., {et~al.} 2024,
  \apjl, 965, L21, \dodoi{10.3847/2041-8213/ad391f}

\bibitem[{{Kulkarni} {et~al.}(2019){Kulkarni}, {Worseck}, \&
  {Hennawi}}]{kulkarni19}
{Kulkarni}, G., {Worseck}, G., \& {Hennawi}, J.~F. 2019, \mnras, 488, 1035,
  \dodoi{10.1093/mnras/stz1493}

\bibitem[{{Labb{\'e}} {et~al.}(2023){Labb{\'e}}, {van Dokkum}, {Nelson},
  {Bezanson}, {Suess}, {Leja}, {Brammer}, {Whitaker}, {Mathews}, {Stefanon}, \&
  {Wang}}]{labbe22}
{Labb{\'e}}, I., {van Dokkum}, P., {Nelson}, E., {et~al.} 2023, \nat, 616, 266,
  \dodoi{10.1038/s41586-023-05786-2}

\bibitem[{{Lai} {et~al.}(2024){Lai}, {Onken}, {Wolf}, {Bian}, \& {Fan}}]{lai24}
{Lai}, S., {Onken}, C.~A., {Wolf}, C., {Bian}, F., \& {Fan}, X. 2024, arXiv
  e-prints, arXiv:2405.10721, \dodoi{10.48550/arXiv.2405.10721}

\bibitem[{{Larson} {et~al.}(2023){Larson}, {Finkelstein}, {Kocevski},
  {Hutchison}, {Trump}, {Arrabal Haro}, {Bromm}, {Cleri}, {Dickinson},
  {Fujimoto}, {Kartaltepe}, {Koekemoer}, {Papovich}, {Pirzkal}, {Tacchella},
  {Zavala}, {Bagley}, {Behroozi}, {Champagne}, {Cole}, {Jung}, {Morales},
  {Yang}, {Zhang}, {Zitrin}, {Amor{\'\i}n}, {Burgarella}, {Casey}, {Ch{\'a}vez
  Ortiz}, {Cox}, {Chworowsky}, {Fontana}, {Gawiser}, {Grazian}, {Grogin},
  {Harish}, {Hathi}, {Hirschmann}, {Holwerda}, {Juneau}, {Leung}, {Lucas},
  {McGrath}, {P{\'e}rez-Gonz{\'a}lez}, {Rigby}, {Seill{\'e}}, {Simons}, {de La
  Vega}, {Weiner}, {Wilkins}, {Yung}, \& {Ceers Team}}]{larson23}
{Larson}, R.~L., {Finkelstein}, S.~L., {Kocevski}, D.~D., {et~al.} 2023, \apjl,
  953, L29, \dodoi{10.3847/2041-8213/ace619}

\bibitem[{{Laureijs} {et~al.}(2020){Laureijs}, {Racca}, {Mellier}, {Musi},
  {Brouard}, {B{\"o}enke}, {Gaspar Venancio}, {Maiorano}, {Short}, {Strada},
  {Altieri}, {Buenadicha}, {Dupac}, {Gomez Alvarez}, {Hoar}, {Kohley},
  {Vavrek}, {Rudolph}, {Schmidt}, {Amiaux}, {Aussel}, {Berth{\'e}}, {Cropper},
  {Cuillandre}, {Dabin}, {Dinis}, {Nakajima}, {Maciaszek}, {Scaramella}, {da
  Silva}, {Tereno}, {Williams}, {Zacchei}, {Azzollini}, {Bernardeau},
  {Brinchmann}, {Brockley-Blatt}, {Castander}, {Cimatti}, {Conselice}, {Ealet},
  {Fosalba}, {Gillard}, {Guzzo}, {Hoekstra}, {Hudelot}, {Jahnke}, {Kitching},
  {Miller}, {Mohr}, {Percival}, {Pettorino}, {Rhodes}, {Sanchez}, {Sauvage},
  {Serrano}, {Teyssier}, {Weller}, \& {Zoubian}}]{laureijs20}
{Laureijs}, R., {Racca}, G.~D., {Mellier}, Y., {et~al.} 2020, in Society of
  Photo-Optical Instrumentation Engineers (SPIE) Conference Series, Vol. 11443,
  Space Telescopes and Instrumentation 2020: Optical, Infrared, and Millimeter
  Wave, ed. M.~{Lystrup} \& M.~D. {Perrin}, 114430F, \dodoi{10.1117/12.2563145}

\bibitem[{{Li} {et~al.}(2024){Li}, {Silverman}, {Shen}, {Volonteri}, {Jahnke},
  {Zhuang}, {Scoggins}, {Ding}, {Harikane}, {Onoue}, \& {Tanaka}}]{li24}
{Li}, J., {Silverman}, J.~D., {Shen}, Y., {et~al.} 2024, arXiv e-prints,
  arXiv:2403.00074.
\newblock \doarXiv{2403.00074}

\bibitem[{{Lusso} {et~al.}(2015){Lusso}, {Worseck}, {Hennawi}, {Prochaska},
  {Vignali}, {Stern}, \& {O'Meara}}]{lusso15}
{Lusso}, E., {Worseck}, G., {Hennawi}, J.~F., {et~al.} 2015, \mnras, 449, 4204,
  \dodoi{10.1093/mnras/stv516}

\bibitem[{{Madau} {et~al.}(2024){Madau}, {Giallongo}, {Grazian}, \&
  {Haardt}}]{madau24}
{Madau}, P., {Giallongo}, E., {Grazian}, A., \& {Haardt}, F. 2024, arXiv
  e-prints, arXiv:2406.18697, \dodoi{10.48550/arXiv.2406.18697}

\bibitem[{{Madau} \& {Haardt}(2015)}]{madau15}
{Madau}, P., \& {Haardt}, F. 2015, \apjl, 813, L8,
  \dodoi{10.1088/2041-8205/813/1/L8}

\bibitem[{{Maiolino} {et~al.}(2023){Maiolino}, {Scholtz}, {Curtis-Lake},
  {Carniani}, {Baker}, {de Graaff}, {Tacchella}, {{\"U}bler}, {D'Eugenio},
  {Witstok}, {Curti}, {Arribas}, {Bunker}, {Charlot}, {Chevallard},
  {Eisenstein}, {Egami}, {Ji}, {Jones}, {Lyu}, {Rawle}, {Robertson},
  {Rujopakarn}, {Perna}, {Sun}, {Venturi}, {Williams}, \&
  {Willott}}]{maiolino23lf}
{Maiolino}, R., {Scholtz}, J., {Curtis-Lake}, E., {et~al.} 2023, arXiv
  e-prints, arXiv:2308.01230, \dodoi{10.48550/arXiv.2308.01230}

\bibitem[{{Maiolino} {et~al.}(2024{\natexlab{a}}){Maiolino}, {Scholtz},
  {Witstok}, {Carniani}, {D'Eugenio}, {de Graaff}, {{\"U}bler}, {Tacchella},
  {Curtis-Lake}, {Arribas}, {Bunker}, {Charlot}, {Chevallard}, {Curti},
  {Looser}, {Maseda}, {Rawle}, {Rodr{\'\i}guez del Pino}, {Willott}, {Egami},
  {Eisenstein}, {Hainline}, {Robertson}, {Williams}, {Willmer}, {Baker},
  {Boyett}, {DeCoursey}, {Fabian}, {Helton}, {Ji}, {Jones}, {Kumari},
  {Laporte}, {Nelson}, {Perna}, {Sandles}, {Shivaei}, \&
  {Sun}}]{maiolino23gnz11}
{Maiolino}, R., {Scholtz}, J., {Witstok}, J., {et~al.} 2024{\natexlab{a}},
  \nat, 627, 59, \dodoi{10.1038/s41586-024-07052-5}

\bibitem[{{Maiolino} {et~al.}(2024{\natexlab{b}}){Maiolino}, {Risaliti},
  {Signorini}, {Trefoloni}, {Juodzbalis}, {Scholtz}, {Uebler}, {D'Eugenio},
  {Carniani}, {Fabian}, {Ji}, {Mazzolari}, {Bertola}, {Brusa}, {Bunker},
  {Charlot}, {Comastri}, {Cresci}, {DeCoursey}, {Egami}, {Fiore}, {Gilli},
  {Perna}, {Tacchella}, \& {Venturi}}]{maiolino24xray}
{Maiolino}, R., {Risaliti}, G., {Signorini}, M., {et~al.} 2024{\natexlab{b}},
  arXiv e-prints, arXiv:2405.00504, \dodoi{10.48550/arXiv.2405.00504}

\bibitem[{{Makan} {et~al.}(2021){Makan}, {Worseck}, {Davies}, {Hennawi},
  {Prochaska}, \& {Richter}}]{makan21}
{Makan}, K., {Worseck}, G., {Davies}, F.~B., {et~al.} 2021, \apj, 912, 38,
  \dodoi{10.3847/1538-4357/abee17}

\bibitem[{{Makan} {et~al.}(2022){Makan}, {Worseck}, {Davies}, {Hennawi},
  {Prochaska}, \& {Richter}}]{makan22}
---. 2022, \apj, 927, 175, \dodoi{10.3847/1538-4357/ac524a}

\bibitem[{{Mascia} {et~al.}(2024){Mascia}, {Pentericci}, {Calabr{\`o}},
  {Santini}, {Napolitano}, {Arrabal Haro}, {Castellano}, {Dickinson}, {Ocvirk},
  {Lewis}, {Amor{\'\i}n}, {Bagley}, {Bhatawdekar}, {Cleri}, {Costantin},
  {Dekel}, {Finkelstein}, {Fontana}, {Giavalisco}, {Grogin}, {Hathi},
  {Hirschmann}, {Holwerda}, {Jung}, {Kartaltepe}, {Koekemoer}, {Lucas},
  {Papovich}, {P{\'e}rez-Gonz{\'a}lez}, {Pirzkal}, {Trump}, {Wilkins}, \&
  {Yung}}]{mascia24}
{Mascia}, S., {Pentericci}, L., {Calabr{\`o}}, A., {et~al.} 2024, \aap, 685,
  A3, \dodoi{10.1051/0004-6361/202347884}

\bibitem[{{Matsuoka} {et~al.}(2023){Matsuoka}, {Onoue}, {Iwasawa}, {Strauss},
  {Kashikawa}, {Izumi}, {Nagao}, {Imanishi}, {Akiyama}, {Silverman}, {Asami},
  {Bosch}, {Furusawa}, {Goto}, {Gunn}, {Harikane}, {Ikeda}, {Inayoshi},
  {Ishimoto}, {Kawaguchi}, {Kikuta}, {Kohno}, {Komiyama}, {Lee}, {Lupton},
  {Minezaki}, {Miyazaki}, {Murayama}, {Nishizawa}, {Oguri}, {Ono}, {Oogi},
  {Ouchi}, {Price}, {Sameshima}, {Sugiyama}, {Tait}, {Takada}, {Takahashi},
  {Takata}, {Tanaka}, {Toba}, {Wang}, \& {Yamashita}}]{Matsuoka23}
{Matsuoka}, Y., {Onoue}, M., {Iwasawa}, K., {et~al.} 2023, \apjl, 949, L42,
  \dodoi{10.3847/2041-8213/acd69f}

\bibitem[{{Matthee} {et~al.}(2024){Matthee}, {Naidu}, {Brammer}, {Chisholm},
  {Eilers}, {Goulding}, {Greene}, {Kashino}, {Labbe}, {Lilly}, {Mackenzie},
  {Oesch}, {Weibel}, {Wuyts}, {Xiao}, {Bordoloi}, {Bouwens}, {van Dokkum},
  {Illingworth}, {Kramarenko}, {Maseda}, {Mason}, {Meyer}, {Nelson}, {Reddy},
  {Shivaei}, {Simcoe}, \& {Yue}}]{matthee24}
{Matthee}, J., {Naidu}, R.~P., {Brammer}, G., {et~al.} 2024, \apj, 963, 129,
  \dodoi{10.3847/1538-4357/ad2345}

\bibitem[{{McGreer} {et~al.}(2018){McGreer}, {Fan}, {Jiang}, \&
  {Cai}}]{McGreer18}
{McGreer}, I.~D., {Fan}, X., {Jiang}, L., \& {Cai}, Z. 2018, \aj, 155, 131,
  \dodoi{10.3847/1538-3881/aaaab4}

\bibitem[{{Menci} {et~al.}(2014){Menci}, {Gatti}, {Fiore}, \&
  {Lamastra}}]{menci14}
{Menci}, N., {Gatti}, M., {Fiore}, F., \& {Lamastra}, A. 2014, \aap, 569, A37,
  \dodoi{10.1051/0004-6361/201424217}

\bibitem[{{Micheva} {et~al.}(2017){Micheva}, {Iwata}, \& {Inoue}}]{micheva17}
{Micheva}, G., {Iwata}, I., \& {Inoue}, A.~K. 2017, \mnras, 465, 302,
  \dodoi{10.1093/mnras/stw1329}

\bibitem[{{Mu{\~n}oz} {et~al.}(2024){Mu{\~n}oz}, {Mirocha}, {Chisholm},
  {Furlanetto}, \& {Mason}}]{munoz24}
{Mu{\~n}oz}, J.~B., {Mirocha}, J., {Chisholm}, J., {Furlanetto}, S.~R., \&
  {Mason}, C. 2024, arXiv e-prints, arXiv:2404.07250.
\newblock \doarXiv{2404.07250}

\bibitem[{{Naidu} {et~al.}(2022){Naidu}, {Oesch}, {van Dokkum}, {Nelson},
  {Suess}, {Brammer}, {Whitaker}, {Illingworth}, {Bouwens}, {Tacchella},
  {Matthee}, {Allen}, {Bezanson}, {Conroy}, {Labbe}, {Leja}, {Leonova},
  {Magee}, {Price}, {Setton}, {Strait}, {Stefanon}, {Toft}, {Weaver}, \&
  {Weibel}}]{naidu22}
{Naidu}, R.~P., {Oesch}, P.~A., {van Dokkum}, P., {et~al.} 2022, \apjl, 940,
  L14, \dodoi{10.3847/2041-8213/ac9b22}

\bibitem[{{Niida} {et~al.}(2020){Niida}, {Nagao}, {Ikeda}, {Akiyama},
  {Matsuoka}, {He}, {Matsuoka}, {Toba}, {Onoue}, {Kobayashi}, {Taniguchi},
  {Furusawa}, {Harikane}, {Imanishi}, {Kashikawa}, {Kawaguchi}, {Komiyama},
  {Shirakata}, {Terashima}, \& {Ueda}}]{niida20}
{Niida}, M., {Nagao}, T., {Ikeda}, H., {et~al.} 2020, \apj, 904, 89,
  \dodoi{10.3847/1538-4357/abbe11}

\bibitem[{{Oesch}(2020)}]{oesch20}
{Oesch}, P. 2020, in Uncovering Early Galaxy Evolution in the ALMA and JWST
  Era, ed. E.~{da Cunha}, J.~{Hodge}, J.~{Afonso}, L.~{Pentericci}, \&
  D.~{Sobral}, Vol. 352, 12--12, \dodoi{10.1017/S1743921320001064}

\bibitem[{{Onken} {et~al.}(2022){Onken}, {Wolf}, {Bian}, {Fan}, {Hon},
  {Raithel}, {Tisserand}, \& {Lai}}]{onken22}
{Onken}, C.~A., {Wolf}, C., {Bian}, F., {et~al.} 2022, \mnras, 511, 572,
  \dodoi{10.1093/mnras/stac051}

\bibitem[{{Padmanabhan} \& {Loeb}(2023)}]{padmanabhan23}
{Padmanabhan}, H., \& {Loeb}, A. 2023, \apjl, 958, L7,
  \dodoi{10.3847/2041-8213/ad09ac}

\bibitem[{{Pan} {et~al.}(2022){Pan}, {Jiang}, {Fan}, {Wu}, \& {Yang}}]{pan22}
{Pan}, Z., {Jiang}, L., {Fan}, X., {Wu}, J., \& {Yang}, J. 2022, \apj, 928,
  172, \dodoi{10.3847/1538-4357/ac5aab}

\bibitem[{{P{\'e}rez-Gonz{\'a}lez} {et~al.}(2024){P{\'e}rez-Gonz{\'a}lez},
  {Barro}, {Rieke}, {Lyu}, {Rieke}, {Alberts}, {Williams}, {Hainline}, {Sun},
  {Pusk{\'a}s}, {Annunziatella}, {Baker}, {Bunker}, {Egami}, {Ji}, {Johnson},
  {Robertson}, {Rodr{\'\i}guez Del Pino}, {Rujopakarn}, {Shivaei}, {Tacchella},
  {Willmer}, \& {Willott}}]{perezgonzalez24}
{P{\'e}rez-Gonz{\'a}lez}, P.~G., {Barro}, G., {Rieke}, G.~H., {et~al.} 2024,
  \apj, 968, 4, \dodoi{10.3847/1538-4357/ad38bb}

\bibitem[{{Planck Collaboration} {et~al.}(2020){Planck Collaboration},
  {Aghanim}, {Akrami}, {Ashdown}, {Aumont}, {Baccigalupi}, {Ballardini},
  {Banday}, {Barreiro}, {Bartolo}, {Basak}, {Battye}, {Benabed}, {Bernard},
  {Bersanelli}, {Bielewicz}, {Bock}, {Bond}, {Borrill}, {Bouchet}, {Boulanger},
  {Bucher}, {Burigana}, {Butler}, {Calabrese}, {Cardoso}, {Carron},
  {Challinor}, {Chiang}, {Chluba}, {Colombo}, {Combet}, {Contreras}, {Crill},
  {Cuttaia}, {de Bernardis}, {de Zotti}, {Delabrouille}, {Delouis}, {Di
  Valentino}, {Diego}, {Dor{\'e}}, {Douspis}, {Ducout}, {Dupac}, {Dusini},
  {Efstathiou}, {Elsner}, {En{\ss}lin}, {Eriksen}, {Fantaye}, {Farhang},
  {Fergusson}, {Fernandez-Cobos}, {Finelli}, {Forastieri}, {Frailis},
  {Fraisse}, {Franceschi}, {Frolov}, {Galeotta}, {Galli}, {Ganga},
  {G{\'e}nova-Santos}, {Gerbino}, {Ghosh}, {Gonz{\'a}lez-Nuevo}, {G{\'o}rski},
  {Gratton}, {Gruppuso}, {Gudmundsson}, {Hamann}, {Handley}, {Hansen},
  {Herranz}, {Hildebrandt}, {Hivon}, {Huang}, {Jaffe}, {Jones}, {Karakci},
  {Keih{\"a}nen}, {Keskitalo}, {Kiiveri}, {Kim}, {Kisner}, {Knox},
  {Krachmalnicoff}, {Kunz}, {Kurki-Suonio}, {Lagache}, {Lamarre}, {Lasenby},
  {Lattanzi}, {Lawrence}, {Le Jeune}, {Lemos}, {Lesgourgues}, {Levrier},
  {Lewis}, {Liguori}, {Lilje}, {Lilley}, {Lindholm}, {L{\'o}pez-Caniego},
  {Lubin}, {Ma}, {Mac{\'\i}as-P{\'e}rez}, {Maggio}, {Maino}, {Mandolesi},
  {Mangilli}, {Marcos-Caballero}, {Maris}, {Martin}, {Martinelli},
  {Mart{\'\i}nez-Gonz{\'a}lez}, {Matarrese}, {Mauri}, {McEwen}, {Meinhold},
  {Melchiorri}, {Mennella}, {Migliaccio}, {Millea}, {Mitra},
  {Miville-Desch{\^e}nes}, {Molinari}, {Montier}, {Morgante}, {Moss}, {Natoli},
  {N{\o}rgaard-Nielsen}, {Pagano}, {Paoletti}, {Partridge}, {Patanchon},
  {Peiris}, {Perrotta}, {Pettorino}, {Piacentini}, {Polastri}, {Polenta},
  {Puget}, {Rachen}, {Reinecke}, {Remazeilles}, {Renzi}, {Rocha}, {Rosset},
  {Roudier}, {Rubi{\~n}o-Mart{\'\i}n}, {Ruiz-Granados}, {Salvati}, {Sandri},
  {Savelainen}, {Scott}, {Shellard}, {Sirignano}, {Sirri}, {Spencer},
  {Sunyaev}, {Suur-Uski}, {Tauber}, {Tavagnacco}, {Tenti}, {Toffolatti},
  {Tomasi}, {Trombetti}, {Valenziano}, {Valiviita}, {Van Tent}, {Vibert},
  {Vielva}, {Villa}, {Vittorio}, {Wandelt}, {Wehus}, {White}, {White},
  {Zacchei}, \& {Zonca}}]{planck20}
{Planck Collaboration}, {Aghanim}, N., {Akrami}, Y., {et~al.} 2020, \aap, 641,
  A6, \dodoi{10.1051/0004-6361/201833910}

\bibitem[{{Puchwein} {et~al.}(2019){Puchwein}, {Haardt}, {Haehnelt}, \&
  {Madau}}]{puchwein19}
{Puchwein}, E., {Haardt}, F., {Haehnelt}, M.~G., \& {Madau}, P. 2019, \mnras,
  485, 47, \dodoi{10.1093/mnras/stz222}

\bibitem[{{Reichardt} {et~al.}(2021){Reichardt}, {Patil}, {Ade}, {Anderson},
  {Austermann}, {Avva}, {Baxter}, {Beall}, {Bender}, {Benson}, {Bianchini},
  {Bleem}, {Carlstrom}, {Chang}, {Chaubal}, {Chiang}, {Chou}, {Citron},
  {Moran}, {Crawford}, {Crites}, {de Haan}, {Dobbs}, {Everett}, {Gallicchio},
  {George}, {Gilbert}, {Gupta}, {Halverson}, {Harrington}, {Henning}, {Hilton},
  {Holder}, {Holzapfel}, {Hrubes}, {Huang}, {Hubmayr}, {Irwin}, {Knox}, {Lee},
  {Li}, {Lowitz}, {Luong-Van}, {McMahon}, {Mehl}, {Meyer}, {Millea}, {Mocanu},
  {Mohr}, {Montgomery}, {Nadolski}, {Natoli}, {Nibarger}, {Noble}, {Novosad},
  {Omori}, {Padin}, {Pryke}, {Ruhl}, {Saliwanchik}, {Sayre}, {Schaffer},
  {Shirokoff}, {Sievers}, {Smecher}, {Spieler}, {Staniszewski}, {Stark},
  {Tucker}, {Vanderlinde}, {Veach}, {Vieira}, {Wang}, {Whitehorn},
  {Williamson}, {Wu}, \& {Yefremenko}}]{reichardt20}
{Reichardt}, C.~L., {Patil}, S., {Ade}, P.~A.~R., {et~al.} 2021, \apj, 908,
  199, \dodoi{10.3847/1538-4357/abd407}

\bibitem[{{Romano} {et~al.}(2019){Romano}, {Grazian}, {Giallongo}, {Cristiani},
  {Fontanot}, {Boutsia}, {Fiore}, \& {Menci}}]{romano19}
{Romano}, M., {Grazian}, A., {Giallongo}, E., {et~al.} 2019, \aap, 632, A45,
  \dodoi{10.1051/0004-6361/201935550}

\bibitem[{{Schindler} {et~al.}(2019{\natexlab{a}}){Schindler}, {Fan},
  {McGreer}, {Yang}, {Wang}, {Green}, {Fynbo}, {Krogager}, {Green}, {Huang},
  {Kadowaki}, {Patej}, {Wu}, \& {Yue}}]{Sch19a}
{Schindler}, J.-T., {Fan}, X., {McGreer}, I.~D., {et~al.} 2019{\natexlab{a}},
  \apj, 871, 258, \dodoi{10.3847/1538-4357/aaf86c}

\bibitem[{{Schindler} {et~al.}(2019{\natexlab{b}}){Schindler}, {Fan}, {Huang},
  {Yue}, {Yang}, {Hall}, {Wenzl}, {Hughes}, {Litke}, \& {Rees}}]{Sch19b}
{Schindler}, J.-T., {Fan}, X., {Huang}, Y.-H., {et~al.} 2019{\natexlab{b}},
  \apjs, 243, 5, \dodoi{10.3847/1538-4365/ab20d0}

\bibitem[{{Schindler} {et~al.}(2023){Schindler}, {Ba{\~n}ados}, {Connor},
  {Decarli}, {Fan}, {Farina}, {Mazzucchelli}, {Nanni}, {Rix}, {Stern},
  {Venemans}, \& {Walter}}]{schindler23}
{Schindler}, J.-T., {Ba{\~n}ados}, E., {Connor}, T., {et~al.} 2023, \apj, 943,
  67, \dodoi{10.3847/1538-4357/aca7ca}

\bibitem[{{Scholtz} {et~al.}(2023){Scholtz}, {Maiolino}, {D'Eugenio},
  {Curtis-Lake}, {Carniani}, {Charlot}, {Curti}, {Silcock}, {Arribas}, {Baker},
  {Bhatawdekar}, {Boyett}, {Bunker}, {Chevallard}, {Circosta}, {Eisenstein},
  {Hainline}, {Hausen}, {Ji}, {Ji}, {Johnson}, {Kumari}, {Looser}, {Lyu},
  {Maseda}, {Parlanti}, {Perna}, {Rieke}, {Robertson}, {Rodr{\'\i}guez Del
  Pino}, {Sun}, {Tacchella}, {{\"U}bler}, {Venturi}, {Williams}, {Willmer},
  {Willott}, \& {Witstok}}]{scholtz23}
{Scholtz}, J., {Maiolino}, R., {D'Eugenio}, F., {et~al.} 2023, arXiv e-prints,
  arXiv:2311.18731, \dodoi{10.48550/arXiv.2311.18731}

\bibitem[{{Shin} {et~al.}(2022){Shin}, {Im}, \& {Kim}}]{shin22}
{Shin}, S., {Im}, M., \& {Kim}, Y. 2022, \apj, 937, 32,
  \dodoi{10.3847/1538-4357/ac854b}

\bibitem[{{Singha} {et~al.}(2024){Singha}, {Sarmiento}, {Malhotra}, {Rhoads},
  {Yung}, {Wang}, {Zheng}, {Lin}, {Kim}, {Kang}, \& {Harish}}]{singha24}
{Singha}, M., {Sarmiento}, J., {Malhotra}, S., {et~al.} 2024, arXiv e-prints,
  arXiv:2406.18730, \dodoi{10.48550/arXiv.2406.18730}

\bibitem[{{Spina} {et~al.}(2024){Spina}, {Bosman}, {Davies}, {Gaikwad}, \&
  {Zhu}}]{spina24}
{Spina}, B., {Bosman}, S. E.~I., {Davies}, F.~B., {Gaikwad}, P., \& {Zhu}, Y.
  2024, arXiv e-prints, arXiv:2405.12273, \dodoi{10.48550/arXiv.2405.12273}

\bibitem[{{Stevans} {et~al.}(2014){Stevans}, {Shull}, {Danforth}, \&
  {Tilton}}]{Stenans14}
{Stevans}, M.~L., {Shull}, J.~M., {Danforth}, C.~W., \& {Tilton}, E.~M. 2014,
  \apj, 794, 75, \dodoi{10.1088/0004-637X/794/1/75}

\bibitem[{{Trinca} {et~al.}(2023){Trinca}, {Schneider}, {Maiolino}, {Valiante},
  {Graziani}, \& {Volonteri}}]{trinca23}
{Trinca}, A., {Schneider}, R., {Maiolino}, R., {et~al.} 2023, \mnras, 519,
  4753, \dodoi{10.1093/mnras/stac3768}

\bibitem[{{Trinca} {et~al.}(2022){Trinca}, {Schneider}, {Valiante}, {Graziani},
  {Zappacosta}, \& {Shankar}}]{trinca22}
{Trinca}, A., {Schneider}, R., {Valiante}, R., {et~al.} 2022, \mnras, 511, 616,
  \dodoi{10.1093/mnras/stac062}

\bibitem[{{{\"U}bler} {et~al.}(2023){{\"U}bler}, {Maiolino}, {Curtis-Lake},
  {P{\'e}rez-Gonz{\'a}lez}, {Curti}, {Perna}, {Arribas}, {Charlot}, {Marshall},
  {D'Eugenio}, {Scholtz}, {Bunker}, {Carniani}, {Ferruit}, {Jakobsen}, {Rix},
  {Rodr{\'\i}guez Del Pino}, {Willott}, {Boeker}, {Cresci}, {Jones}, {Kumari},
  \& {Rawle}}]{ubler23}
{{\"U}bler}, H., {Maiolino}, R., {Curtis-Lake}, E., {et~al.} 2023, \aap, 677,
  A145, \dodoi{10.1051/0004-6361/202346137}

\bibitem[{{Volonteri} {et~al.}(2023){Volonteri}, {Habouzit}, \&
  {Colpi}}]{volonteri23}
{Volonteri}, M., {Habouzit}, M., \& {Colpi}, M. 2023, \mnras, 521, 241,
  \dodoi{10.1093/mnras/stad499}

\bibitem[{{Worseck} {et~al.}(2019){Worseck}, {Davies}, {Hennawi}, \&
  {Prochaska}}]{worseck19}
{Worseck}, G., {Davies}, F.~B., {Hennawi}, J.~F., \& {Prochaska}, J.~X. 2019,
  \apj, 875, 111, \dodoi{10.3847/1538-4357/ab0fa1}

\bibitem[{{Worseck} {et~al.}(2014){Worseck}, {Prochaska}, {O'Meara}, {Becker},
  {Ellison}, {Lopez}, {Meiksin}, {M{\'e}nard}, {Murphy}, \&
  {Fumagalli}}]{worseck14}
{Worseck}, G., {Prochaska}, J.~X., {O'Meara}, J.~M., {et~al.} 2014, \mnras,
  445, 1745, \dodoi{10.1093/mnras/stu1827}

\bibitem[{{Wyithe} \& {Bolton}(2011)}]{wyithe11}
{Wyithe}, J. S.~B., \& {Bolton}, J.~S. 2011, \mnras, 412, 1926,
  \dodoi{10.1111/j.1365-2966.2010.18030.x}

\bibitem[{{Xie} {et~al.}(2020){Xie}, {De Lucia}, {Hirschmann}, \&
  {Fontanot}}]{Xie20}
{Xie}, L., {De Lucia}, G., {Hirschmann}, M., \& {Fontanot}, F. 2020, \mnras,
  498, 4327, \dodoi{10.1093/mnras/staa2370}

\bibitem[{{Xie} {et~al.}(2017){Xie}, {De Lucia}, {Hirschmann}, {Fontanot}, \&
  {Zoldan}}]{Xie17}
{Xie}, L., {De Lucia}, G., {Hirschmann}, M., {Fontanot}, F., \& {Zoldan}, A.
  2017, \mnras, 469, 968, \dodoi{10.1093/mnras/stx889}

\bibitem[{{Xie} {et~al.}(2024){Xie}, {De Lucia}, {Fontanot}, {Hirschmann},
  {Bah{\'e}}, {Balogh}, {Muzzin}, {Vulcani}, {Baxter}, {Forrest}, {Wilson},
  {Rudnick}, {Cooper}, \& {Rescigno}}]{Xie24}
{Xie}, L., {De Lucia}, G., {Fontanot}, F., {et~al.} 2024, \apjl, 966, L2,
  \dodoi{10.3847/2041-8213/ad380a}

\bibitem[{{Xu} {et~al.}(2023){Xu}, {Ouchi}, {Nakajima}, {Harikane}, {Isobe},
  {Ono}, {Umeda}, \& {Zhang}}]{xu23}
{Xu}, Y., {Ouchi}, M., {Nakajima}, K., {et~al.} 2023, arXiv e-prints,
  arXiv:2310.06614, \dodoi{10.48550/arXiv.2310.06614}

\bibitem[{{Yang} {et~al.}(2016){Yang}, {Wang}, {Wu}, {Fan}, {McGreer}, {Bian},
  {Yi}, {Yang}, {Ai}, {Dong}, {Zuo}, {Green}, {Jiang}, {Wang}, {Wang}, \&
  {Yue}}]{yang16}
{Yang}, J., {Wang}, F., {Wu}, X.-B., {et~al.} 2016, \apj, 829, 33,
  \dodoi{10.3847/0004-637X/829/1/33}

\bibitem[{{Yung} {et~al.}(2024){Yung}, {Somerville}, {Finkelstein}, {Wilkins},
  \& {Gardner}}]{yung24}
{Yung}, L.~Y.~A., {Somerville}, R.~S., {Finkelstein}, S.~L., {Wilkins}, S.~M.,
  \& {Gardner}, J.~P. 2024, \mnras, 527, 5929, \dodoi{10.1093/mnras/stad3484}

\bibitem[{{Zhu} {et~al.}(2022){Zhu}, {Becker}, {Bosman}, {Keating},
  {D'Odorico}, {Davies}, {Christenson}, {Ba{\~n}ados}, {Bian}, {Bischetti},
  {Chen}, {Davies}, {Eilers}, {Fan}, {Gaikwad}, {Greig}, {Haehnelt},
  {Kulkarni}, {Lai}, {Pallottini}, {Qin}, {Ryan-Weber}, {Walter}, {Wang}, \&
  {Yang}}]{zhu22}
{Zhu}, Y., {Becker}, G.~D., {Bosman}, S. E.~I., {et~al.} 2022, \apj, 932, 76,
  \dodoi{10.3847/1538-4357/ac6e60}

\bibitem[{{Zhu} {et~al.}(2024){Zhu}, {Becker}, {Bosman}, {Cain}, {Keating},
  {Nasir}, {D'Odorico}, {Ba{\~n}ados}, {Bian}, {Bischetti}, {Bolton}, {Chen},
  {D'Aloisio}, {Davies}, {Davies}, {Eilers}, {Fan}, {Gaikwad}, {Greig},
  {Haehnelt}, {Kulkarni}, {Lai}, {Puchwein}, {Qin}, {Ryan-Weber}, {Satyavolu},
  {Spina}, {Walter}, {Wang}, {Wolfson}, \& {Yang}}]{zhu24}
---. 2024, \mnras, 533, L49, \dodoi{10.1093/mnrasl/slae061}

\end{thebibliography}
\bibliographystyle{aasjournal}

\end{document}